\documentclass[11pt]{article}
\usepackage[margin=2cm]{geometry}
\usepackage{amsmath,amssymb,amsthm,
	    natbib,
	    tabularx,
	    graphicx,
	    booktabs,
	    color,
	    xspace,    
	  }
\usepackage{array}
\newcolumntype{L}[1]{>{\raggedright\let\newline\\\arraybackslash\hspace{0pt}}m{#1}}
\newcolumntype{C}[1]{>{\centering\let\newline\\\arraybackslash\hspace{0pt}}m{#1}}
\newcolumntype{R}[1]{>{\raggedleft\let\newline\\\arraybackslash\hspace{0pt}}m{#1}}
\usepackage{caption}

\usepackage{setspace}
\usepackage{textcomp}
\usepackage[mathscr]{eucal}
\usepackage{float}
\usepackage[colorlinks=true,citecolor=blue]{hyperref}
\usepackage{authblk}
\usepackage{indentfirst}
\usepackage{bbm}
\usepackage{mathtools,enumerate}

\newcommand{\uF}{F^{\vee}}
\newcommand{\lF}{F^{\wedge}}
\newcommand{\vX}{\vec{X}}
\newcommand{\vY}{\vec{Y}}
\newcommand{\vu}{\vec{u}}

\newcommand{\vv}{\vec{v}}
\newcommand{\Spm}{\mathcal{S}_d}

\newcommand{\R}{\mathbb{R}}

\renewcommand{\P}{\mathbb{P}}
\newcommand{\A}{\mathcal{A}}
\newcommand{\IN}{\vec{i}_n}
\newcommand{\FR}{\mathfrak{F}(F_1,\dots,F_d)}
\newcommand{\FRH}{\mathfrak{F}_d(F,\dots,F)}
\newcommand{\FRT}{\mathfrak{F}_2(F_1,F_2)}
\renewcommand{\qed}{~\hspace*{\fill}$\square$\par\medskip}
\renewcommand{\vec}[1]{\boldsymbol{#1}}

\newcommand{\rg}{\overset{\text{r}}{\sim}}
\newcommand{\cx}{ \Sigma_{cx}}
\newcommand{\E}{\mathbb{E}}

\newcommand{\N}{\mathbb{N}}

\newcommand{\lcx}{\leq_{\text{cx}}}

\newcommand{\var}{\text{var}}

\renewcommand{\P}{\mathbb{P}}

\newcommand{\Id}{\text{\upshape{Id}}}
\newcommand{\idc}{\text{\upshape{I}}}

\newcommand{\mycaptionminipage}[3][3=c,usedefault]{%
    \begin{minipage}[#3]{#1}%
        \ifthenelse{\equal{#3}{b}}{\captionsetup{aboveskip=0pt}}{}
        \ifthenelse{\equal{#3}{t}}{\captionsetup{belowskip=0pt}}{}
        \vspace{0pt}\centering\captionsetup{width=\textwidth} 
        #2%
    \end{minipage}%
}%
\def\laweq{\sim}

\newcommand{\cov}{\mathrm{cov}}

\newcommand{\id}{\text{I}}
\newtheorem{theorem}{Theorem}
\newtheorem{corollary}[theorem]{Corollary}

\newtheorem{proposition}[theorem]{Proposition}
\theoremstyle{definition}
\newtheorem{definition}{Definition}[section]
\newtheorem{example}{Example}[section]

\newtheorem{remark}{Remark}[section]

\numberwithin{equation}{section} \numberwithin{theorem}{section}

\newcommand{\upperRomannumeral}[1]{\uppercase\expandafter{\romannumeral#1}}
\title{Extremal dependence concepts}
\author[1]{Giovanni Puccetti}
\author[2]{Ruodu Wang}
\affil[1]{\small{\textit{Department of Economics, Management and Quantitative Methods,
 University of Milano, 20122 Milano, Italy}}}
\affil[2]{\small{\textit{Department of Statistics and Actuarial Science, University of Waterloo, Waterloo, ON N2L3G1, Canada}}}

\date{\small Journal version published in \emph{Statistical Science}, 2015, Vol. 30, No. 4, 485--517\\
Some corrections made in May 2020,  June 2020 and March 2025 by RW\footnote{Benjamin C\^ot\'e is thanked for some corrections.}}

\begin{document}
\maketitle
\numberwithin{equation}{section}
\begin{abstract}
The probabilistic characterization of the relationship between {two or more} random variables calls for a notion of dependence. Dependence modeling leads to mathematical and statistical challenges and  recent developments in extremal dependence concepts have drawn a lot of attention  to probability and its applications in several disciplines.
The aim of this paper is to review various concepts  of extremal positive and negative dependence, including several recently established results,
 reconstruct their history, link them to probabilistic optimization problems, and provide a list of open questions in this area.
While the concept of extremal positive dependence is agreed upon for random vectors of arbitrary dimensions, various notions of extremal negative dependence arise when more than two random variables are involved.
We review existing popular concepts of extremal negative dependence given in literature and introduce a novel notion, which in a general sense includes the existing ones as particular cases.
Even if much of the literature on dependence is   focused on positive dependence, we show that negative dependence plays an equally important role in the solution of many optimization problems.
While the most popular tool used nowadays to model dependence is that of a copula function, in this paper
we use the equivalent concept of a set of rearrangements. This is not only for historical reasons.
Rearrangement functions describe the relationship between random variables in a completely deterministic way, allow a deeper understanding of dependence itself, and have several advantages on the approximation of solutions in a broad class of optimization problems.
\end{abstract}
\noindent
{\small \\
Key words: Dependence Modeling, Rearrangement, Copulas, Comonotonicity, Countermonotonicity, Pairwise Countermonotonicity, Joint Mixability, $\Sigma$-countermonotonicity.   \\ \\
AMS 2010 Subject Classification: 60E05 (primary), 60E15, 91B30.
}
\\
\begin{center}
\textbf{Acknowlegments}
\end{center}\small

The authors are grateful to the Editor, the Associate Editor and four anonymous referees for their comprehensive reviews of an earlier version of this manuscript. The authors 
benefited from various discussions with Carole Bernard, Hans B\"uhlmann, Fabrizio Durante, Paul Embrechts, Harry Joe, Jan-Frederik Mai, Louis-Paul Rivest, Matthias Scherer and Johanna Ziegel. 

Giovanni Puccetti acknowledges the grant under the call PRIN 2010-2011 from MIUR within the project \emph{Robust decision making in markets and organisations}.
Ruodu Wang acknowledges support from the Natural Sciences
 and Engineering Research Council of Canada (NSERC Grant No. 435844). Both authors thank RiskLab and the FIM at ETH Zurich for support leading up to this research.
\\
\begin{center}
We acknowledge an intellectual debt to our colleague and friend  Ludger R\"uschendorf for his outstanding contribution to the field of Dependence Modeling and we would like to dedicate this paper to him.
\end{center}
\newpage

\section{Dependence as a set of rearrangements}

In the mathematical modelling of a random phenomenon or experiment, the quantity of interest is described by a measurable function $X: \Omega \to \R$  from a pre-assigned   atomless probability space $(\Omega, \A, \P)$ to some other measurable space, which will be chosen as the real line in what follows. This $X$ is called a \emph{random variable}. A random variable, if considered as an individual entity, is univocally described by its law (distribution)
$$F(x):=\P(X \leq x), \; x \in \R.$$ In the remainder, $X \laweq F$ indicates that $X$ has distribution $F$ while $X \laweq Y$ means that the random variables $X$ and $Y$ have the same law. We denote by $L^p,~ p\in[0,\infty)$ the set of random variables in $(\Omega, \A, \P)$ with finite $p$-th moment and by $L^\infty$ the set of bounded random variables. The notation $U[0,1]$ denotes the uniform distribution on the unit interval, while   $\idc(A)$   denotes the indicator function of the set $A \subset \A$. Throughout, we use the terms ``increasing" versus ``strictly increasing" for functions, and all functions  are assumed to be Borel measurable.
Most of the results stated in this paper have been given in the literature in different forms (even if in some cases we provide a self-contained proof), whereas Sections \ref{se:PND} and \ref{sec:4} contain original results.

Exploring the relationship between \emph{two or more} random variables
is crucial to stochastic modeling in numerous applications and requires a much more challenging statistical analysis.
Typically, a number of $d \geq 2$ random variables $X_1,\dots,X_d:  \Omega \to \R$ are gathered into a random vector $\vec{X}:=(X_1,\dots,X_d): \Omega \to \R^d$. A full model description of $(X_1,\dots,X_d)$ can be provided in the form of its  \emph{joint} distribution function
$$F(x_1,\dots,x_d):=\P(X_1 \leq x_1, \dots, X_d \leq x_d), \; x_1,\dots,x_d \in \R. $$
In this case, we keep the notation $\vec{X} \laweq F$ and the univariate distributions $F_j(x):=\P(X_j \leq x), j=1,\dots,d,$ are referred to as the \emph{marginal distributions} of $F$.
When $d \geq 2$, the full knowledge of the individual models $F_1,\dots,F_d$ is not sufficient to determine the joint distribution $F$. In fact, the set $\FR$ of all possible distributions $F$ sharing the same marginals $F_1,\dots,F_d$ typically contains infinitely (uncountably) many elements. $\FR$ is called a \emph{Fr\'echet class}.
We also say that a Fr\'echet class $\FR$ supports a random vector $\vec{X}$ if the distribution of $\vec{X}$ is in $\FR$; equivalently we write $\vX \in_d \FR$ if $X_j\laweq F_j$, $j=1,\dots,d$. More details on the set $\FR$ can be found in~\citet[Chapter~3]{hJ97}.

In order to isolate a single element in $\FR$ one needs to establish the dependence relationship among a set of given marginal distributions. In what follows, we use the notion of a \emph{rearrangement} to describe dependence among a set of random variables.

\begin{definition}
Let $f,g: [0,1] \to [0,1]$ be measurable functions. Then $g$ is called a rearrangement of $f$, denoted by $g \rg f$ if $g$ and $f$ have the same distribution function under $\lambda$, the restriction of Lebesgue measure to $[0,1]$. Formally, $g \rg f$ if and only if
$$ \lambda[ g \leq v ]= \lambda[ f \leq v ] \, \text{ for all $\, v \in [0,1]$}.$$
\end{definition}

Given a measurable function $f:[0,1] \to [0,1]$, there always exists a decreasing rearrangement $f_* \rg f$ and an increasing rearrangement $f^* \rg f$, defined by
$$
f_*(u):=F^{-1}(1-u), \text{ and } f^*(u):=F^{-1}(u),
$$
where $F(v):=\lambda \{u : f(u)  \leq v\}$. In the above equation and throughout the paper, the \emph{quasi-inverse} $F^{-1}$ of a distribution function $F: A \subset \R \to [0,1]$ is defined as
\begin{equation}\label{eq:quasi-inverse}
F^{-1}(u):=\inf \left\{x\in A:F(x)\geq u \right\},~ u \in (0,1],
\end{equation}
and $F^{-1}(0):=\inf \left\{x\in A:F(x)> 0 \right\}$.

\begin{figure}[h]
\begin{center}
\scalebox{.54}{\includegraphics{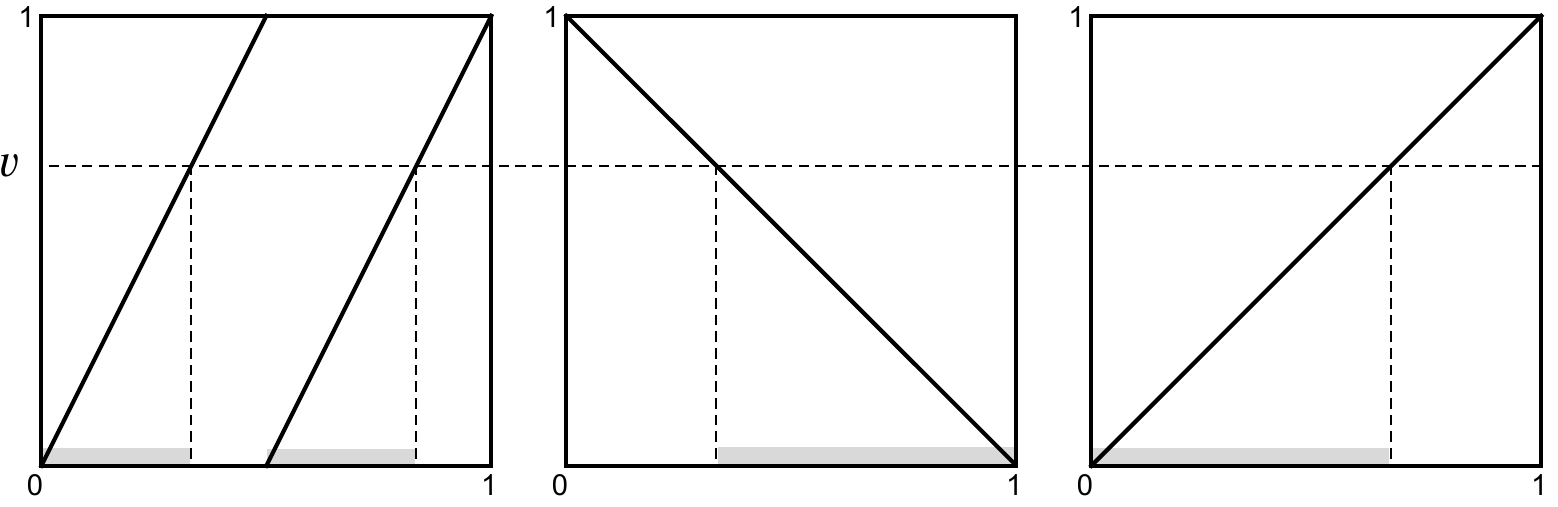}}
\caption{\small  A function $f:[0,1] \to [0,1]$ (left), its decreasing rearrangement $f_*$ (centre) and its increasing rearrangement $f^*$ (right). The grey areas represent the sets $\{f \leq v\}$ (left) ,$\{f_* \leq v\}$ (center) and $\{f^* \leq v\}$ (right) which all have the same $\lambda$-measure for any $v \in [0,1]$.}
\label{fi:1}
\end{center}
\end{figure}

In Figure~\ref{fi:1} we illustrate a function $f:[0,1] \to [0,1]$ (left) together with its decreasing (center) and increasing (right) rearrangements. Note that any rearrangement function $f$ in Figure~\ref{fi:1} is itself a  rearrangement of $\Id$, the identity function on $[0,1]$. We have that $f \rg \,\Id$ if and only if $f(U) \laweq U[0,1]$ for any $U \laweq U[0,1]$. In some of the literature, rearrangements are known under the name of \emph{measure-preserving transformations}; see for instance \citet{raV79} and \citet{DS12}.

It is well known that a random variable $X_j$ with distribution $F_j$ has the same law as the random variable $F_j^{-1}(U)$, where $U \laweq U[0,1]$. This of course remains true if one replaces $U$ with $f(U)$, $f \rg \Id$.
Analogously, each component $X_j$ of a  random vector $(X_1,\dots,X_d)$ has the same law as $F_j^{-1} \circ f_j(U_j)$, for some $f_j \rg \,\Id$ and $U_j\sim U[0,1]$.
For $d \geq 2$, different $d$-tuples of rearrangements $f_1,\dots,f_d$ generate random vectors with the same marginal distributions but different interdependence among their components. Conversely, any dependence among the univariate components of a $d$-dimensional random vector can be generated by using a suitable set of $d$ rearrangements. The following theorem reveals the non-trivial fact that the random variables $U_j$ can be replaced by a single random variable $U$.

\begin{theorem}\label{pr:1}
The following statements hold.
\begin{enumerate}[(a)]
\item If $f_1,\dots,f_d$ are $d$ rearrangements of $\, \Id$, and  $F_1,\dots,F_d$ are $d$ univariate distribution
functions, then
$$
\left(F_1^{-1} \circ f_1(U), \dots, F_d^{-1} \circ f_d(U) \right)
$$
is a random vector with marginals $F_1,\dots,F_d$.
\item Conversely, assume $(X_1,\dots,X_d)$ is a random vector
with joint distribution $F$ and marginal distributions $F_1,\dots,F_d$.
Then there exist $d$ rearrangements $f_1,\dots,f_d$ of $\,\Id$ such that
\begin{equation}\label{eqofpr:1}
(X_1,\dots,X_d) \laweq \left(F_1^{-1} \circ f_1(U), \dots, F_d^{-1} \circ f_d(U) \right),
\end{equation}
where $U$ is any $U[0,1]$ random variable.
\end{enumerate}
\end{theorem}
\emph{Proof of (a).} Since $f_j \rg \, \Id$, $f_j(U)$ is uniformly distributed on $[0,1]$ and, consequently,  $F_j^{-1} \circ f_j(U)$ has distribution $F_j$. As a result, the random vector $\left(F_1^{-1} \circ f_1 (U), \dots, F_d^{-1} \circ f_d (U)\right)$ has marginal distributions $F_1,\dots,F_d$.
\\

\emph{Proof of (b) in case $F_1,\dots,F_d$ are continuous.} Let $U \laweq U[0,1]$. Without  loss of generality, we take $U$ as a random variable on $([0,1], \mathcal{B}, \lambda)$, where $\mathcal{B}$ denotes the Borel $\sigma$-algebra of $[0,1]$. Let $(X_1,\dots,X_d)$ be a random vector having joint distribution $F$ with marginal distributions $F_1,\dots,F_d$. Let $C$ be the distribution of $(F_1(X_1),\dots,F_d(X_d))$. Since $F_j(X_j) \laweq U[0,1]$  if $F_j$ is continuous, $C$ is a distribution on $[0,1]^d$ with $U[0,1]$ marginals. Let $V_C$ be the measure on $[0,1]^d$ induced by $C$.

The idea of the proof is to find a one-to-one measurable mapping $f:([0,1],\lambda) \to ([0,1]^d,V_C) $ which is measure-preserving and whose inverse is also measure-preserving. In fact, we need $f$ such that for every $B^d \in \mathcal{B}([0,1]^d)$, the Borel $\sigma$-algebra of $[0,1]^d$, and for every $B \in \mathcal{B}$ we have that
$$V_C(B^d)=\lambda \circ f^{-1}(B^d) \; \text{ and } \; \lambda(B)=V_C\circ f(B).$$
In order to define such $f$, we first take a one-to-one measurable function $\phi:[0,1]^d \to [0,1]$ such that $\phi^{-1}$ is also measurable. The existence of such $\phi$ is implied by Theorem~2.12 in~\citet{kP67}.
Let $G$ be the distribution function associated to the measure $V_C \circ \phi^{-1}$ and define $f:[0,1] \to [0,1]^d$ as  $f=\phi^{-1} \circ G^{-1}$.   We have that $G^{-1}(U) \laweq G$, implying that $f(U)=\phi^{-1} \circ G^{-1}(U) \laweq C$. If $f(U)=(f_1(U),\dots,f_d(U)) \laweq C$ then $\left(F_1^{-1} \circ f_1(U), \dots, F_d^{-1} \circ f_d(U) \right) \laweq F$. To conclude the proof, it remains to show that the above defined $f_j$'s are rearrangements of $\Id$, but this is directly implied by the fact that  $(f_1(U),\dots,f_d(U)) \laweq C$ and the marginals of $C$ are uniform.
\\

\emph{Proof of (b) in case $F_1,\dots,F_d$ are arbitrary.} If the $F_j$'s have jumps one can proceed as for continuous marginals by replacing $F_j$ by $\hat{F}_j$ defined as
$$
\hat{F}_j(x)=F_j(x-)+\left(F_j(x+)-F_j(x-)\right)U_x,
$$
where $U_x$ are uniformly distributed on $[0,1]$ and independent for all (countable) discontinuity points $x$ of $F_j$. In fact, instead of the distribution $F_j$, one uses in the proof above its \emph{distributional transform} as defined in~\citet{lR09}: the value of $F_j$ is randomized over the length of the jumps. \qed

\begin{remark}\label{re:1}
We make the following remarks about Theorem~\ref{pr:1}.
\begin{enumerate}[(i)]
\item The representation in~\eqref{eqofpr:1} is equivalent to the one given in Theorem~5.1 in~\citet{wW76} and Lemma~1 in~\citet{lR83}.  The proof of  Lemma~1 in~\citet{lR83} refers the reader to~Lemma 2.7 in~\citet{wW76}, which is based on Theorem 2.12 in~\citet{kP67}. The proof of~\citet{wW76} uses similar arguments and is based on Sklar's theorem. It is shown in~\citet{vR52} and~\citet{kP67} that two Borel subsets of complete separable metric spaces are isomorphic if and only if they have the same cardinality. This allows for the identification of an isomorphism $\phi:[0,1]^d \to [0,1]$ as in the above proof.
Apart from this last mentioned key result, the proof of Theorem~\ref{pr:1} presented here is self contained.
\item A still different proof of Theorem~\ref{pr:1} using the language of copulas (see Definition~\ref{de:copula}) can be found in \citet[Theorem~3.1]{KMS08}. As stressed in the latter reference, the set of $d$ rearrangements in~\eqref{eqofpr:1} is unique up to a rearrangement of $\, \Id$. In fact, \eqref{eqofpr:1} holds true  even if $f_1,\dots,f_d$ are replaced  by $f_1\circ \psi,\dots,f_d \circ \psi$, where $\psi \rg \Id$.

\item The notation $C$ for the distribution of the vector $(F_1(X_1),\dots,F_d(X_d))$ in the above proof is not unintended: $C$ is a \emph{copula} under the terminology introduced in Definition~\ref{de:copula} below.
\item Theorem~\ref{pr:1} holds true also in the case that a different definition of quasi-inverse is used in~\eqref{eq:quasi-inverse}. Quasi-inverses are generalizations of the inverse of a function that are defined even when the function is not strictly monotone. For a distribution function $F$ and $y \in [0,1]$ let  $F^{\leftarrow}(y)=\{x  : F(x) = y \}$. If $F$ is strictly increasing, then the cardinality of $F^{\leftarrow}(y)$ is always a singleton and one can simply set $F^{\leftarrow}(y) := F^{-1}(y)$. If the cardinality of $F^{\leftarrow}(y)$ is more than one, one has to somehow choose between the various elements of $F^{\leftarrow}(y)$, thus allowing for different notions of quasi-inverse which all coincide except on at most a countable set of discontinuities. The notion of quasi-inverse used in this paper and defined in~\eqref{eq:quasi-inverse} is the left-continuous one; see also~\citet{EH13} for a comprehensive investigation of its properties.
\end{enumerate}
\end{remark}

On the basis of Theorem~\ref{pr:1}, it is natural to identify the structure of dependence among the components of a random vector with a set of $d$ rearrangements of the identity function on $[0,1]$. An equivalent concept used to model the structure of dependence in a random vector is the notion of a \emph{copula} function. Since their introduction in the late 50s, copulas (or copul\ae) have gained a lot of popularity in several fields of applied probability and statistics like hydrology, finance, insurance and reliability theory.
Especially in quantitative risk management, copulas present a widely used tool for market and credit risk, risk aggregation, portfolio selection, etc. Textbook introductions to copulas can be found in~\citet{hJ97, hJ14},~\citet{rN06} and~\citet{DS15} while more application-oriented references are~\citet{MNFE05} and~\citet{CTA}.
\begin{definition}\label{de:copula}
A copula $C$ is a distribution function on $[0,1]^d$ with $U[0,1]$ marginals.
\end{definition}

Using Theorem~\ref{pr:1}, we can immediately see that the notion of a copula is equivalent to a set of $d$ rearrangements of the identity function on $[0,1]$. The following corollary of Theorem~\ref{pr:1} is essentially a rewriting of
Theorem~3.1 in~\citet{KMS08}.

\begin{corollary}\label{co:1}
The function $C$ is a copula if and only if there exists a set of $d$ rearrangements $f_1,\dots,f_d$ of $\,\Id$ such that
\begin{equation}\label{eq:corgmrep}
(f_1(U),\dots,f_d(U)) \laweq C,
\end{equation}
where $U \laweq U[0,1]$. We also note that the representation of a copula via $d$ rearrangements $f_1,\dots,f_d$ as in~\eqref{eq:corgmrep} is not unique as we have
$$
(f_1(U),\dots,f_d(U)) \laweq (f_1\circ \psi(U),\dots,f_d\circ \psi(U)).
$$
for any rearrangement $\psi$ of $\Id$. Consequently, when $f_1$ in~\eqref{eq:corgmrep} is one-to-one, we can always set $f_1= \Id$.

\end{corollary}

Even if the equivalent concept of a rearrangement has been used to model dependence much earlier than the introduction of copulas (see the Historical Remark at the end of Section~\ref{se:ppd}), nowadays copulas are considered a \emph{standard} tool to model dependence at least in the above mentioned fields.
The popularity of copula-based models is mainly due to their mathematical interpretation
which is fully captured by Sklar's theorem.

\begin{theorem}[Sklar's theorem]
Given a copula $C$ and $d$ univariate marginals $F_1,\dots,F_d$, one can always define a distribution function $F$ on $\mathbb{R}^d$ having these marginals by
\begin{equation}\label{eq:sklar}
F(x_1,\dots,x_d)= C(F_1(x_1),\dots,F_d(x_d)), \; x_1,\dots,x_d \in \mathbb{R}.
\end{equation}
Conversely, it is always possible to find a copula $C$ coupling
the marginals $F_j$ of a fixed joint distribution $F$ through the above expression~\eqref{eq:sklar}.
For continuous marginal distributions, this copula in \eqref{eq:sklar} is unique.
\end{theorem}

Because of its importance in applied probability and statistics, Sklar's theorem has received a lot of attention and has been proved several times with different techniques. In our opinion, the most elegant proof of the theorem is the one provided in~\citet{lR09} based on distributional transforms. Sklar's theorem was first announced, but not proved, in~\citet{aS59}; for the two-dimensional case $d=2$, a complete proof only appeared in~\citet{SS74}. For a complete history of Sklar's theorem (and all its proofs as well as a new one) see~\citet{DSS12}.

The equality~\eqref{eq:sklar} illustrates a way of isolating  the description of the dependence structure, given by a copula function $C$, from the distributions $F_1,\dots,F_d$ of the marginal components of a random vector. Via Sklar's theorem, the mathematical construction, statistical estimation and the simulation of a complex multivariate model were made more accessible to the broader audience (see, for instance, some of the earliest applied papers~\citet{CR99} and~\citet{EMNS02}). Various methodologies exist for estimating dependence parameters in a family of copulas; see for instance Chapter 6 in~\citet{MS14}. On the other side, copulas possess a number of deficiencies, especially when they are used in higher dimensions; see~\citet{tM06} and~\citet{MS13}.

In the remainder of this paper, we will use rearrangements to model the structure of dependence of random vectors. This is not only for historical reasons. Looking at dependence as a set of deterministic functions has several advantages for the solution of some specific optimization problems and allows for obtaining a deeper understanding of the dependence itself.

\begin{figure}[h]
\begin{center}
\scalebox{.36}{\includegraphics{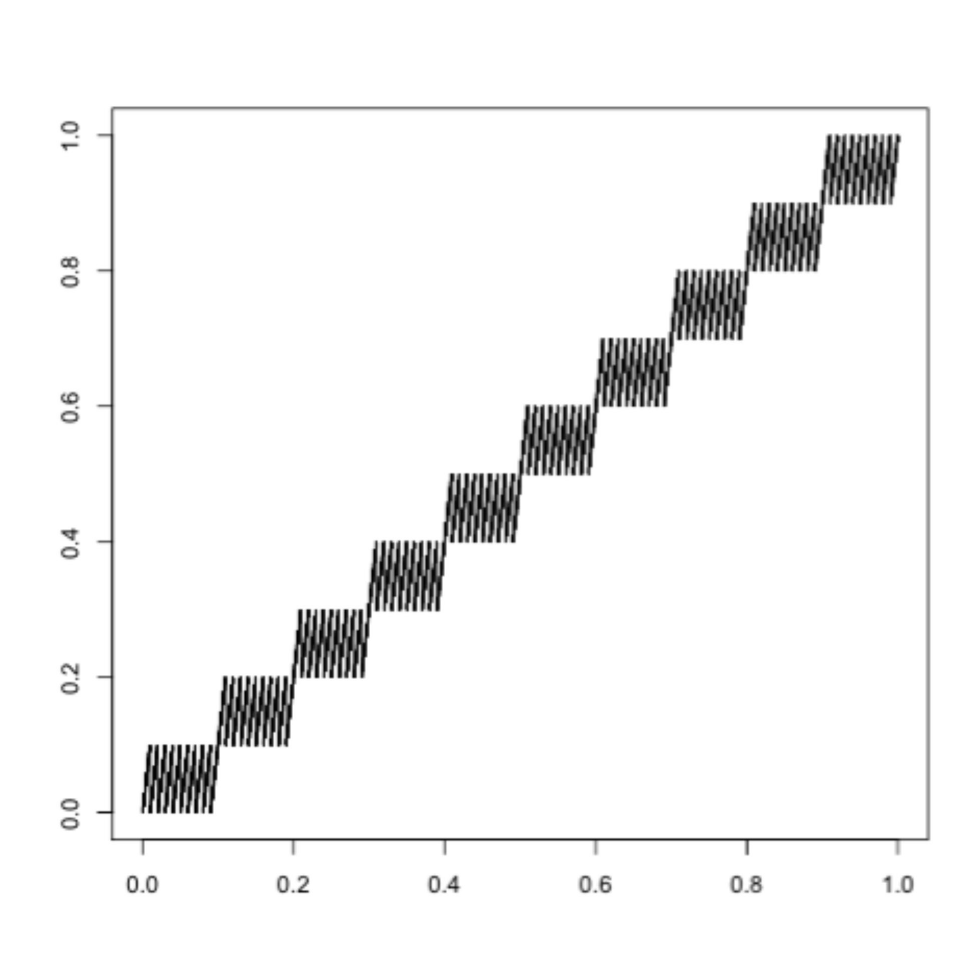}\includegraphics{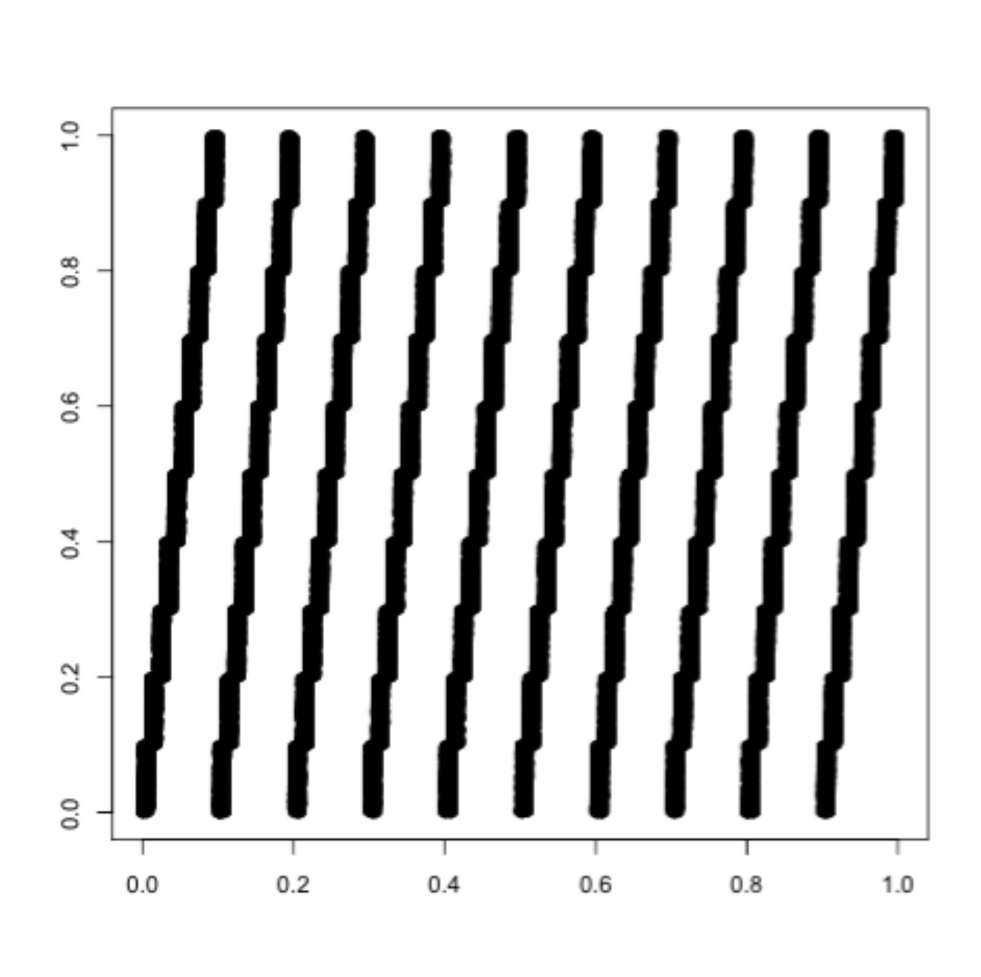}}
\caption{\small A set of two rearrangements $f_1$ (left) and $f_2$ (right) defining a two-dimensional  random vector with independent components.}\label{fi:indeprearrg}
\end{center}
\end{figure}

\begin{remark}
A direct consequence of Theorem~\ref{pr:1} is that any random vector can be seen as a deterministic function of a single random factor. In principle, in order to generate (simulate) an observation for a $d$-variate random vector, we need only to sample a point from the unit interval. This last assertion includes a random vector with independent components as a particular case. For example if we write $u \in [0,1]$ in decimal form, e.g.  $u=0.u_1u_2u_3\dots$ (in case $u$ has more than one representation we choose the one with infinitely many 0's), define $f_1(u)=0.u_1u_3\dots$ and $f_2(u)=0.u_2u_4\dots$.  For $U \sim U[0,1]$, $f_1(U)$ and $f_2(U)$  are then  independent and $U[0,1]$-distributed random variables.
These rearrangement functions $f_1$ and $f_2$ are illustrated in Figure~\ref{fi:indeprearrg}.
\end{remark}

By~\eqref{eqofpr:1}, each component of an arbitrary random vector can be seen as a function of a common random factor. However, this does not imply that the knowledge of a single component implies the knowledge of the others. For example, take the random vector $(X_1,X_2):=(U,f(U))$ where $f$ is the rearrangement given in Figure~\ref{fi:1} (left) and $U \laweq U[0,1]$. The second random component $X_2$ is \emph{completely dependent} on $X_1$ (i.e. takes a.s. only one value for each value of $X_1$), but not vice versa. This occurs because the rearrangement $f$ is not one-to-one. There exists an interesting class of one-to-one rearrangements that, under the copula taxonomy, are known under the name of \emph{shuffle of Min}.

\begin{definition}\label{de:som}
A copula $C$ is a \emph{shuffle of  Min} if there exist $d$ one-to-one, piecewise continuous rearrangements $f_1,\dots,f_d$ of $\,\Id$ such that
$
(f_1(U),\dots,f_d(U)) \laweq C,
$
where $U \laweq U[0,1]$.
\end{definition}

Shuffle of Mins were originally introduced in~\citet{MST92b} in the two-dimensional case as copulas having as support a suitable rearrangement of the mass distribution of a particular copula, called the Min copula (see~\eqref{eq:mincopula} below) -- hence the name. The multivariate definition given here is based on Corollary~2.3
of~\citet{DS12} and clearly illustrates that shuffle of Min's express a special type of dependence, called \emph{mutually complete dependence} in~\citet{hL63}, under which each component of a random vector is completely dependent on any of the others. The requirement of piecewise continuity of the rearrangements in Definition~\ref{de:som} is introduced only for historical reasons to match the bivariate definition given in~\citet{MST92b}, but it is not really necessary.

Mutually completely dependent discrete random vectors can be represented in terms of a matrix.
For a given $(n \times d)$-matrix $\vec{X}=(x_{i,j})$, we define $\mathcal{P}(\vec{X})$ as the set of all $(n \times d)$-matrices obtained from $\vec{X}$ by rearranging the elements within a number of its columns in a different order, that is
$$
\mathcal{P}(\vec{X})=\left\{\tilde{\vec{X}}=(\tilde{x}_{i,j}) :\tilde{x}_{i,j}=x_{\pi_j(i),j},~\pi_1,\dots, \pi_d \text{ are permutations of $\{1,\dots,n\}$}  \right\}.
$$
We call each matrix in $\mathcal{P}(\vec{X})$ a \emph{rearrangement matrix}.

\begin{figure}[t]
\begin{center}
\scalebox{.18}{\includegraphics{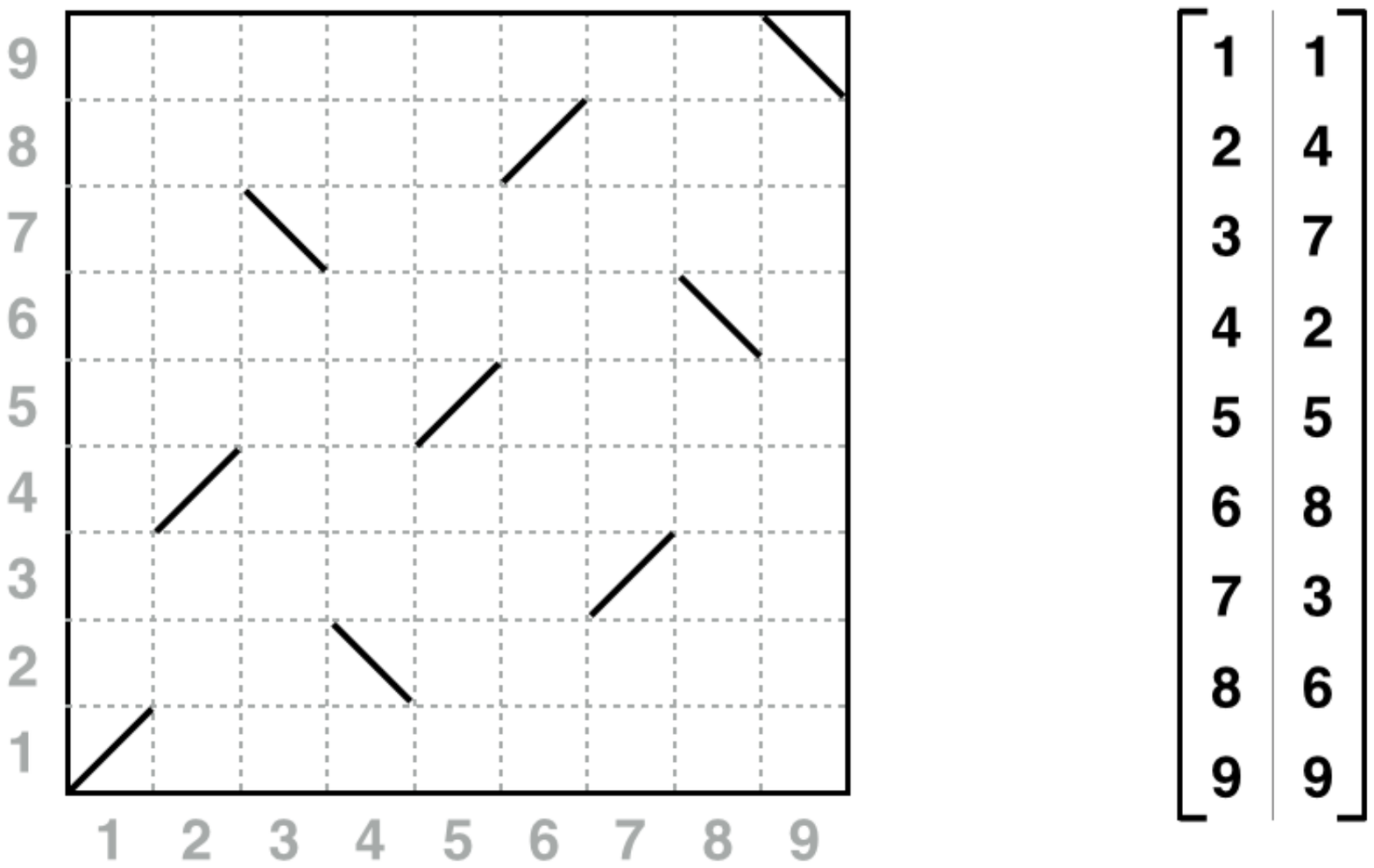}}
\caption{\small \small The support of the copula (left) of the third element of the sequence, as described in~~\citet{KS78}, approximating a independent pair of random variables. On the right part of the figure, we provide a rearrangement matrix representing a discrete bivariate distribution with the same copula and marginal distributions uniformly distributed over the first nine integers.}
\label{fi:comonotonic.}
\label{fi:indeprearrg2}
\end{center}
\end{figure}

Any rearrangement matrix $\vec{\tilde{X}} \in \mathcal{P}(\vec{X})$ can be seen as the support of a discrete, $d$-variate distribution giving probability mass $1/n$ to each one of its $n$ row vectors. Under this view, any such $\vec{\tilde{X}}$ has the same marginal distributions $F_1,\dots,F_d$, where for each $j$, $F_j$ is uniformly distributed over the $n$ real values $x_{i,j},~ 1 \leq i \leq n$, assumed distinct for convenience. Therefore, any rearrangement matrix represents a different dependence structure coupling the fixed discrete marginal distributions $F_j$.
In particular, each $\vec{\tilde{X}}$  has a copula belonging to the class of shuffles of Min and represents a mutually complete dependence between its marginal components.
The class of shuffle of Min copulas has been proved to be dense in the class of copulas endowed for instance with the $L^{\infty}$-norm, and this result again does not need the  continuity assumption in Definition~\ref{de:som}.  In fact, any copula can be considered as a generalization to the infinite-dimensional space of such rearrangement matrices (see for instance~\citet{KMMS06}).
Equivalently stated, any dependence structure can be approximated by the copula of a rearrangement matrix for $n$ large enough and, in particular, this result implies that any pair of independent random variables can be approximated by a sequence of pairs of mutually completely dependent random variables. An early example of this fact can be found in~\citet{KS78}, where the approximation sequence is explicitly given (the copula of the third element of the sequence and the corresponding rearrangement matrix are illustrated in Figure~\ref{fi:indeprearrg2}).
The matrix representation described above and the corresponding density result turn out to be extremely useful to approximate the solution of a broad variety of optimization problems in Section~\ref{sec:4}.
For more details on the link between the idea of a rearrangement and copulas as dependence structures, we refer to~\citet{lR83}. For a review of known results on the approximation of copulas via shuffles of Mins and via the more general concept of \emph{shuffle of copulas}, see~\citet{DS12} . We remark that the $L^\infty$-norm between copulas is sometimes argued as not being a natural norm between probability measures. More interesting types of convergence are investigated in~\citet{DS12} and~\citet{ST14}. For a insight on not necessarily bijective measure-preserving transformations, we refer to~\citet{TS13}.

\paragraph{Scope of the paper}
In what follows, we review various concepts of extremal positive and negative dependence.  The term \emph{extremal} used in the title does not refer to the field of \emph{multivariate extreme value theory} (MEVT), which is not the focus of this paper.
Indeed, so-called extreme value copulas (such as the Gumbel family of copulas) can be used to model strong positive dependence; see for instance~\citet{GS10}. However they are not capable of modeling any negative dependence, as shown in~\citet{MO83}. This is a consequence of the significant mathematical asymmetry between  extremal positive dependence and  extremal negative dependence, as we shall illustrate in Sections \ref{se:ppd} and \ref{se:pndhd}.
Furthermore, in this paper we focus on \emph{concepts of dependence} rather than \emph{statistical methods for dependence}; however many examples useful in statistics will be provided along the way.

\clearpage
\section{Extremal positive dependence}\label{se:ppd}
Much of the literature on dependence modeling is focused on the notion of an  \emph{extremal positive dependence structure}.
The word \emph{extremal} used in this paper refers to dependence structures leading to extremal values under certain criteria which will be
specified later.
Extremal positive dependence concepts are typically defined by requiring that all the components of a random vector behave similarly, e.g. can be expressed as increasing functions of a common factor.
This scenario can be interpreted as the ordinary perception of a catastrophe or  extreme natural event:
the intensity of an earthquake/tsunami, a flooding, a famine, a war or an epidemic can be seen as a single random variable which affects in the same direction  people, properties and economical factors confined to the same geographic area. The higher the magnitude of the catastrophe, the higher the damage for \emph{all} the individuals involved. Analogously, in a financial market all assets might be influenced by a unique economic shock (e.g. a terroristic attack) and react similarly.
Random variables resembling this type of behaviour are called \emph{commonly monotonic}: high values for one of them imply high values for all the remaining and vice versa; in one word: \emph{comonotonic}.

\begin{definition}\label{de:comvec}
A random vector $(X_1,\dots,X_d)$  is said to be \emph{comonotonic}
if there exists a single rearrangement $f \rg \,\Id$ such that
$$
(X_1,\dots,X_d) \laweq \left(F_1^{-1} \circ f(U), \dots, F_d^{-1} \circ f(U) \right),
$$
where $U \laweq U[0,1]$. As $f(U) \laweq U[0,1]$, the rearrangement function $f$ can always be chosen as $f=\Id$. Thus, the components of a $\R^d$-valued comonotonic random vector are a.s. increasing functions of a common random factor $U$.
\end{definition}

Comonotonic random vectors represent the solution of a wide class of optimization problems. In particular, they are well known to maximize the expectation of a supermodular function  under a condition of mean-compatibility over the set $\FR$. 

\begin{definition}
A function $c:\R^d \to \R$ is \emph{supermodular} if
\begin{equation}\label{eq:supermodular}
c(\vu \land \vv) + c(\vu \lor \vv) \geq c(\vu) +c(\vv),\ \text{for
all } \vu,\vv \in \R^d,
\end{equation}
where $\vu \land \vv$ is the component-wise
minimum of $\vu$ and $\vv$, and $\vu \lor \vv$ is the component-wise maximum of $\vu$ and $\vv$.
If~\eqref{eq:supermodular} holds with a strict inequality for all unordered couples of distinct  $\vu,\vv \in \R^d$, then
the function $c$ is \emph{strictly supermodular}. Simple examples of supermodular functions include $c(\vec{x})=f(x_1+\dots+x_d)$
for $f$ convex, and $c(\vec{x})=\prod_{j=1}^{d}x_j$. The reader is referred to \citet[Chapter 6.D]{MOB11} for more examples and properties in the class  $\Spm$ of supermodular functions.
A supermodular function $c$ is \emph{mean-compatible} with $(X_1,\dots,X_d)$ if 
$
|c(\vec{x})|\le \sum_{j=1}^d b_j(x_j) 
$
for some positive functions $b_1,\dots,b_d$ such that $\E[b_j(X_j)]<\infty$ for each $j$.
\end{definition}
Clearly, any bounded supermodular function is mean-compatible with any $(X_1,\dots,X_d)$.
We omit ``with $(X_1,\dots,X_d)$'' when the marginals are clear. 

\begin{theorem}\label{th:comch}
For a random vector $(X_1,\dots,X_d)$ with joint distribution function $F$, the following statements {\upshape{(a)-(e)}} are equivalent:
\begin{enumerate}[\upshape(a)]
\item   $(X_1,\dots,X_d)$ is comonotonic;
\item $F$ is given by
\begin{equation}\label{eq:comdf}
F(x_1,\dots,x_d)=\uF_d(x_1,\dots,x_d) :=\min \{F_1(x_1),\dots,F_d(x_d) \}, \; x_1,\dots,x_d \in  \R,
\end{equation}
where $F_j$ is the marginal distribution of $X_j$, $j=1,\dots,d$;
\item $F \ge G$ on $\R^d$ for all $G\in \FR$; 
\item  for all supermodular functions $c: \R^d \to \R$ that are mean-compatible with $(X_1,\dots,X_d)$, we have that
\begin{equation}\label{eq:maxc}
\E \left[ c(X_1,\dots,X_d) \right]=\sup \left\{ \E \left[
c(Y_1,\dots,Y_d) \right]: Y_j \laweq X_j,~ j=1,\dots,d \right\};
\end{equation} 
\item  there exists a strictly supermodular function $c: \R^d \to \R$ that is mean-compatible with $(X_1,\dots,X_d)$ such that   \eqref{eq:maxc} holds.
\end{enumerate} 
\end{theorem}
\begin{proof}  (a) $\Leftrightarrow$ (b): follows from elementary probability. A self-contained proof can be found in Theorem~2 in the paper~\citet{DDGKV02}, or  in separate parts in~\citet{lR80}.
(b) $\Leftrightarrow$ (c) can be shown using a standard argument as used in \citet{wH40}.
(c) $\Rightarrow$ (d) follows from Theorem~5 in~\citet{aT80}; see the Historical Remark below for a complete history of this result. 
(d) $\Rightarrow$ (c) follows by choosing the supermodular function $\mathbf x\mapsto \idc_{\{\mathbf x\le \mathbf y\}}$ for each $\mathbf y\in \R^d$.
(d) $\Rightarrow$ (e) is obvious. 
(e) $\Rightarrow$ (c): this can be easily proven by discrete approximation and reduction in the discrete case to a classical discrete rearrangement theorem of~\citet{HLP34}; we give more details in the Historical Remark below. In the discrete case, if $(X_1,\dots,X_d)$ is not comonotonic then we can change the order of two elements (while keeping the others) thus obtaining a larger value of the target function by strict supermodularity. 
\end{proof}

\begin{remark}\label{rem:21} We make the following remarks about Theorem~\ref{th:comch}.
\begin{enumerate}[(i)]
\item It is quite easy to show that point (d) in Theorem~\ref{th:comch} cannot be extended to non-supermodular functionals; see  the counterexample given in the proof of Theorem~2.5 in~\citet{PuS10}. Note from (c) and (e) in Theorem~\ref{th:comch} that the maximization of the expectation of supermodular functions and the maximization of the joint distribution function are equivalent within a  Fr\'echet class; see \cite{aT80}.
\item Any function $c:\R^d \rightarrow \R$ which can be expressed as $c(x_1,\dots,x_d)=f(x_1+\cdots+x_d)$, where $f:\R\rightarrow \R$ is a convex function, is supermodular.
This choice of $c$ relates to many applications of particular interest in Finance, Economics and Insurance, where the sum $X_1+\cdots+X_d$ is often interpreted as an aggregation or a risk pooling, and $f$ can be chosen so as to determine risk measurement, utility or insurance premiums. Comonotonic random vectors maximize the expectation of such functions over a Fr\'echet class, and hence they are typically viewed to have the \emph{most dangerous dependence structure} for individual components in a portfolio. Later in Section \ref{se:pndhd} we will show that this property is crucial for characterizing extremal negative dependence concepts, where a minimizer of the expectation of all supermodular functions does not exist in general.
\end{enumerate}
\end{remark}

\begin{figure}[h]
\begin{center}
\scalebox{.26}{\includegraphics{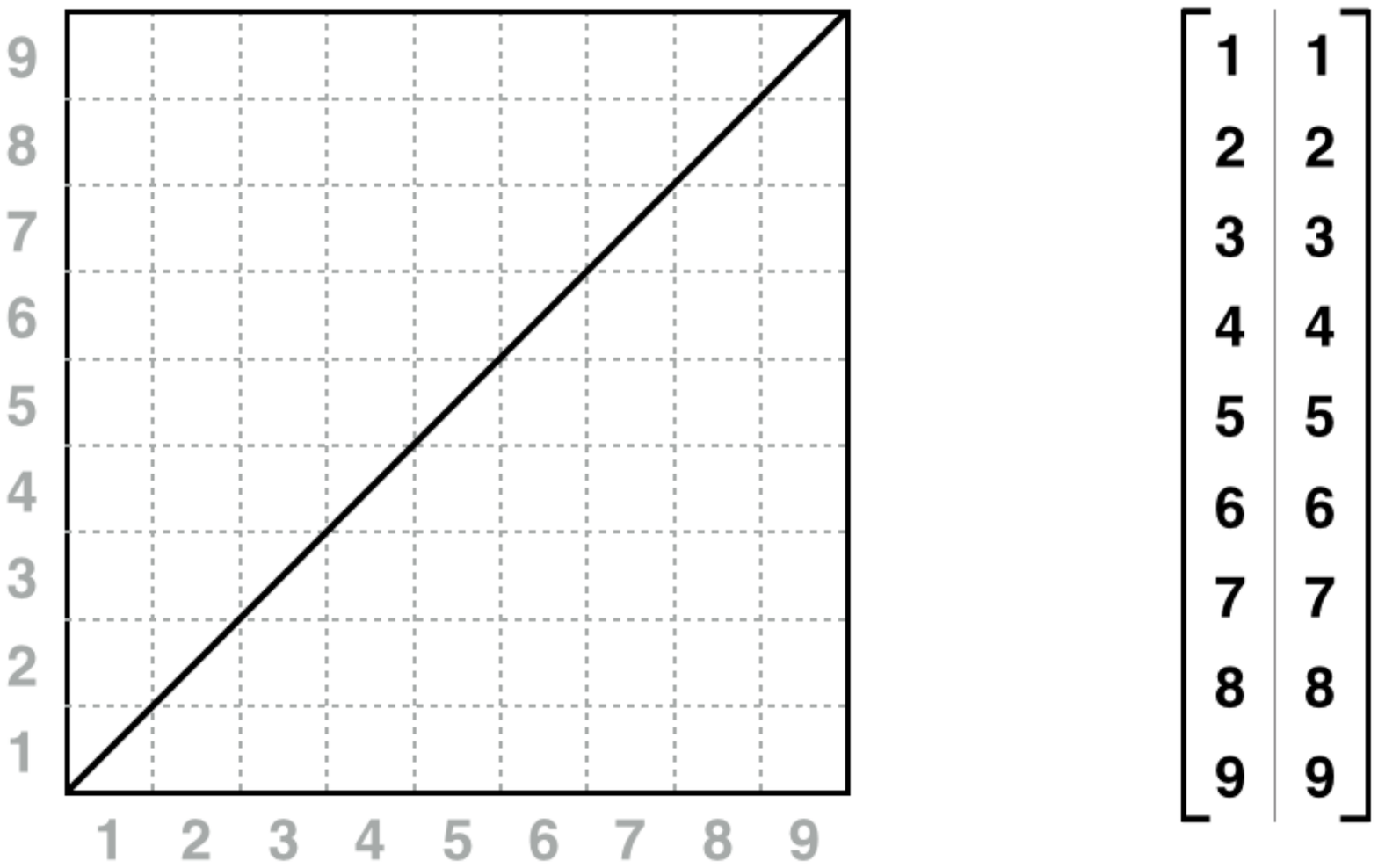}}
\caption{\small The support of the Min copula $M_2$ (left) and a
rearrangement matrix (right) representing a discrete bivariate distributions with copula $M_2$ and marginal distributions uniformly distributed over the first nine integers.}
\label{fi:comonotonic}
\end{center}
\end{figure}

\begin{remark}[The copula $M_d$]
According to Definition~\ref{de:comvec}, a comonotonic dependence structure is represented by a set of identical rearrangements. From~\eqref{eq:comdf}, it is also evident that a random vector is comonotonic if and only if it has copula $M_d$, where $M_d$ is the so-called \emph{Min} copula defined as
\begin{equation}\label{eq:mincopula}
M_d(u_1, \dots ,u_d)=\min\{u_1,\dots,u_d\}.
\end{equation}
The Min copula represents a benchmark in statistical modelling as it is the copula representing perfect positive dependence. Its support consists of the main diagonal of the unit square and, being itself a (trivial) shuffle of Min, it is a copula of any rearrangement matrix having all the columns similarly ordered; see Figure~\ref{fi:comonotonic}.
The most commonly applied families of parametric copulas such as the Clayton, Frank, Gumbel and Gaussian families include the Min copula $M_d$ as a limiting case, see for instance Table~4.1 in~\citet{rN06}.
\end{remark}

\clearpage
Denote by $\mathcal{X}^+$ the set of all a \textbf{comonotonic} random vectors having marginal distributions $F_1,\dots,F_d$. The following properties hold.
\\
\\
\fbox{\parbox[center]{6.6in}{
\begin{enumerate}[\quad]
\item \emph{Existence.} $\mathcal{X}^+ \neq \emptyset$ for any choice of $F_1,\dots,F_d$.
\item \emph{Uniqueness in law.} $\vX^+ \laweq \uF_d$ for any $\vX^+ \in \mathcal{X}^+$.
\item \emph{Maximization of supermodular functions.} Given a mean-compatible supermodular   $c \in \Spm$, we have    $$\E \left[ c(\vX^+) \right]=\sup \left\{ \E \left[c(\vX) \right]: \vX \in_d \FR \right\},$$ for any $\vX^+ \in \mathcal{X}^+$.
\end{enumerate}
}}
\\

\subsubsection*{Historical Remark}

It was already observed in~\citet{wH40} (an English translation is available in~\citet{wH94})
that random vectors having law $\uF_d$ always exist and maximize pairwise correlations over $\FR$ for $d=2$  and~\citet{mF51}. The terminology
\emph{comonotonic random variables} is found in~\citet{mY87} and~\citet{dS89} within the theory of expected utility even if the term \emph{comonotonic}, referred to generic functionals, was already present in~\citet{dS86}. The strictly related ideas of a \emph{monotonic operator} and a \emph{monotonic set} were pioneered  in~\citet{gM62} and~\citet{eZ60}. In particular, in \citet{gM64} one can find the first proof that the support of a comonotonic random vector is contained in a monotonic set, which directly implies point (b) in Theorem~\ref{th:comch}. The stochastic orderings implied by supermodular and convex functions in Finance and Actuarial Science have received considerable interest in the last decade, see for instance the papers~\citet{DDGKV02,DVGKTV06}.

The fact that a comonotonic random vector maximizes a (mean-compatible) supermodular function of random variables with given marginals actually goes back to (Theorem)~368 in the milestone book \citet{HLP34}, where it is proved that the scalar product of two vectors is maximal when the components of the two vectors are similarly ordered (e.g. they are monotonic in the same sense). An extension of this inequality to an arbitrary number of vectors was  given in~\citet{hdR52}.
 In (Theorem)~378 of~\citet{HLP34}, the authors prove the analogous inequality for rearrangements of functions, that is
\begin{equation}\label{eq:hlp}
 \int_0^1 f(x) \, g(x) dx \leq  \int_0^1 f^*(x) \, g^*(x) dx , \;
\end{equation}
where $f^*$ and $g^*$ are the increasing rearrangements of $f,g:[0,1] \to [0,1]$.

\citet{gL53} extended~\eqref{eq:hlp} to
\begin{equation}\label{eq:lorentz}
 \int_0^1 c(f_1(x),\dots,f_d(x)) dx \leq   \int_0^1 c(f_1^*(x),\dots,f_d^*(x)) dx,
 \end{equation}
for any mean-compatible supermodular function $c$. The discrete version of~\eqref{eq:lorentz} was given in~\citet{dL70} for convex functions of a sum and in~\citet{pD72} for general supermodular functions.

The translation of~\eqref{eq:lorentz} into the language of optimization problems over the set $\FR$ was given independently in a number of papers under different regularity conditions and using different nomenclature for the class of supermodular functionals.
The solution of the sup problem in~\eqref{eq:maxc} for $d=2$ has been first provided in~\citet{CSS76} (where supermodular functionals are called \emph{quasi-monotone}), \citet{aT80} (\emph{$n$-positive}; an early version of this paper dates back to 1975), \citet{wW76} (\emph{superadditive}), \citet{lR80} ($\Delta$-\emph{monotone}, which are equivalent to supermodular functions for $d=2$) and in a slight different form in \citet{MN79} (\emph{Shur}). Theorem~\ref{th:comch} in arbitrary dimensions $d$ is already present in~\citet{aT80} but also independently (and more elegantly) given in~\citet{lR83} (L-\emph{superadditive}).

The fact that comonotonic random vectors are maximal wrt to the supermodular order was also rediscovered independently in Actuarial Science; see~\citet{wrH86}. The paper \citet{KDVGD02} contains a geometry proof of the maximal convex sum property of comonotonic random vectors, which later on inspired some relevant work about optimal asset allocations.
In Mathematical Finance, the relationships between comonotonicity and risk
measures has been one of the very important aspects of the theory of comonotonicity.
For instance, \cite{sK01} showed that all law-determined coherent (sublinear) risk measures can be represented as the supremum of risk measures which are additive over comonotonic random variables. \cite{KCC10} contains a characterization of comonotonicity via maximum values of distortion risk measures.

\clearpage

\section{Extremal negative dependence}\label{se:pndhd}

If the role of comonotonic dependence as a benchmark in the modeling of catastrophes and as an  optimizer for the class of supermodular functions is well accepted, we will show that also the concept of negative dependence is equally important and has been historically given less weight mainly due to its difficult extension to higher dimensions, i.e. $d>2$.
In dimension $d=2$ we define (see Section~\ref{se:counter}) an extremally negatively dependent random vector, called a \emph{countermonotonic} random vector, via the requirement that its components are oppositely ordered. Similarly to comononotonic vectors, bivariate countermonotonic random vectors are always supported in any Fr\'echet class, have a unique law and minimize the expectation of (mean-compatible) supermodular functions.

Unfortunately, the definition of countermonotonicity and the implied properties cannot be trivially extended in dimensions $d>2$ and therefore alternative negative dependence concepts are called for. Section~\ref{se:MER} is dedicated to  the concept of \emph{pairwise countermonotonicity}, studied in the milestone paper~\citet{gDA72}. Pairwise countermonotonic vectors represent a
natural extension of countermonotonicity to higher dimensions, but can be defined only under quite restrictive assumptions. Only recently, a more general and practical notion of negative dependence, called \emph{joint mixability} (see Section~\ref{se:jointly mixable}), has been introduced; and this with a focus on the sum $X_1+\dots+X_d$.  Pairwise coutermonotonicity and joint mixability can be seen as particular cases of the novel concept of $\Sigma$\emph{-countermonotonicity} which we will introduce in Section~\ref{se:PND}.
We will illustrate the concepts presented in this section via pedagogical examples from
multivariate normal distributions.

\subsection{Countermonotonicity}\label{se:counter}
In dimension $d=2$ we define a \emph{countermonotonic} random vector on the requirement that its two components are oppositely ordered, e.g. high values for the first imply low values for the second and vice versa.

\begin{definition}\label{de:counter}
A random vector $(X_1,X_2)$  is said to be \emph{countermonotonic}
if there exists a rearrangement $f \rg \,\Id$ such that
\begin{equation}\label{eq:ppd2}
(X_1,X_2) \laweq \left(F_1^{-1} \circ f(U), F_2^{-1} \circ (1-f(U)) \right),
\end{equation}
where $U \laweq U[0,1]$.
\end{definition}
As the rearrangement $f$ in~\eqref{eq:ppd2} can always be taken as $f=\,\Id$, in a countermonotonic random vector the first (second) component  is almost surely  an increasing (decreasing) function of a common random factor $U$. Similarly to comonotonic random vectors, countermonotonic random vectors minimize the expectation of supermodular functions over the class of all random vectors having the same marginals.

\begin{theorem}\label{th:comcoh}
For a random vector $(X_1,X_2)$ with joint distribution function $F$, the following statements {\upshape{(a)-(d)}} are equivalent:
\begin{enumerate}[\upshape(a)]
\item $(X_1,X_2)$  is countermonotonic;
\item $F$ is given by
\begin{equation}\label{eq:conmdf}
F(x_1,x_2)=\lF_2(x_1,x_2) :=\max \{F_1(x_1)+F_2(x_2)-1,0 \}, \; x_1,x_2 \in  \R,
\end{equation}
where $F_j$ is the marginal distribution of $X_j$, $j=1,2$;
\item $F \le G$ on $\R^2$ for all $G\in \mathfrak F_2(F_1,F_2)$; 
 \item  for all supermodular functions $c: \R^2 \to \R$ that are mean-compatible with $(X_1,X_2)$, we have that
\begin{equation}\label{eq:minc}
\E \left[ c(X_1,X_2) \right]=\inf \left\{ \E \left[
c(Y_1,Y_2) \right]: Y_j \laweq X_j, j=1,2\right\};
\end{equation} 
\item  there exists a strictly supermodular function $c: \R^2 \to \R$ that is mean-compatible with $(X_1,X_2)$ such that \eqref{eq:minc} holds.
\end{enumerate}
\end{theorem}

The proof of Theorem~\ref{th:comcoh} is completely analogous to the one for Theorem~\ref{th:comch} as,
when $d=2$, both propositions are a direct consequence of the following well known facts. First, given two bivariate joint distributions
$F$ and $G$ we have
\begin{equation}\label{eq:tchen}
F \leq G \,  \quad \text{ if and only if } \;\int c \, dF \leq \int c \, dG, \text{ for any mean-compatible supermodular  $c:\R^2 \to \R$.}
\end{equation}
The equivalence~\eqref{eq:tchen} can be easily derived from~\citet{aT80} and basically follows from the fact that the class of bivariate supermodular functions can be written as the convex cone generated by indicator functions of the type $f(\vec{x})=\idc\{\vec{x} \leq \vec{t}\}$ for some $\vec{t}\in\R^2$; see  Theorem~2 in~\citet{lR80} considering that, when $d=2$, $\Delta$-monotone functions correspond to supermodular functions. The equivalence  \eqref{eq:tchen} can also be stated and extended under the language of stochastic orderings; the interested reader can start for instance from (9.A.18) in~\citet{SS07}.

Second, $\FR$  is well known from~\citet{wH40} and \citet{mF51} to have a smallest and a largest element when $d=2$.
 Formally, for  any $F \in \FRT$ we have that
\begin{equation}\label{eq:frechetbounds}
\lF_2 \leq F \leq \uF_2,
\end{equation}
where the smallest element is the distribution~\eqref{eq:conmdf} of any countermonotonic random vector and the largest element is the distribution~\eqref{eq:comdf} of any two-dimensional comonotonic random vector having marginals $F_1$ and $F_2$. From~\eqref{eq:tchen} and~\eqref{eq:frechetbounds}, it readily follows that the expectation of a supermodular function of a bivariate random vector is maximized (resp. minimized) under a comonotonic (resp. countermonotonic) law.

The equivalence in~\eqref{eq:tchen} is no longer true in higher dimensions $d>2$ for the class of supermodular functions (a counterexample has been provided in~\citet{MS00}) but holds true for the smaller (when $d>2$) class of so-called $\Delta$-monotone functions; see Theorem~3 in \citet{lR80}, and also \citet{lR04} for a characterization of $\Delta$-monotone functions.
However, even for $\Delta$-monotone functionals the extension of Theorem~\ref{th:comcoh}(c) to arbitrary dimensions is not possible as $\FR$ does not admit in general a smallest element when $d>2$. More precisely,
the inequality
\begin{equation}\label{eq:frechet}
\lF_d \leq F \leq \uF_d ,
\end{equation}
where $\lF_d :=\max \{F_1(x_1)+\dots+F_d(x_d)-d+1,0 \}$,
holds true (and cannot be improved; see~\citet{lR81}) for any $F \in \FR$ but   $\lF_d$   might fail to be a distribution function when $d>2$.
The extremal distributions $\lF_d$ and $\uF_d$ are also called the \emph{lower} and, respectively, \emph{upper Fr\'echet-Hoeffding bound} in honour of the two scholars; see Remark 2.1 in~\citet{lR13} on this. The notion of countermonotonic random variables associated with a reduction of their correlation was already presented under the term \emph{antithetic variates} in~\citet{HM56}.

\begin{remark}[The copula $W$]
According to Definition~\ref{de:counter},  a countermonotonic dependence structure is represented by a set of oppositely ordered rearrangements. From~\eqref{eq:conmdf}, it is also evident that a vector is countermonotonic if and only if it has copula $W$, where $W$ is defined as
$$
W(u_1,u_2)=\max\{u_1+u_2-1,0\}.
$$
The copula $W$ is therefore the copula representing perfect negative dependence. Its support consists of the secondary diagonal of the unit square and, being a shuffle of Min (roughly speaking, it is a horizontal reflection of the Min copula), it is a copula of any rearrangement matrix having two columns being oppositely ordered; see Figure~\ref{fi:comonotonic}.
Parametric families of copulas interpolating between the copula $W$ and the copula $M_2$, and also including the independence copula as a particular case are called \emph{comprehensive}. Examples of comprehensive families of copulas are the Frank copula defined in~\citet[(4.2.5)]{rN06} and the bivariate Gaussian copula, which is defined as the copula of a bivariate normal distribution.
\end{remark}

\begin{example}[Bivariate normal distribution]
Assume that the random vector $(X_1,X_2)$ follows a bivariate normal distribution $N_2(\vec{\mu},\vec{\Sigma})$, where $\vec{\mu}$  is the vector of means, and
$$
\vec{\Sigma}= \left(
           \begin{array}{cc}
             \sigma_1^2 & \sigma_{12}  \\
             \sigma_{12} & \sigma_2^2  \\
           \end{array}
         \right)
$$
is the positive semidefinite covariance matrix. The standard deviations $\sigma_j  \geq 0, j=1,2,$ are assumed to be fixed, that is the marginal distributions of the vector are given. The covariance parameter $\sigma_{12}$, is allowed to vary under the constraint that $\vec{\Sigma}$ is positive semidefinite, that is
$$
-\sigma_1\sigma_2 \leq \sigma_{12} \leq\sigma_1\sigma_2.
$$
Within this parametric model, the extremal positive dependence structure  is attained when
$\sigma_{12}$ is maximized, that is when $\sigma_{12}=\sigma_1\sigma_2$. In this case, $(X_1,X_2)$ is  comonotonic and has copula $M_2$.
The extremal {negative} dependence structure  is attained when
$\sigma_{12}$ is instead minimized, that is when $\sigma_{12}=-\sigma_1\sigma_2$; in this case, $(X_1,X_2)$ is  countermonotonic and has copula $W$. Both in the comonotonic and countermonotonic case, the bivariate normal model represents a singular distribution ($\vec{\Sigma}$ is not invertible).
\end{example}

\begin{figure}[h]
\begin{center}
\scalebox{.26}{\includegraphics{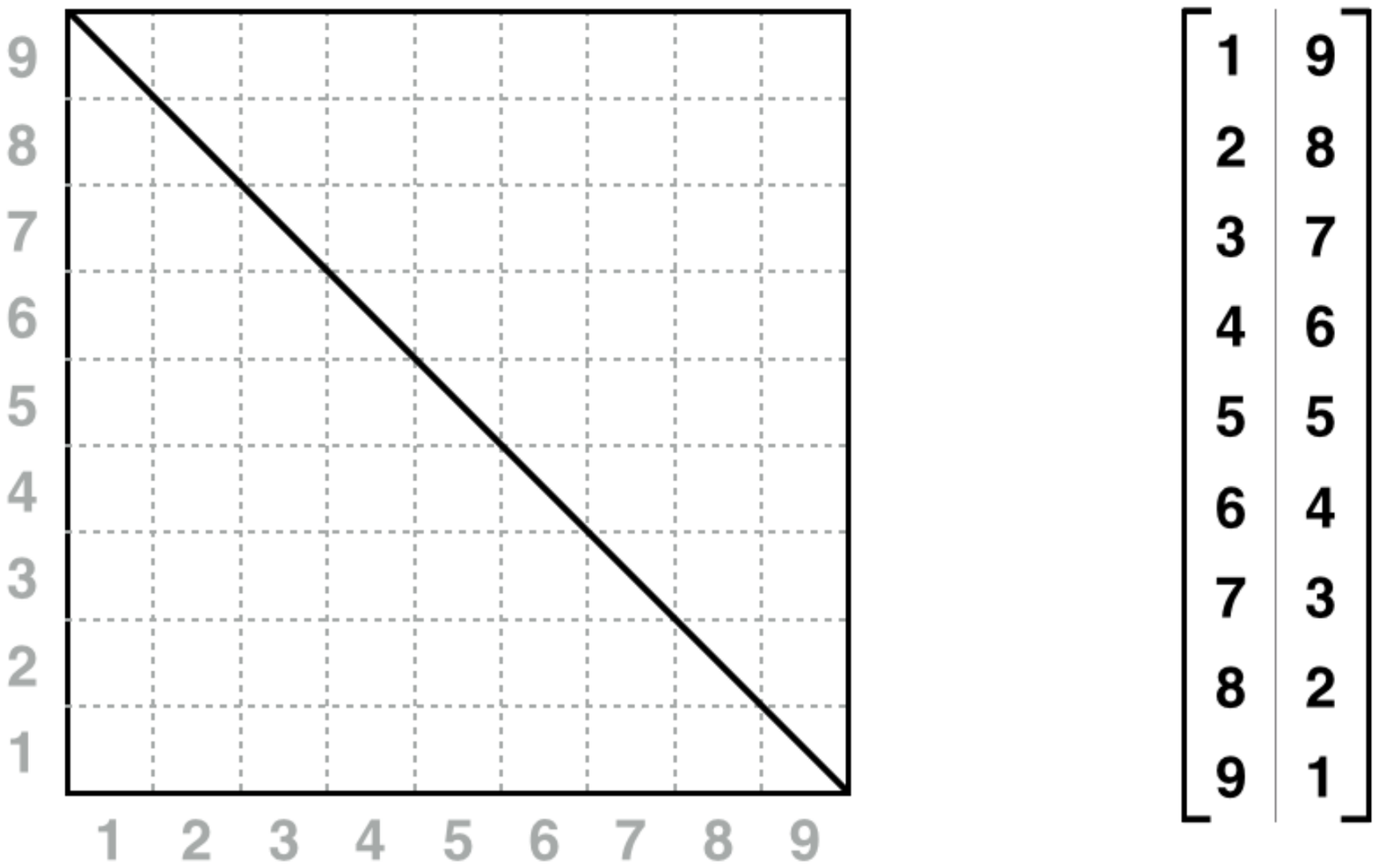}}
\caption{\small The support of the copula $W$ (left) and a
rearrangement matrix (right) representing a discrete bivariate distribution with copula $W$ and marginal distributions uniformly distributed over the first nine integers}
\label{fi:countermonotonic}
\end{center}
\end{figure}

\subsection{Pairwise countermonotonicty}\label{se:MER}

Similarly to extremal positive dependence, the idea of \emph{extremal negative dependence} has been historically associated to the notion of \emph{minimal} correlation.
Following the discussion carried out in Section~\ref{se:ppd}, we can state three desirable properties that set $\mathcal{X}^-$ of all negatively dependent random vectors having marginal distributions $F_1,\dots,F_d$ should satisfy.
\\
\\
\fbox{\parbox[center]{6.6in}{
\begin{enumerate}[\quad]
\item \textbf{(E)} \emph{Existence.} $\mathcal{X}^- \neq \emptyset$ for any choice of $F_1,\dots,F_d$.
\item \textbf{(U)} \emph{Uniqueness in law.} $\vX^- \laweq \lF_d$ for any $\vX^- \in \mathcal{X}^-$.
\item \textbf{(M)} \emph{Minimization of supermodular functions.} Given a mean-compatible supermodular   $c \in \Spm$, we have   $$\E \left[ c(\vX^-) \right]=\inf \left\{ \E \left[c(\vX) \right]: \vX \in_d \FR \right\},$$ for any $\vX^- \in \mathcal{X}^-$.
\end{enumerate}
}}
\\
\\

In dimension $d=2$, a countermonotonic random vector satisfies \textbf{(E)},\textbf{(U)} and \textbf{(M)} hence countermonotonicity is the natural notion of extremal negative dependence to use. Unfortunately, when $d>2$ there does not exist any concept of negative dependence satisfying all three requirements listed above. In arbitrary dimensions $d$, it is still true that a vector having law $\lF_d$ minimizes the expectation of any supermodular function, but it was shown in~\citet{gDA72} that such a vector only exists under very special assumptions, hence not satisfying  \textbf{(E)}.
We call such exceptional cases pairwise countermonotonic random vectors.

\begin{definition}
A random vector $(X_1,\dots,X_d)$ is said to be \emph{pairwise countermonotonic} if all its bivariate projections
$(X_i,X_j)$, $i \neq j$, are countermonotonic random vectors.
\end{definition}

Pairwise countermonotonicity is the most intuitive extension of the concept
of countermonotonicity in higher dimensions. The name pairwise countermonotonicity was however not introduced in~\citet{gDA72}, which was the first paper to give conditions for the existence of a $d$-variate distribution attaining the lower Fr\'echet bound $\lF_d$.
Pairwise countermonotonicity has been also studied in actuarial science under different names, in particular with respect to the minimisation of the so-called \emph{stop-loss premium} for a (re-)insurance policy.
The first actuarial paper that studied the \emph{safest dependence structure} for two-point distributions has been~\citet{HW99};
\citet{DD99} were the first who systematically developed (probabilistic) properties and characterizations of  \emph{mutually exclusive risks} in a more general setting.  Finally \citet{KCCL14} generalized many of the results of the two papers mentioned above.

From the definition, it is straightforward (see Lemma~1 in~\citet{gDA72}) that the distribution of a pairwise countermonotonic vector has to be the lower Fr\'echet bound $\lF_d$. However, $\lF_d \in \FR$ holds true (and hence a pairwise countermonotonic random vector exists) only under very restrictive assumptions on the marginals.
Indeed, already in~\citet{gDA59} it is shown that if $U,V,Z$ are continuous random variables with $(U,V)$ and $(V,Z)$ countermonotonic random vectors, then $(U,Z)$ has to be comonotonic (only continuity of $U$ is actually needed).
We have $\lF_d \in \FR$ only in the case in which all marginal distributions $F_j$ have a jump  at their essential infimums or all at their essential supremums. The following proposition combines Lemma~2 and Theorem~3 in~\citet{gDA72}.
\\
\\

\begin{proposition}\label{pr:dallagalio}
Assume $d \geq 3$ and that at least three among the $F_j$'s are non degenerate (otherwise we go back to the case $d=2$). We have that  $\lF_d \in \FR$ if and only if either
\begin{equation}\label{eq:da1}
\sum_{j=1}^d [1-F_j(F_j^{-1}(0))] \leq 1,
\end{equation}
or
\begin{equation}\label{eq:da2}
\sum_{j=1}^d F_j(F_j^{-1}(1)-) \leq 1.
\end{equation}
If~\eqref{eq:da1} is satisfied, then a random vector $(X_1,\dots, X_d)$ is pairwise countermonotonic iff it has a.s. at most one component strictly bigger than its essential infimum, that is
$$
P(X_i>F_i^{-1}(0), X_j>F^{-1}_j(0))=0 \text{ for $i\ne j$}.
$$
If~\eqref{eq:da2} is satisfied, then a random vector $(X_1,\dots, X_d)$ is pairwise countermonotonic iff it has a.s. at most one component strictly smaller than its essential supremum, that is
$$
P(X_i<F_i^{-1}(1), X_j<F^{-1}_j(1))=0 \text{ for $i\ne j$}.
$$
\end{proposition}


\begin{figure}
  \begin{minipage}[l]{0.2\textwidth}
   \scalebox{.70}{ \includegraphics[width=\textwidth]{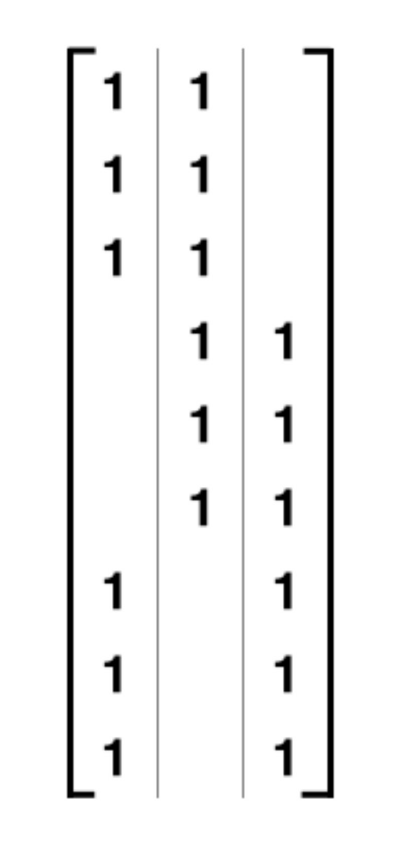}}
  \end{minipage}\hfill
  \begin{minipage}[r]{0.8\textwidth}
    \caption{\small A rearrangement matrix representing a discrete bivariate distributions with
a pairwise countermonotonic dependence structure. The ones in matrix represents the essential infimum of the corresponding marginal distributions. Since each row of the matrix takes probability 1/9, according to Proposition~\ref{pr:dallagalio}, the number of ones must be at least 18. The same argument holds by interpreting the ones in the matrix as essential supremums.
    } \label{fi:pwcounter}
  \end{minipage}
\end{figure}

\begin{figure}[h]
\begin{center}
\scalebox{.42}{\includegraphics{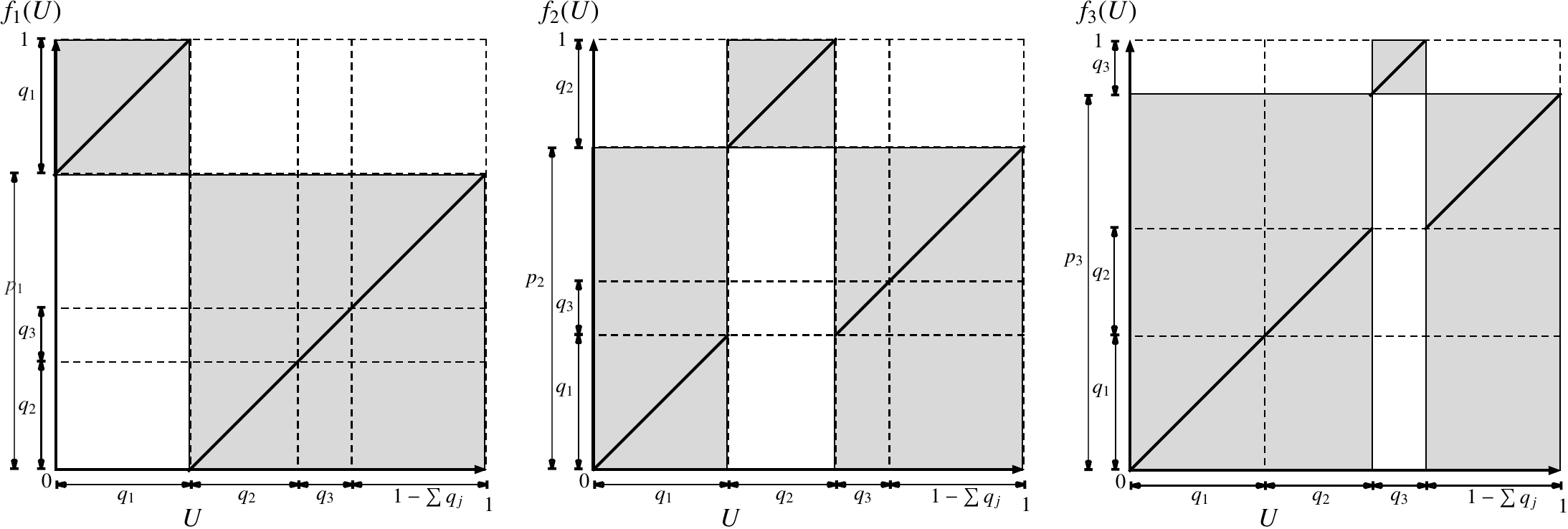}}
\caption{\small A set of three rearrangements $f_1,f_2,f_3$ defining a pairwise countermonotonic random vector. In the figure we set
$q_j:=[1-{F}_j(F_j^{-1}(0))]$ and $p_j:=1-q_j$, $j=1,\dots,d$. The compatibility condition in~\eqref{eq:da1}, i.e.
$\sum q_j \leq 1$, is assumed to be satisfied.}
\label{fi:merfi}
\end{center}
\begin{center}
\scalebox{.42}{\includegraphics{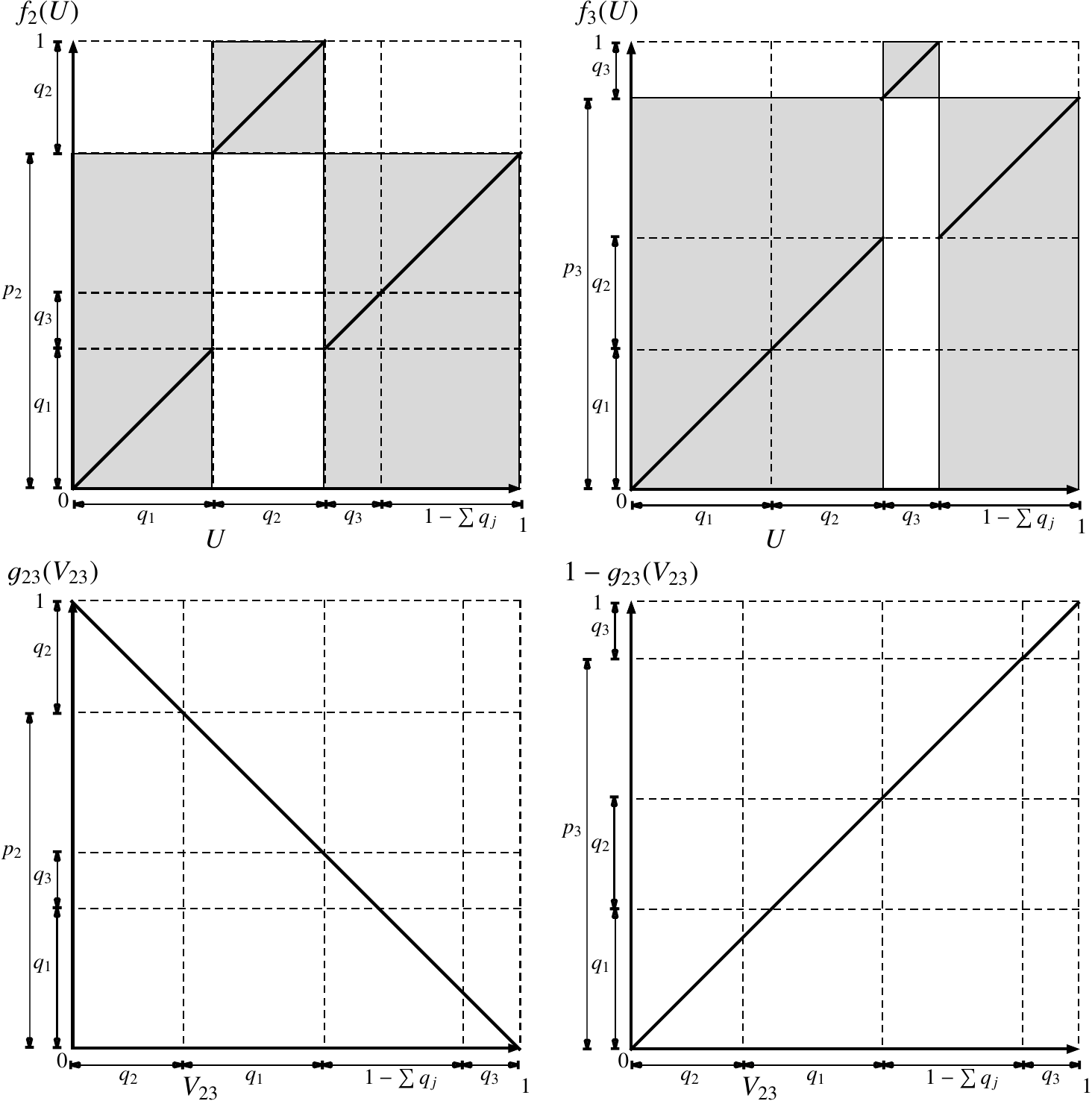}}
\caption{\small A possible choice for $g_{23}(V_{23})$ in the representation~\eqref{eq:forthefigure}.}
\label{fi:merfi2}
\end{center}
\end{figure}

Figure~\ref{fi:pwcounter} illustrates the necessity and sufficiency of the conditions in Proposition~\ref{pr:dallagalio} for discrete marginal distributions.
It is also pedagogical to see how marginals with jumps allow the building of pairwise countermonotonicity avoiding counterexamples like the one in~\citet{gDA59}.
Assume that the marginal distributions $F_1,\dots,F_d$ satisfy  condition~\eqref{eq:da1} (an analogous example can be built for marginals satisfying~\eqref{eq:da2}).
In Figure~\ref{fi:merfi} we give a set of three rearrangements $f_1,f_2,f_3$ under which the vector
$$
(X_1,X_2,X_3):=\left(F_1^{-1}\circ f_1(U),F_2^{-1}\circ f_2(U),F_3^{-1}\circ f_3(U)\right),
$$
$U \laweq U[0,1]$, is pairwise countermonotonic.
For each pair $(f_i,f_j)$ it is possible to find a new rearrangement $g_{ij}$ and a
random variable $V_{ij} \laweq U[0,1]$ such that
\begin{equation}\label{eq:forthefigure}
\left(F_i^{-1}\circ g_{ij}(V_{ij}), F_j^{-1}\circ (1-g_{ij})(V_{ij})\right) \laweq \left(F_i^{-1}\circ f_i(U), F_j^{-1}\circ f_j(U)\right).
\end{equation}
As a consequence $(X_i,X_j)$ is countermonotonic for $i \neq j$. In Figure~\ref{fi:merfi2} we show a possible choice for $g_{23}(V_{23})$, where the uniform random variable $V_{23}$ is illustrated as a
rearrangement of the unit interval. The construction of pairwise countermonotonicity is made possible in Figure~\ref{fi:merfi} and~\ref{fi:merfi2} because jumps at the essential infimum of the distributions allow the choice of the rearrangement function arbitrarily within each grey rectangle in Figure~\ref{fi:merfi}.

A pairwise countermonotonic random vector enjoys all the properties of Theorem~\ref{th:comcoh} in arbitrary dimension $d$.

\begin{theorem}\label{th:merth}
For a random vector $(X_1,\dots,X_d)$ with joint distribution function $F$, the following statements {\upshape{(a)-(d)}} are equivalent:
\begin{enumerate}[\upshape(a)]
\item $(X_1,\dots,X_d)$ is pairwise countermonotonic;
\item $F$ is given by
\begin{equation*}
F(x_1,\dots,x_d)=\lF_d(x_1,\dots,x_d) = \max \{F_1(x_1)+\dots+F_d(x_d)-d+1,0 \}, \; x_1,\dots,x_d \in  \R,
\end{equation*}
where $F_j$ is the marginal distribution of $X_j$, $j=1,\dots,d$;
\item $F \le G$ on $\R^d$ for all $G\in \FR$. 
\item  for all supermodular functions $c: \R^d \to \R$, we have that
\begin{equation}\label{eq:minc2}
\E \left[ c(X_1,\dots,X_d) \right]=\inf \left\{ \E \left[
c(Y_1,\dots,Y_d) \right]: Y_j \laweq X_j, j=1,\dots,d\right\}.
\end{equation} 
\item[\upshape(e)] there exists a strictly supermodular function $c: \R^d \to \R$ that is mean-compatible with $(X_1,\dots,X_d)$ such that  \eqref{eq:minc2} holds.
\end{enumerate}

\end{theorem}
\emph{Proof.} 
(a) $\Leftrightarrow$ (b) follows from~\citet{gDA72}. (b) $\Leftrightarrow$ (c) follows from the point-wise attainability of the lower Fr\'echet-Hoeffding bound. (a) $\Rightarrow$ (d)   is proved for $\Delta$-monotone functions in Theorem~5 in \citet{lR80}, for supermodular functions with mean-compability in Theorem~12 in~\citet{DD99}. The fact that (d) does not require mean-compatibility is shown in Theorem 7 of \cite{cote2025convex}. 
(d) $\Rightarrow$ (c) follows by choosing the supermodular function $\mathbf x\mapsto \idc_{\{\mathbf x\le \mathbf y\}}$ for each $\mathbf y\in \R^d$.
(d) $\Rightarrow$ (e) is obvious.
(e) $\Rightarrow$  (a) follows from a standard rearrangement argument. \qed

Denote by $\mathcal{X}^P$ the set of all pairwise countermonotonic random vectors having marginal distributions $F_1,\dots,F_d$.
From the above theorem,  two fundamental properties follow.
\\
\\
\fbox{\parbox[center]{6.6in}{
\begin{enumerate}[\quad]
\item \textbf{(U)} \emph{Uniqueness in law.} $\vX^P \laweq \lF_d$ for any $\vX^P \in \mathcal{X}^P$.
\item \textbf{(M)} \emph{Minimization of supermodular functions.} Given a mean-compatible supermodular   $c \in \Spm$, we have    $$\E \left[ c(\vX^P) \right]=\inf \left\{ \E \left[c(\vX) \right]: \vX \in_d \FR \right\},$$ for any $\vX^P \in \mathcal{X}^P$.
\end{enumerate}
}}

\begin{remark}
In  \citet{gDA72}, the author did not introduce the term \emph{pairwise countermonotonicity}, but equivalently investigated the conditions under which the lower Fr\'echet bound $\lF_d$ is a well-defined distribution function in arbitrary dimension  $d$. Consistenly with the expository scope of this paper, we however find it more appropriate to identify all vectors having distribution $\lF_d$ as being pairwise countermonotonic.
\end{remark}

\begin{remark}
The existence of multivariate probability measures with given margins and other constraints was more generally studied in \cite{nV62}  and in \cite{vS65} where an elegant duality theorem (Theorem 7 of that paper) was established; see also Section~1.6 in \citet{lR13}.
\end{remark}

\subsection{Joint mixability}\label{se:jointly mixable}

Requiring that an extremally negatively dependent vector satisfies properties  (\textbf{U})  and (\textbf{M}) poses strong constraints on (\textbf{E}) when $d>2$, producing a definition of extremal negative dependence of very restricted applicability.
Proposition~\ref{pr:dallagalio} implies for instance that any Fr\'echet class supporting pairwise countermonotonic vectors does not contain vectors with continuous marginal components.
Consequently, any $d$-variate normal model does not include pairwise countermonotonicity
for $d >2$.

\begin{example}[Multivariate normal distribution]\label{ex:multnorm}
Assume that the random vector $(X_1,X_2,X_3)$ follows a three-variate normal distribution $N_3(\vec{0},\vec{\Sigma})$, where $\vec{0}$ is the a vector of zeros, and
$$
\vec{\Sigma}=\left(
           \begin{array}{ccc}
             \sigma_1^2 & \sigma_{12} & \sigma_{13} \\
             \sigma_{12} & \sigma_2^2 & \sigma_{23} \\
             \sigma_{13} & \sigma_{23}  & \sigma_3^2 \\
           \end{array}
         \right).
$$
is a positive semidefinite covariance matrix. The standard deviations $\sigma_j >0,~ j=1,2,3,$ are assumed to be fixed, that is the marginal distributions of the vector are given. The covariance parameters $\sigma_{12}$, $\sigma_{13}$ and $\sigma_{23}$ are allowed to vary under the constraint that $\vec{\Sigma}$ is positive semidefinite.
Straightforward constraints for $\sigma_{12}$, $\sigma_{13}$ and $\sigma_{23}$ are that
$$
-\sigma_i\sigma_j \leq \sigma_{ij} \leq\sigma_i\sigma_j,\; 1\leq i<j\leq 3.
$$
Within this parametric model, the extremal positive dependence structure  is attained when the pairwise correlations are individually maximized, that is when $\sigma_{ij}=\sigma_i\sigma_j$, $1\leq i<j\leq 3$. In this case $(X_1,X_2,X_3)$ is  comonotonic and has copula $M_3$.

Finding values of $(\sigma_1,\sigma_2,\sigma_3)$ yielding an extremal \emph{negative} dependence structure is a much trickier question. Indeed, the pairwise covariance parameters cannot achieve their respective smallest values $\sigma_{ij}=-\sigma_i\sigma_j$ within the same model.
A trivial lower bound bound on the variance (var) of the standardized marginal models gives
  $$\var\left(\frac {X_1}{\sigma_1}+\frac{X_2}{\sigma_2}+\frac{X_3}{\sigma_3}\right)\ge 0 ~~\Rightarrow ~~  \frac{\sigma_{12}}{\sigma_1\sigma_2}+\frac{\sigma_{13}}{\sigma_1\sigma_3}+\frac{\sigma_{23}}{\sigma_2\sigma_3}\ge -\frac32.$$
Consequently, there does not seem to exist a univocally defined set of correlation parameters representing the most negative dependence structure for this multivariate normal model.  Keeping this example in mind, we will explore different notions of extremal negative dependence.
\end{example}

In order to define a more practical notion of perfect negative dependence in dimensions $d>2$, we need to relax
our requirements. In the remainder of the paper, instead of requiring a negatively dependent random vector to be the minimizer of \emph{any} supermodular function, we focus only on those supermodular functions $c$ that can be expressed as $c(\vec{x})=f(x_1+\dots+x_d)$ for some convex function $f$. This is a strict restriction as for instance the expectation of the product and the variance of the sum of uniformly distributed random variables are minimized by different dependence structures; see~\citet{WW11}. This consideration of optimization problems is also of practical interest; see also the discussion in Remark \ref{rem:21}.  The relevant concept here is that of convex order.

\begin{definition}
We say that a random variable $X$ is smaller than $Y$ in \emph{convex order}, denoted by $X \lcx Y$, if
\begin{equation*}
\E[f(X)] \leq \E[f(Y)], \mbox{ for all convex functions } f:\R \to \R \text{ such that the expectations exist}.
\end{equation*}
\end{definition}

A straightforward consequence of $X \lcx Y$ is that $\E[X]=\E[Y]$ and $\E[X^2] \leq \E[Y^2]$ given that they exist.  However, convex order dominance is stronger than having the same mean and a larger variance and is related to the concept of so-called \emph{majorization}  of $d$-valued vectors. When two random variables have the same finite mean, as within the set $\FR$ with  finite-mean marginals, convex order is equivalent to \emph{increasing convex} order as defined in~\citet{MS02}. Comprehensive references regarding the link between comonotonicity, convex order, rearrangements and majorization of vectors are~\citet{MOB11} and~\citet{lR13}.

\begin{definition}
We say that $\vX=(X_1,\dots,X_d)$ is a \emph{$\cx$-smallest} element in $\FR$ if $\vX \in_d \FR$ and
$$
\sum_{j=1}^{d} X_j \lcx \sum_{j=1}^{d} Y_j,\text{ for any } \vY=(Y_1,\dots,Y_d) \in_d \FR.
$$
\end{definition}
From the definition of convex order, it directly follows that a $\cx$-smallest element $\vec{X} \in \FR$ satisfies
\begin{enumerate}[]
\item (\textbf{M1}): $
\E \left[ f(X_1+\dots+X_d) \right]=\inf \left\{ \E \left[f(Y_1+\dots+Y_d) \right]: \vY \in_d \FR,~  \E \left[f(Y_1+\dots+Y_d) \right] \mbox{~exists}   \right\},
$
for any convex function $f$ such that $\E \left[ f(X_1+\dots+X_d) \right]$ is properly defined.
\end{enumerate}

Theorems~\ref{th:comcoh} and~\ref{th:merth} immediately imply that countermonotonic (for $d=2$) and pairwise countermonotonic random vectors (when they exist) are $\cx$-smallest elements of the corresponding Fr\'echet classes.
The paper~\citet{DDGKV02} contains an elegant geometric proof of the maximal convex sum property of comonotonic random vectors, which later on inspired some relevant research about optimal asset allocations.
We will now define negatively dependent random vectors based on the weaker requirement (\textbf{M}').
The restriction to supermodular functions which can be expressed as convex functions of a sum is quite intuitive as the sum is the most natural aggregating operator and $\cx$-smallest elements are still minimizers for a broad class of functionals including for instance the variance of the sum.

 Unfortunately, not all Fr\'echet classes admit a $\cx$-smallest element; see Example~3.1 in~\citet{BJW13}. However, it is still possible to define a much wider applicable notion of extremal negative dependence, which has been recently introduced in the literature under the name of \emph{joint mixability}.

\begin{definition}\label{de:jointly mixable}
A random vector $(X_1,\dots,X_d)$ is said to be a \emph{joint mix} if
$$
P\left( X_1+\dots +X_d = k \right)=1,
$$
for some $k\in \R$.
\end{definition} 

\begin{example}[Multivariate normal distribution, continued] \label{ex32}
Simple examples of joint mixes include normal random vectors with special covariance matrices; see \cite{WW14}.
We now show that a joint mix $(X_1,X_2,X_3)$ having  the three-variate normal distribution described in Example \ref{ex:multnorm} exists if and only if
\begin{equation}\label{eq:ex32q11}
2\max_{1 \leq i \leq 3} \sigma_i\le \sigma_1+\sigma_2+\sigma_3.
\end{equation}
Without loss of generality we assume $\sigma_1\ge \sigma_2\ge \sigma_3>0$.
If $(X_1,X_2,X_3)$ has law $N_3(\vec{0},\vec{\Sigma})$,
we can write
\begin{equation}
\label{eq:ex32q1}
\left(
    \begin{array}{c}
      X_1 \\
      X_2 \\
      X_3 \\
    \end{array}
  \right) =\left(
                    \begin{array}{ccc}
                      a_{11} & 0 & 0 \\
                      a_{21} & a_{22} & 0 \\
                      a_{31} & a_{32} & a_{33} \\
                    \end{array}
                  \right)\left(
    \begin{array}{c}
      Z_1 \\
      Z_2 \\
      Z_3 \\
    \end{array}
  \right),
\end{equation}
where $Z_1,Z_2,Z_3$ are iid standard normal random variables.
 Since $\var(X_1)=\sigma_1^2$, we can take $a_{11}=\sigma_1$.
 First we suppose that $(X_1,X_2,X_3)$ is a joint mix.
  From $X_1+X_2+X_3=0$ we obtain that $a_{11}+a_{21}+a_{31}=0$. 
Note that $|a_{21}|\le \sigma_2\le \sigma_1$ and $|a_{31}|\le \sigma_3\le \sigma_1$. From $a_{11}+a_{21}+a_{31}=0$, it follows that $\sigma_1\le \sigma_2+\sigma_3$ which is \eqref{eq:ex32q11}.
 Now suppose that \eqref{eq:ex32q11} holds.
 Take $\sigma_{12}=\frac{1}{2}(\sigma_3^2-\sigma_1^2-\sigma_2^2),$
$\sigma_{13}=\frac{1}{2}(\sigma_2^2-\sigma_1^2-\sigma_3^2),$ and
$\sigma_{23}=\frac{1}{2}(\sigma_1^2-\sigma_2^2-\sigma_3^2).$
We can verify that the matrix
$$\left(
           \begin{array}{ccc}
             \sigma_1^2 & \frac{1}{2}(\sigma_3^2-\sigma_1^2-\sigma_2^2) & \frac{1}{2}(\sigma_2^2-\sigma_1^2-\sigma_3^2) \\
              \frac{1}{2}(\sigma_3^2-\sigma_1^2-\sigma_2^2) & \sigma_2^2 & \frac{1}{2}(\sigma_1^2-\sigma_2^2-\sigma_3^2) \\
             \frac{1}{2}(\sigma_2^2-\sigma_1^2-\sigma_3^2) & \frac{1}{2}(\sigma_1^2-\sigma_2^2-\sigma_3^2)  & \sigma_3^2 \\
           \end{array}
         \right)
$$
is positive semi-definite if and only if $\sigma_1\le \sigma_2+\sigma_3$.
It is easy to see that if $(X_1,X_2,X_3)$ has law $N_3(\vec{0},\vec{\Sigma})$, then
$$\var(X_1+X_2+X_3)=\sigma_1^2+\sigma_2^2+\sigma_3^2 + 2\sigma_{12}+2\sigma_{13}+2\sigma_{23}=0,$$
that is, $(X_1,X_2,X_3)$ is a joint mix. 


 This indicates that, in a multivariate normal model, a joint mix is supported if and only if the variances
of the marginal components are homogeneous enough.
This conclusion can be analogously extended to the class of elliptical distributions; see Theorem \ref{th:jointly mixable3} (c) below.
\end{example}

\begin{example}[Survey sampling]
The problem of constructing $d$ dependent variables with a constant sum occurs in \emph{survey sampling}.
In a survey sampling context, $d$ Bernoulli random variables (with possibly different success probabilities) are associated to $d$ units in a finite population.
Each Bernoulli variable takes the value 1 if the corresponding unit is drawn in the sample and 0 otherwise.  Constructing a sample design with a fixed sample size $k$ is equivalent to
constructing a joint mix for possibly inhomogeneous Bernoulli random variables.
Many solutions to the problem of designing unequal probability survey sampling designs with fixed sample size $k$ have been published, see for instance~\citet{HB80} and~\citet{BH83}.
A paper describing the method implemented in the \texttt{SAS  SURVEYSELECT} procedure is~\citet{kV68}.
\end{example}

\begin{figure}[h!]
\begin{center}
\scalebox{.22}{\includegraphics{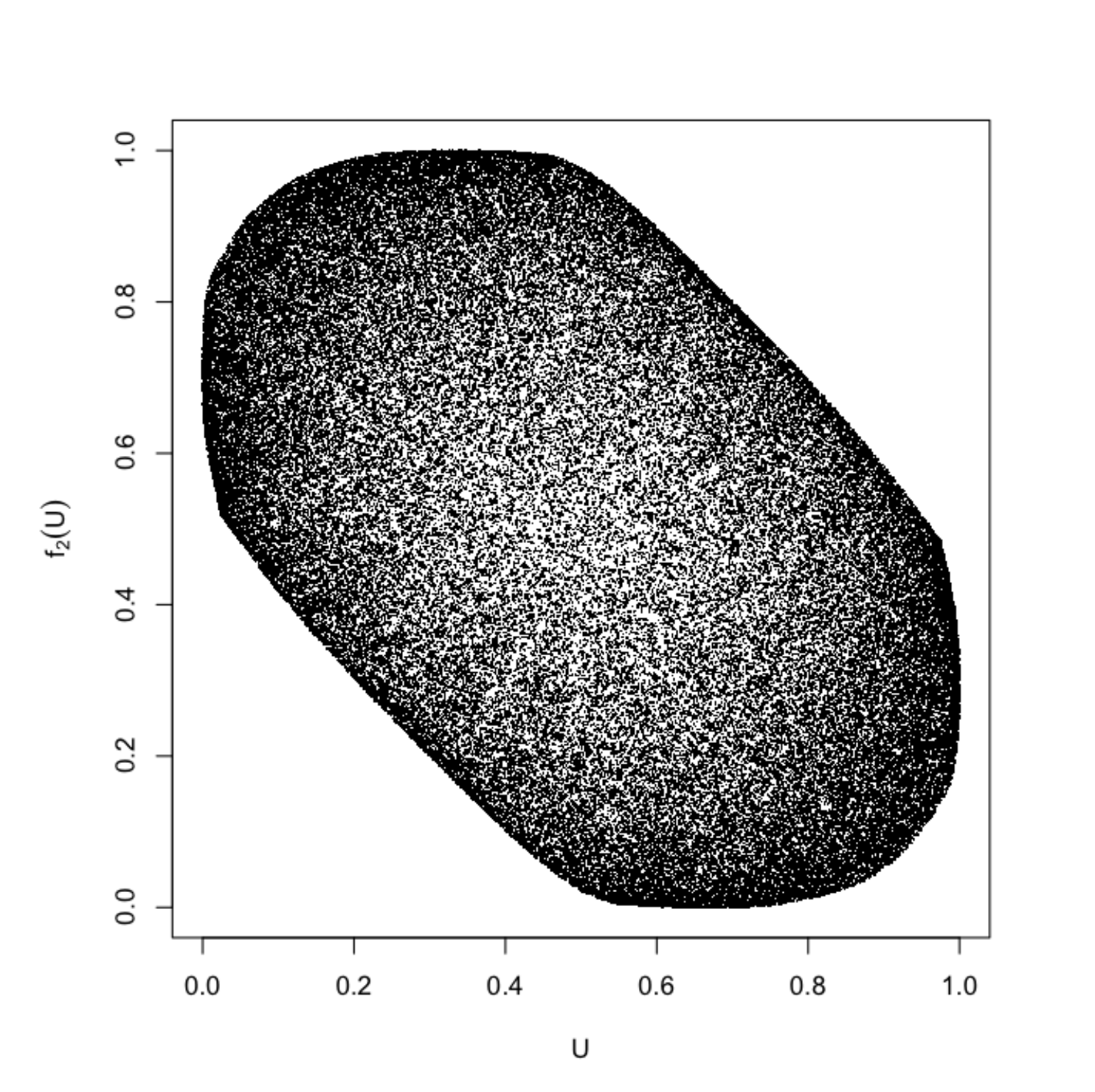}\includegraphics{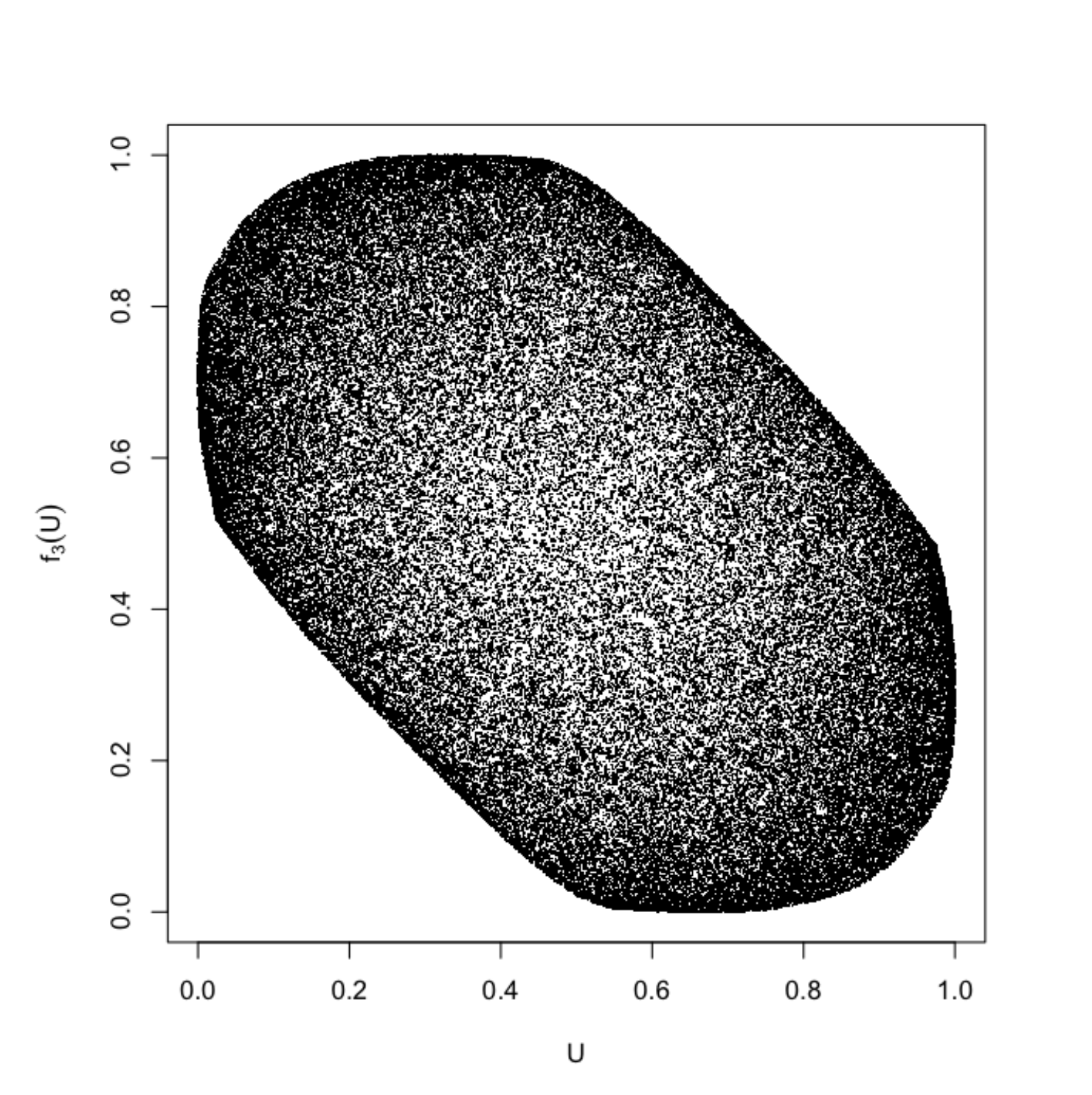}\includegraphics{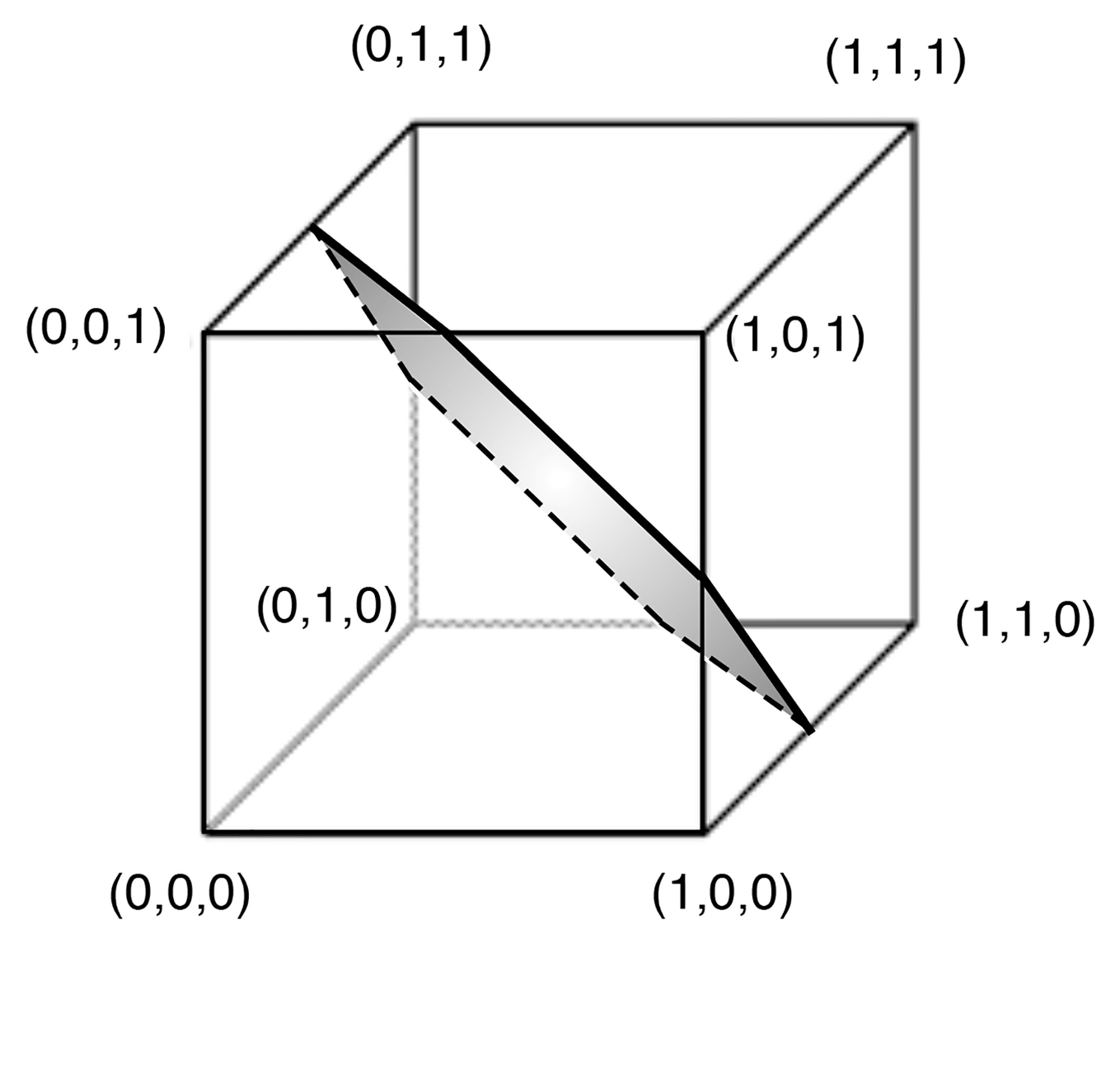}}
\caption{\small Left and middle part of the Figure: a set of rearrangements $f_1,f_2,f_3$ under which the sum of three uniform random variables is equal to $k=1.5$ with probability one ($f_1$ is not shown as it can always be taken as the identity function). These rearrangements define a 3-dimensional distribution (right) which is (not uniformly) distributed on the simplex $\{(u_1,u_2,u_3): u_1+u_2+u_3=1.5\}\subset [0.1]^3$. A different set of rearrangements of the unit interval with constant sum can be found in~\citet[Example~3]{GR81}.}
\label{fi:rgnprod0}
\end{center}
\end{figure}

In  Figure~\ref{fi:rgnprod0}  we show the dependence structure of a joint mix with $U[0,1]$  marginals.
Similarly to a pairwise countermonotonic random vector, a joint mix might fail to be supported in a fixed Fr\'echet class.
For instance, a bivariate random vector $(X_1,X_2)$ is a joint mix if and only if $X_1=k-X_2$ a.s. for some constant $k$, thus if and only if its marginal components are symmetric with respect to $k$.
Thus it is natural to investigate whether a Fr\'echet class supports a joint mix.

\clearpage

\begin{definition}
A $d$-tuple of distributions $(F_1,\dots,F_d)$ is said to be \emph{jointly mixable} if $\FR$ supports a joint mix.
Equivalently, $(F_1,\dots,F_d)$ is jointly mixable if and only if
there exist $d$ rearrangements $f_1,\dots,f_d \rg \Id$ and $k\in \R$ such that
$$
P\left( F_1^{-1}\circ f_1(U)+\dots +F_d^{-1}\circ f_d(U) = k \right)=1,
$$
where $U \laweq U[0,1]$. The constant $k$ is called a \emph{joint center} of $(F_1,\dots,F_d)$.
\end{definition}

Denote now by $\mathcal{X}^J$ the set of all joint mixes with marginal distributions $F_1,\dots,F_d$ having finite first moment.
From the definition of a joint mix and Jensen's inequality,
the following properties follow.
\\
\\
\fbox{\parbox[center]{6.6in}{
\begin{enumerate}[\quad]
\item \textbf{(U1)} \emph{Uniqueness in law for the sum.} The distribution of $(X_1^J + \dots + X_d^J)$ is degenerate at the joint center for any $\vX^J \in \mathcal{X}^J$.
\item \textbf{(M1)} \emph{$\cx$-minimality.} Given a convex function $f$ such that $\E \left[ f(X_1^J+\dots+X_d^J) \right]$ exists, we have  
$$
\E \left[ f(X_1^J+\dots+X_d^J) \right]=\inf \left\{ \E \left[f(X_1+\dots+X_d) \right]: \vX \in_d \FR,~ \E \left[ f(X_1+\dots+X_d) \right] \mbox{~exists} \right\},
$$
for any $\vX^J \in \mathcal{X}^J$.
\end{enumerate}
}}
\\
\\

\begin{figure}[h]
\begin{center}
\scalebox{.40}{\includegraphics{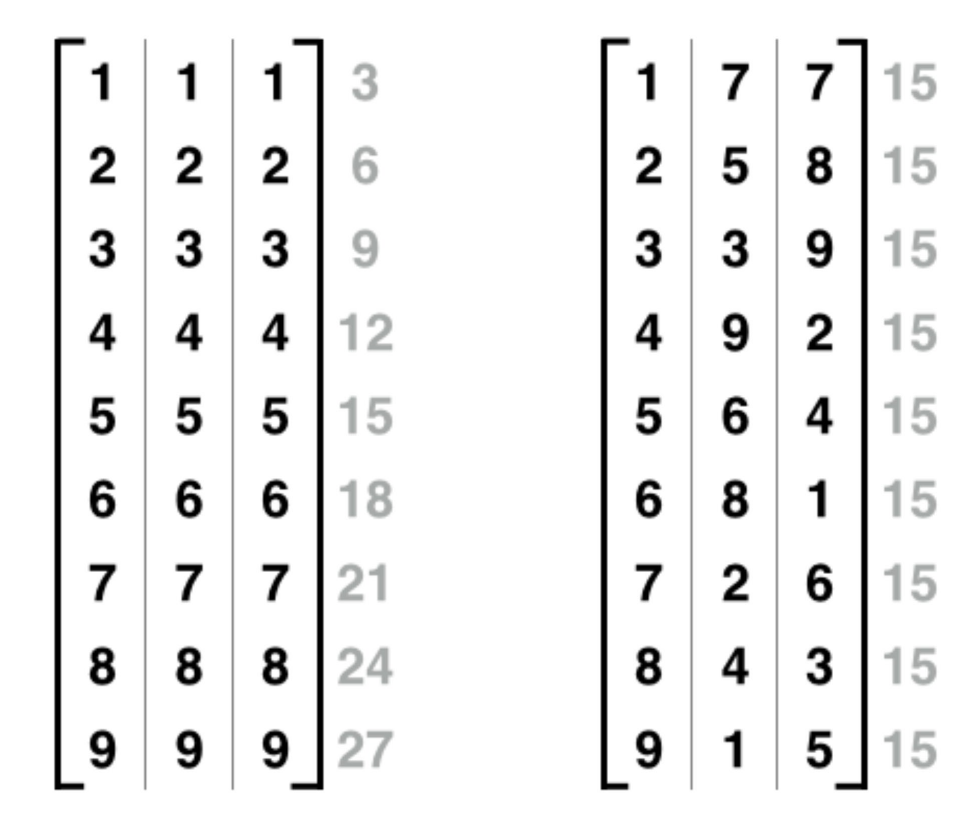}}
\caption{\small Rearrangement matrices representing: a discrete bivariate distribution with
a (left) comonotonic  dependence structure and a (right) joint mix
with the same marginal distributions. While comonoticity maximizes the variance of the sum of the marginal components (reported outside the matrix),  a joint mix attains the corresponding minimum.}
\label{fi:jj}
\end{center}
\end{figure}

Joint mixability represents a concept of negative dependence. For instance, it is clear that in dimension $d=2$ a joint mix is countermonotonic (the converse does not hold). In arbitrary dimensions,
property (\textbf{M1}) implies that a joint mix having marginal components with finite mean is a $\cx$-smallest element in the corresponding Fr\'echet class. For instance, a joint mix therefore attains the smallest possible variance for the sum of its marginal components; see Figure~\ref{fi:jj} where a representation of a joint mix  in terms of rearrangement matrix is given and compared with comonotonicity.
Even if the law of a joint mix might not be unique, property
\textbf{(U1)} states that the law of the sum of the components of any joint  mix is unique.

If  a joint mix satisfies generally weaker versions of properties \textbf{(U)} and  (\textbf{M}), what can we say about the existence of a joint mix? For $d \ge 3$ and a given $d$-tuple of distributions $(F_1,\dots,F_d)$, it is generally an open question to identify whether a joint mix is supported by $(F_1,\dots,F_d)$.  It should be noted that the marginal distributions of a joint mix cannot be one-sided (e.g. $F_j^{-1}(0) > -\infty$  for all $1\le j\le d$ and  $F_j^{-1}(1) = \infty$ for some $j$); see Proposition 2.1(7) in~\citet{WW11}.
Below, we list some existing results in their most general form.
Proposition~\ref{th:jointly mixable1} and Theorem~\ref{th:jointly mixable3} below are given in \citet{WW14}. In the following $d \in \mathbb{N}$, although
the cases $d \leq 2$ are trivial. We say a function $||\cdot||:L^1\to [0,\infty]$ is a \emph{law-determined norm} if $||\cdot||$ satisfies: (i) $||X||=0$ if and only if $X=0$ a.s.;  (ii) $||\lambda X||=|\lambda|\cdot||X||$ for all $\lambda \in \R$ and $X \in L^1$; (iii) $||X+Y||\le ||X||+||Y||$ for all $X,Y \in L^1$; (iv) $||X||=||Y||$ if $X\laweq Y$. Note that here we allow $||\cdot||$ to take value in $+\infty$ and hence it is not a proper norm in classic functional analysis.

\begin{proposition}[Necessary conditions for joint mixability]\label{th:jointly mixable1}
For $j=1,\dots,d,$ let $\mu_j$ be the mean of $F_j$, $a_j=\sup\{x:F_j(x)=0\}$, $b_j=\inf\{x:F_j(x)=1\}$ and $l_j=b_j-a_j$.
If the $d$-tuple of distributions $(F_1,\dots,F_d)$ is jointly mixable, and $\mu_1,\dots,\mu_d$ are finite, then the following inequalities hold:
\begin{enumerate}[(a)]
\item (Mean inequality)
\begin{equation}\sum_{j=1}^d a_j+ \max_{j=1,\dots,d}l_j\le \sum_{j=1}^d \mu_j\le\sum_{j=1}^d b_j-\max_{j=1,\dots,d}l_j.\label{meancd}\end{equation}
\item(Norm inequality)
\begin{equation}\sum_{j=1}^d ||X_j-\mu_j||\ge 2\max_{j=1,\dots,d} ||X_j-\mu_j||,\label{polycd}\end{equation}
where $X_j\sim F_j$, $j=1,\dots,d$ and  $||\cdot||$ is any law-determined norm on $L^1$.\end{enumerate}
\end{proposition}
As special cases of Proposition~\ref{th:jointly mixable1}, the following conditions hold if $(F_1,\dots,F_d)$ is jointly mixable:
\begin{equation}\sum_{j=1}^d l_j\ge 2\max_{j=1,\dots,d} l_j,\label{lcd}\end{equation}
and
\begin{equation}\sum_{j=1}^d \sigma_j\ge 2\max_{j=1,\dots,d} \sigma_j.\label{varcd}\end{equation}
where $\sigma_j^2$ is the variance of $F_j$, $j=1,\dots,d$. All the above quantities are not necessarily finite. The conditions~\eqref{lcd} and~\eqref{varcd} are usually easier than~\eqref{polycd} to check, and sometimes they can also be sufficient, as stated below.



\begin{theorem}[Sufficient conditions for joint mixability]~\label{th:jointly mixable3}
\begin{enumerate}[(a)]
\item Suppose $F_1,\dots,F_d$ are $d$ distributions with decreasing densities on their respective supports. Then the $d$-tuple $(F_1,\dots,F_d)$ is jointly mixable if and only if the mean inequality \eqref{meancd} is satisfied.
\item  Suppose $F_1,\dots,F_d$ are distributions with unimodal-symmetric densities, and mode 0. Let $f_j(x)$ be the density function of $F_j$ and let $G_j(x)=F_j(x)-xf_j(x)-\frac12$ for $j=1,\dots,d$ and $x\ge 0.$ Then $(F_1,\dots,F_d)$ is jointly mixable  if for all $a\in (0,\frac{1}{2})$,
\begin{equation}\sum_{j=1}^d G_j^{-1}(a)\ge 2\max_{j=1,\dots,d} G_j^{-1}(a).\label{lcdsym}\end{equation}
In particular, suppose $F_1,\dots,F_d$ are unimodal-symmetric distributions from the same location-scale family. Then $(F_1,\dots,F_d)$ is jointly mixable if and only if  \eqref{polycd} holds for some law-determined norm $||\cdot||$.
\item Suppose $F_1,\dots,F_d$ are marginal distributions of a $d$-elliptical distribution. Then $(F_1,\dots,F_d)$ is jointly mixable if and only if  \eqref{polycd} holds for some law-determined norm $||\cdot||$.
\end{enumerate}

\end{theorem}

 For a definition of elliptical distributions see for instance~\citet{MNFE05}.
Further characterization results on joint mixability are available for \emph{homogeneous} Fr\'echet classes of the type $\FRH$ (here we use a subscript $d$ to indicate the dimension of the Fr\'echet class). The homogeneous version of joint mixability is called \emph{complete mixability}.

\begin{definition}\label{de:cm}
A distribution $F$ is said to be $d$-\emph{completely  mixable} ($d$-CM) if $\FRH$ supports a joint mix.
\end{definition}

Proposition \ref{th:jointly mixable4} (a) below is given in \citet[Theorem 8.3.10]{MS02};  (b)-(c) were given in \citet{PWW12}; (d) was given in \citet{PWW13}.

\begin{proposition}~\label{th:jointly mixable4}
\begin{enumerate}[(a)]
\item A $d$-discrete uniform distribution, that is a distribution giving probability mass $1/d$ to each of the $d$ points in its support, is $d$-CM.
\item The binomial distribution $B(d,p/q),p, q\in \mathbb N,$ is $q$-CM.
\item The Cauchy distribution is $d$-CM for $d\ge 2$.
\item Any  continuous distribution with a concave density on a bounded interval $[a,b]$ is $d$-CM for $d\ge 3$.
\item  Any continuous distribution function $F$ on a bounded interval $[a,b]$, $a<b$, having a density $f$ satisfying
\begin{equation}\label{eq:thmain}
f(x) \geq \frac{3}{d(b-a)}, \: \text{ for all } x \in [a,b],
\end{equation}
is $d$-CM.
\end{enumerate}
\end{proposition}

Even if a complete mathematical characterization of the class of jointly mixable distributions remains open,
it is  possible to numerically check whether a $d$-tuple of distribution functions is jointly mixable via the so-called Mixability Detection Procedure (MDP) introduced in~\citet{PW14}.

Joint mixability can help to identify the $\cx$-smallest element even if the Fr\'echet class does not support a joint mix; see \cite{WW11} and \cite{BJW13} for the cases of one-sided, unbounded marginal distributions. The concept directly relates to a class of optimization problems, such as the ones discussed in Section \ref{sec:4}, and Value-at-Risk maximization problems (which were the original motivation behind the concept; see \cite{WPY13}).  The recent developments of sufficient conditions for joint mixability typically involve techniques in probabilistic combinatorics, used for instance in the main results of \cite{WW11}, \cite{PWW12}, \cite{PWW13} and \cite{WW14}. A large class of distributions are \emph{asymptotically mixable}, see \cite{PWW13} and \cite{rW14}. This property makes joint mixability a flexible concept for the study of high-dimensional problems.


\subsubsection*{Historical Remark}
The concept of \emph{risks with a constant sum} goes back to~\citet{GR81}, where the complete mixability of a set of uniform distributions was shown. In Remark~1(b) in~\citet{lR82}, the author conjectures that concentrating a multivariate probability measure on a constant would yield optimal bounds for the distribution function of the sum of the marginal components.
The same notion appears in \citet{RU02}, \citet[Section 8.3.1]{MS02} and \citet{KS06} in the context of variance minimization or as the safest aggregate risk of some random variables.
The term \emph{complete mixability} was actually coined  and developed as a property of distributions in \citet{WW11}, and the term \emph{joint mixability} was introduced in \citet{WPY13}.
Theoretical properties of complete mixability and joint mixability have also been developed recently in \citet{PWW12}, \citet{PWW13}, \citet{PW14} and \citet{WW14}.
Some early work as special cases of Theorem \ref{th:jointly mixable3} are: \citet{RU02} showed the complete mixability of distributions with a unimodal-symmetric density; \citet{WW11} gave a necessary and sufficient condition for the complete mixability of distributions with monotone densities; \citet{WPY13} gave a necessary and sufficient condition for the joint mixability of  tuples of normal distributions; a similar result on the variance reduction of normal distributions can be found in \citet{KS06}.

\subsection{$\Sigma$-countermonotonicity}\label{se:PND}

Joint mixability  is a notion of extremal negative dependence which is far more applicable than pairwise countermonotonicity.
Nevertheless, not all $d$-tuples of distribution functions are jointly mixable and the $\cx$-smallest element in a Fr\'echet class might not   exist or might not be a joint mix.
At this point, it is natural to ask whether there exists a concept of negative dependence in dimensions $d>2$
which is supported in any Fr\'echet class and that includes countermonotonicity, pairwise countermonotonicity and joint mixability
as particular cases. The answer is affirmative: for this, we define the new notion of a \emph{$\Sigma$-countermonotonic} random vector
based on the requirement that the sum of any subset of its components  is countermonotonic with respect to the sum of the
remaining ones. All the results contained in this section are new.

\begin{definition}\label{de:DCM}
A  random vector $\vec{X}$ is said to be \emph{$\Sigma$-countermonotonic} if
for any subset $I \subset \{1,\dots,d\}$, we have that the random variables
$\sum_{j \in I} X_j$ and $\sum_{j \notin I} X_j$ are countermonotonic.
\end{definition}

The terminology $\Sigma$-countermonotonic stresses the sum operator as a basis for our criterion. It can be analogously defined for other operators, such as max, min or product. 

\begin{theorem}\label{th:frechetsupports}
Any Fr\'echet class $\FR$ supports a $\Sigma$-countermonotonic random vector.
\end{theorem}
\begin{proof}  The statement is trivial for $d=1$; we assume $d\ge 2$ in the following. First, we suppose that $F_1,\dots, F_d$ have finite second moments.
By a compactness argument (see for instance \citet{lR83}), there exists $(X_1,\dots,X_d)\in \FR$ such that
\begin{equation}\label{eq:dcm1}
\E \left[(X_1+\dots+X_d)^2 \right]=\inf \left\{ \E \left[
(Y_1+\dots+Y_d)^2 \right]: (Y_1,\dots,Y_d)\in \FR \right\} < \infty.
\end{equation}
We will show that any such $(X_1,\dots,X_d)$ is $\Sigma$-countermonotonic.
For some $k, 1 \leq k \leq d-1,$ define the two random variables
$$
Y_1:=X_1+\dots+X_k \quad  \text{ and } \quad Y_2:=X_{k+1}+\dots+X_d,
$$
and denote by $G_1$, respectively, $G_2$ their laws. We have that
$$
(X_1,\dots,X_d,Y_1,Y_2) \laweq \left(F_1^{-1} \circ f_1(U), \dots, F_d^{-1} \circ f_d(U), G_1^{-1} \circ g_1(U), G_2^{-1} \circ g_2(U) \right),
$$
for some $f_j \rg \,\Id$, $j=1,\dots,d$, $g_1,g_2\rg \, \Id$, $U\laweq U[0,1]$.
Let $Z_1 = g_2^{-1} \circ g_1(1-U)\laweq U[0,1]$. By properties of generalized inverses (see for instance Proposition~1 in~\citet{EH13})  we can write $Z_2:= G^{-1}_2 \circ g_1(1-U)=G_2^{-1}\circ g_2 (Z_1)= F_{k+1}^{-1} \circ f_{k+1}(Z_1)+ \dots+ F_d^{-1} \circ f_d(Z_1).$ Since $Y_1$ and $Z_2$ are countermonotonic and $(X_1,\dots,X_d)\in \FR$ implies $(Y_1,Y_2) \in \mathfrak F_2(G_1, G_2)$, we have
\begin{align}\label{eq:newversion}
\E[(Y_1+Z_2)^2] &= \inf \left\{ \E \left[(\tilde{Y}_1+\tilde{Y}_2)^2 \right]: (\tilde{Y}_1,\tilde{Y}_2)\in \mathfrak F_2(G_1, G_2) \right\}
\notag \\&\leq \inf \left\{ \E \left[(\tilde{X}_1+\dots+\tilde{X}_d)^2\right]: (\tilde{X}_1,\dots,\tilde{X}_d)\in \FR \right\} = \E [(X_1+\dots+X_d)^2].
\end{align}
Furthermore, note that
$$
\left(F_1^{-1} \circ f_1(U), \dots,F_k^{-1} \circ f_k(U), F_{k+1}^{-1} \circ f_{k+1}(Z_1), \dots, F_d^{-1} \circ f_d(Z_1) \right)\in \FR,
$$
implying that
\begin{align}\label{eq:newversion2}
 \E [(X_1+\dots+X_d)^2] =
\inf \left\{ \E \left[(\tilde{X}_1+\dots+\tilde{X}_d)^2\right]: (\tilde{X}_1,\dots,\tilde{X}_d)\in \FR \right\}\leq \E[(Y_1+Z_2)^2].
\end{align}
From~\eqref{eq:newversion} and~\eqref{eq:newversion2}, we finally obtain that
\begin{align*}
\E[(Y_1+Z_2)^2]&=\inf \left\{ \E \left[(\tilde{Y}_1+\tilde{Y}_2)^2 \right]: (\tilde{Y}_1,\tilde{Y}_2)\in \mathfrak F_2(G_1, G_2) \right\}
\\&=\inf \left\{ \E \left[(\tilde{X}_1+\dots+\tilde{X}_d)^2\right]: (\tilde{X}_1,\dots,\tilde{X}_d)\in \FR \right\}= \E [(X_1+\dots+X_d)^2]
\end{align*}
and therefore
$$
 \E [(X_1+\dots+X_d)^2] = \E [(Y_1+Y_2)^2] = \inf \left\{ \E \left[(\tilde{Y}_1+\tilde{Y}_2)^2 \right]: (\tilde{Y}_1,\tilde{Y}_2)\in \mathfrak F_2(G_1, G_2) \right\}.
$$
By Theorem \ref{th:comcoh} (e), we  have that $Y_1=X_1+\dots+X_k$ and $Y_2=X_{k+1}+\dots+X_d$ are countermonotonic. Since $k$ is arbitrary,
we can similarly show that
$\sum_{j \in I} X_j$ and $\sum_{j \notin I} X_j$ are countermonotonic for any $I \subset \{1,\dots,d\}$.

Now, given arbitrary distributions $F_1,\dots, F_d$, for each $j=1,\dots,d$,  let $\{F_{jk},~k\in \N\}$ be a sequence of  distributions with bounded support, such that $F_{jk}\overset{\text{\tiny d}}{\to} F_j$ as $k\rightarrow \infty$. For instance one can choose $F_{jk}(x):=F_j(x)\id_{\{|x|<k\}}+\id_{\{x>k\}}$, $x\in \R$. It follows from the first part of the proof that we can find a sequence of $\Sigma$-countermonotonic random vectors $\vec{X}_k\in \mathfrak{F}(F_{1k},\dots,F_{dk})$, $k\in \N$. Correspondingly, we can find a sequence $C_k,k \in \N$, so that each $C_k$ is a possible copula of $\vec{X}_k$, $k \in \N$.
Since the set of $d$-copulas is compact with respect to the weak topology, there exists a subsequence $C_{k_i},~ i\in \N,$ which converges  weakly to some $C_0$. Let $\vec{X}_0 \in \FR$ be a random vector having law $C_0(F_1,\dots,F_d)$.
The sequence of the joint distributions of the $\vec{X}_{k_i}$'s weakly converge to the joint distribution of $\vec{X}_0$.

Consequently, for a given $\vec{a}\in \{0,1\}^d$, the sequence of the joint distributions of the $(\vec{X}_{k_i}\cdot \vec{a}  ,  \vec{X}_{k_i} \cdot (\vec{1}-\vec{a}))$'s weakly converges to the joint distribution of  $(\vec{X}_0 \cdot \vec{a} ,  \vec{X}_0 \cdot (\vec{1}-\vec{a}))$, where $\vec b \cdot \vec c$ stands for the dot product of vectors $\vec b$ and $\vec c$. Being each $(  \vec{X}_{k_i}\cdot \vec{a},\vec{X}_{k_i} \cdot (\vec{1}-\vec{a}))$ countermonotonic, this finally  implies that $ \vec{X}_0 \cdot \vec{a}$ and $  \vec{X}_0 \cdot (\vec{1}-\vec{a})$ are  countermonotonic. From arbitrariness of $\vec{a}$, we conclude that $\vec{X}_0$ is $\Sigma$-countermonotonic.
 \end{proof}

We now prove that  $\Sigma$-countermonotonicity coincides with countermonotonicity in dimension $d=2 $ and with pairwise countermonotonicity in arbitrary dimensions when the latter is supported. Moreover, a joint mix and/or the $\cx$-smallest element  in a Fr\'echet class (when they exist)  are always $\Sigma$-countermonotonic.
Recall that  we write $\vX \in_d \FR$ if $X_j\laweq F_j$, $j=1,\dots,d$

\begin{theorem}\label{th:frechetsupports2}
Suppose $\vec{X}\in_d \FR$. The following holds.
\begin{enumerate}[(a)]
\item When $d=2$,  $(X_1,X_2)$ is countermonotonic if and only if $(X_1,X_2)$ is $\Sigma$-countermonotonic.
\item Suppose $\FR$ supports a  pairwise countermonotonic random vector, then $\vec{X}$ is pairwise countermonotonic if and only if $\vec{X}$ is $\Sigma$-countermonotonic.
\item Suppose $\FR$ supports a joint mix. If $\vec{X}$ is a joint mix, then $\vec{X}$ is $\Sigma$-countermonotonic.
\item Suppose  $\vec{X}$ is a  $\cx$-smallest element  in a Fr\'echet class, then $\vec{X}$ is $\Sigma$-countermonotonic.
\end{enumerate}
\end{theorem}
\begin{proof}
(a) This follows directly from Definition~\ref{de:DCM}. (b) Assume, without loss of generality, that $F_j^{-1}(0)=0$, $j=1,\dots,d$, and that \eqref{eq:da1} holds, i.e.
\begin{equation}\label{eq:neededede}
\sum_{j=1}^d P(X_j>0) \leq 1.
\end{equation}
If $\vec{X}$ is pairwise countermonotonic, for any $\vec{a} \in \{0,1\}^d$ at most one of $\vec{X}\cdot \vec{a}$ and $ \vec{X} \cdot (\vec{1-a})$ can be strictly positive, then ($\vec{X} \cdot \vec{a},\vec{X}\cdot (\vec{1-a}))$ is pairwise countermonotonic in dimension $d=2$ and hence countermonotonic. Conversely, assume that $\vec{X}$ is $\Sigma$-countermonotonic and write
$X_{-k}:=\sum_{j \neq k}X_j$.
First observe that
$$
P(X_{-k}=0)=1-P(X_{-k}>0)=1-P(\cup_{j \neq k} \{X_j>0\}) \geq 1- \sum_{j \neq k} P(X_j>0) \geq P(X_k>0)=1-P(X_k=0),
$$
where the last inequality follows from~\eqref{eq:neededede}. Hence
\begin{equation}\label{eq:90}
P(X_{-k}=0)+P(X_k=0)-1 \geq 0.
\end{equation}
Using elementary probability we find that
$$
P(X_k>0,X_{-k}>0)=1-P(X_k=0)-P(X_{-k}=0)+P(X_k=0,X_{-k}=0).
$$
Since $X_k$ and $X_{-k}$ are countermonotonic, from~\eqref{eq:conmdf} and using~\eqref{eq:90} we obtain
$$
P(X_k>0,X_{-k}>0)=1-P(X_k=0)-P(X_{-k}=0)+\max\{P(X_k=0)+P(X_{-k}=0)-1,0\}=0
$$
Consequently $P(X_k>0, X_j>0)=0$ for all $j\ne k$, i.e. $\vec{X}$ is pairwise countermonotonic.
(c) If $\vec{X}$ is jointly mixable, then $ \vec{X} \cdot \vec{a}+\vec{X} \cdot (\vec{1-a})=X_1+\dots+X_d=k$ with probability one. Therefore, $\vec{X} \cdot \vec{a}$ and $\vec{X} \cdot (\vec{1-a})$ are countermonotonic and $\vec{X}$ is $\Sigma$-countermonotonic.
(d) This follows similarly from the  proof of Theorem \ref{th:frechetsupports} by replacing $\E[(X_1+\dots+X_d)^2]$ in~\eqref{eq:dcm1} with $\E[f(X_1+\dots+X_d)]$, where $f$ is any strictly convex function.
\end{proof}

Since $\Sigma$-countermonotonicity always exists for any Fr\'echet class, and is equivalent to pairwise countermonotonicity when the latter is supported, we  consider it as a more fundamental concept compared to pairwise countermonotonicity. Denote now by $\mathcal{X}^{\Sigma}$ the set of all $\Sigma$-countermonotonic random vectors having marginal distributions $F_1,\dots,F_d$. Theorems~\ref{th:frechetsupports} and~\ref{th:frechetsupports2} imply the following properties.
\\
\\
\fbox{\parbox[center]{6.6in}{
\begin{enumerate}[\quad]
\item \textbf{(E)} \emph{Existence.} $\mathcal{X}^{\Sigma} \neq \emptyset$ for any choice of $F_1,\dots,F_d$.
\item \textbf{(M2)} \emph{Minimization of supermodular functions.} If $\vec{X}^{\Sigma} \in_d \FR$ is $\cx$-smallest, then $\vec{X}^{\Sigma}\in \mathcal{X}^{\Sigma}$.
\end{enumerate}
}}
\\
\\
\begin{example}[Multivariate normal distributions, continued]
In Example \ref{ex:multnorm} assume, without loss of generality, that $\sigma_1> \sigma_2> \sigma_3>0$.  By Definition~\ref{de:DCM} it is easy to see that $(X_1,X_2,X_3)$ is $\Sigma$-countermonotonic if and only if the following equations hold
\begin{equation}\label{eq:ex31q1}\rho(X_1+X_2,X_3)=\rho(X_1+X_3,X_2)=\rho(X_2+X_3,X_1)=-1,\end{equation}
where $\rho$ is Pearson's correlation coefficient (see~\eqref{eq:pcorr} below).
There are two sets of solutions of $(\sigma_{12},\sigma_{13},\sigma_{23})$ to \eqref{eq:ex31q1}:
\begin{enumerate}
\item  $\sigma_{12}=-\sigma_1\sigma_2$, $\sigma_{13}=-\sigma_1\sigma_3$, and $\sigma_{23}=\sigma_2\sigma_3$. In this case, $(X_1,X_2)$ and $(X_1,X_3)$ are countermonotonic, while $(X_2,X_3)$ is comonotonic. Roughly speaking, the two components $X_2,X_3$ move together oppositely to $X_1$.
\item $\sigma_{12}=\frac{1}{2}(\sigma_3^2-\sigma_1^2-\sigma_2^2)$, $\sigma_{13}=\frac{1}{2}(\sigma_2^2-\sigma_1^2-\sigma_3^2)$, and $\sigma_{23}=\frac{1}{2}(\sigma_1^2-\sigma_2^2-\sigma_3^2)$.
From Example \ref{ex32},  $\vec{\Sigma}$ with this choice of $(\sigma_{12},\sigma_{13},\sigma_{23})$ is positive semidefinite if and only if $\sigma_1\le \sigma_2+\sigma_3$. In that case, $(X_1,X_2,X_3)$ is a joint mix as in Example \ref{ex32}.
\end{enumerate}
\end{example}

\begin{figure}[h]
\begin{center}
\scalebox{.40}{\includegraphics{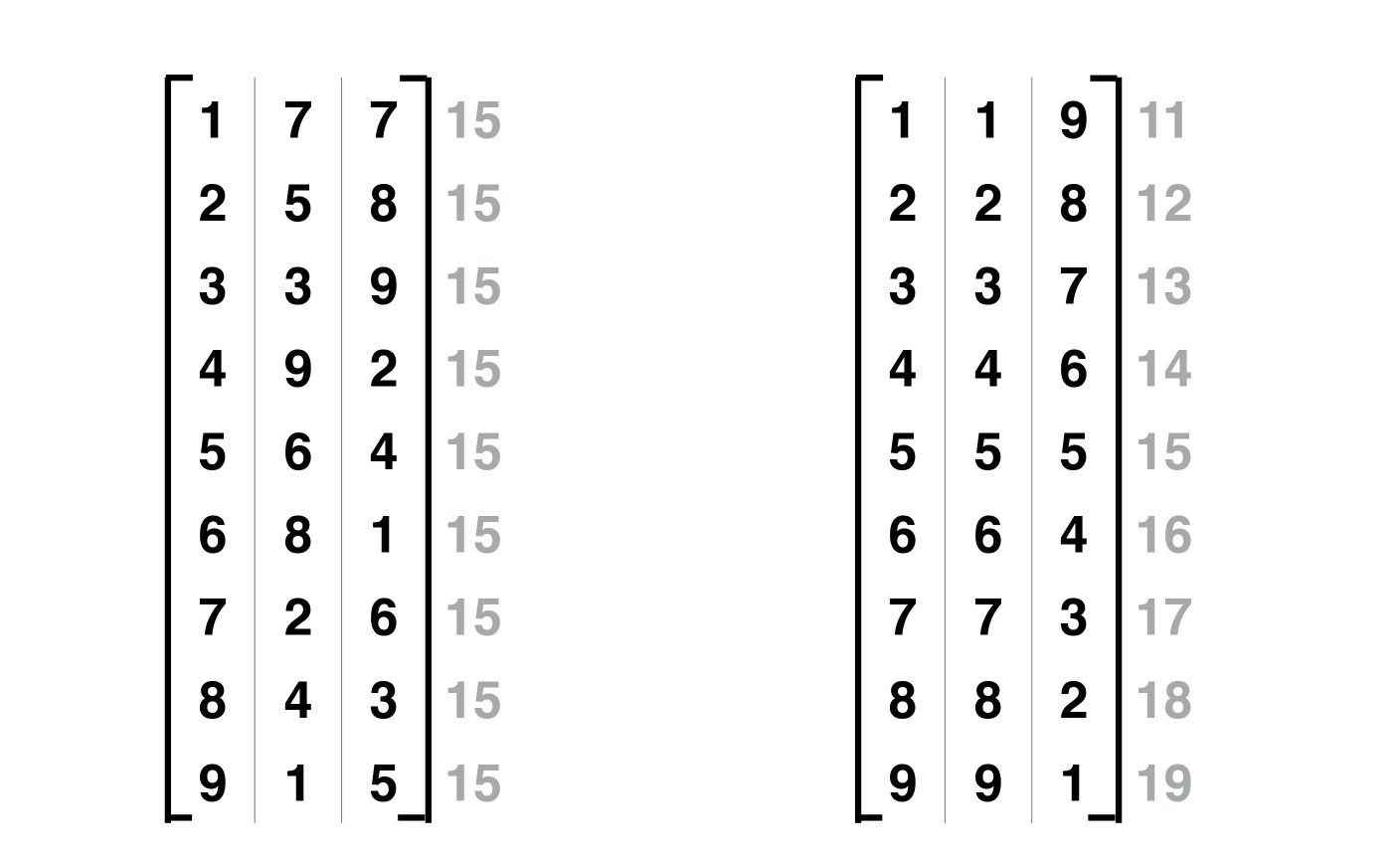}}
\caption{\small Rearrangement matrices representing the three-variate distribution functions of: a joint mix (left) with constant sum of the marginal components (reported outside the matrix);  of a
$\Sigma$-countermonotonic vector (right) with the same marginals. }
\label{fi:sigmacount}
\end{center}
\end{figure}

Even if $\Sigma$-countermonotonic random vectors are supported in any Fr\'echet class,
a single $\Sigma$-countermonotonic random vector might not possess a desired optimality property.
For example let $F_1=F_2=F_3=U[0,1]$.
In~\citet[Example~3]{GR81} the authors give an example of a jointly mixable vector $\vec{U}^*$ with uniform marginals. Being a joint mix, $\vec{U}^*$ is also a $\cx$-smallest element in $\mathfrak{F}(U[0,1],U[0,1],U[0,1])$ and, by Theorem~\ref{th:frechetsupports2}(c), also $\Sigma$-countermonotonic. However, it is straightforward to check that the vector
$$
\vec{U}:=(U,U,1-U) \in_d \mathfrak{F}(U[0,1],U[0,1],U[0,1])
$$
is $\Sigma$-countermonotonic; it is however not a $\cx$-smallest element
in $\mathfrak{F}(U[0,1],U[0,1],U[0,1])$. In Figure~\ref{fi:sigmacount} we give a representation in terms of rearrangement matrices of another $\Sigma$-countermonotonic vector which is not a $\cx$-smallest element in its Fr\'echet class. Despite these counterexamples, it is possible to show that $\Sigma$-countermonotonic random vectors possess a local optimality property which is illustrated by  Proposition \ref{th:localopt}.

\begin{proposition}\label{th:localopt}
Let $\vec{X} \in \FR$ be a $\Sigma$-countermonotonic random vector. For $i=1,\dots,d$, consider the random vector
$\vec{Y}_i:=(X_1,\dots,X_{i-1},Y_i,X_{i+1},\dots,X_d)$, where $Y_i \laweq F_i$.
Then $\vec{Y}_i \in \FR$ and
$$
\sum_{j=1}^{d} X_j \lcx \sum_{j=1}^{d} Y_j.
$$
\end{proposition}
\begin{proof} Denoting $X_{-i}^+:=X_1+\dots+X_d-X_i$, the theorem follows by noting that for any convex function $f$ we  have that
$$
\E[f(X_1+X_2+\dots+X_d)]=\E[f(X_i+X_{-i}^+)] \leq \E[f(Y_i+X_{-i}^+)] =\E[f(Y_1+\dots+Y_d)],
$$
where the above inequality is implied by Theorem~\ref{th:comcoh}(d) since $X_i$ and $X_{-i}^+$ are countermonotonic. \end{proof}

The local property stated in Proposition~\ref{th:localopt} means that the sum of the components of a $\Sigma$-countermonotonic vector $\vec{X}$ is always $\cx$-dominated by any vector obtained by changing a single rearrangement in the dependence structure of  $\vec{X}$. We note that changing more than one rearrangement would not maintain the optimality in Proposition~\ref{th:localopt} as for instance the vector $(U,U,1-U)$ is not $\cx$-dominated by any joint mix $\vec{U}^*$, which can be always obtained from $(U,U,1-U)$ by changing two components. We remark that, although a $\Sigma$-countermonotonic vector is always supported in a Fr\'echet class, it is not trivial to determine its law. A numerical procedure to find $\Sigma$-countermonotonic vectors in a discrete setting can be built analogously to the Rearrangement Algorithm introduced in~\citet{PR12}.

\begin{remark}
In~\citet{LA14} a different notion of extremal negative dependence is introduced; see also~\citet{AHN15} for some further developments.
A $d$-random vector $\vec{X}$ with continuous marginal distributions is said to be \emph{$d$-countermonotonic} if
it is possible to find strictly increasing functions $f_1,\dots,f_d$ such that
$f_1(X_1)+\dots +f_d(X_d)= 1 \text{ with probability one.}$
Equivalently, $\vec{X}$ is $d$-countermonotonic if and only if there exist strictly increasing functions $f_i$, such that $(f_1(X_1),\dots,f_d(X_d))$ is a joint mix. From the definition, it directly follows that a joint mix is always $d$-countermonotonic.

The notion of $d$-countermonotonicity only depends on the
dependence structure of a random vector and not on its marginal distributions.
This means that any random vector sharing the same copula of a  $d$-countermonotonic vector is $d$-countermonotonic; see Lemma~1 in~\citet{LA14}.
Defining a concept of extremal negative dependence not depending on the marginals has some relevant consequences when $d>2$. First of all, $d$-countermonotonicity is a too
general notion; for instance it is easy to see that any vector
$(U,U,\dots,U,1-U)$ is $d$-countermonotonic.
This implies that $d$-countermonotonicity and $\Sigma$-countermonotonicity are  different dependence concepts.
Furthermore, any dependence concept that does not take into account the marginal distributions fails to solve any optimization problems for $d>2$ which depends on the margins. It is easy to show that, under an extra continuity assumption, $d$-countermonotonicity is a weaker notion than $\Sigma$-countermonotonicity.
\end{remark}

\begin{proposition}
If $\vec{X}$ is a $\Sigma$-countermonotonic random vector with continuous marginals and
$\sum_{j\neq i} X_j$ is continuously distributed for some $i$, then $\vec{X}$ is $d$-countermonotonic.
\end{proposition}
\begin{proof}
The result follows by noting that if the marginal distributions $F_1$ and $F_2$ of a $2$-countermonotonic random vector $(X_1,X_2)$ are continuous,
then it is possible to find a strictly increasing function $f$ such that
$f(X_1)+X_2= 1 \text{ with probability one.}$
For instance one can choose $f(X_1):=-F_2^{-1}(1-F_1(X_1))+1$ which is strictly increasing on the range of $X_1$.
If $X_j$ and $\sum_{j\neq i} X_j$ are continuous and countermonotonic by assumption then we can find a strictly increasing function $g$ such that
$g(X_j)+\sum_{j\neq i} X_j= 1 \text{ with probability one,}$ showing that $\vec{X}$ is $d$-countermonotonic.
\end{proof}



\section{Optimization problems}\label{sec:4}

In a situation  where one wants to describe the influence of the dependence structure on a statistical problem, with given marginals of the random vector under study, one considers an optimization problem
over the Fr\'echet class $\FR$ of all joint distributions with given marginals $F_1,\dots,F_d$. We suppress the explicit notation of the marginals and assume that they are fixed throughout this section.
For a given measurable function $c: \R^d \to \R$, an optimization problem over the class of possible dependence structures takes the form
\begin{align}
 M(c):= \sup &\left\{ \int c \, dF: F \in \FR \right\}\label{eq:optmax},~~\mbox{or}~~\\
m(c):=\inf & \left\{ \int c \, dF: F \in \FR \right\}.\label{eq:optmin}
\end{align}
In~\eqref{eq:optmax} and~\eqref{eq:optmin} (and in what follows) the supremum and infimum are meant to be taken over all  $F\in \FR$ such that the integral $\int c\,dF $ is well-defined.  As $\FR$ is a compact set with respect to the weak topology,
the domain of the supremum in~\eqref{eq:optmax} is an interval and the sup is attained under very general boundedness or continuity properties of $c$; see Theorem~2.19 in~\citet{hK84}.  Several different techniques to compute $M(c)$ and $m(c)$ exist. The  functional $\int c \, dF$ is linear in $F$ and has to be optimized over the convex set $\FR$.
 For instance, $M(c)$ can be considered as an infinite dimensional linear optimization problem and, as such, possesses the dual formulation
\begin{align}
 D(c):= \inf \left\{
 \sum_{j=1}^d \int f_j \, dF_j: \int f_j \, dF_j<\infty \; \text{ s.t.}\; \bigoplus f_j \geq c
 \right\} \label{eq:dual},
\end{align}
where $\bigoplus f_j(\vec{x}):=\sum_{j=1}^{d}f_j(x_j)$. 
We will throughout assume mean-compatibility of $c$ to guarantee that such $f_1,\dots,f_d$ exist and $D(c)$ is finite. While we always have
\begin{equation}\label{eq:dual1}
\int c \, dF \leq M(c) \leq D(c) \leq \sum_{j=1}^d \int f_j \, dF_j,
\end{equation}
for any $F \in \FR$ and $f_1,\dots, f_d$ satisfying $\bigoplus f_j \geq c$, the equality $M(c) = D(c)$ holds under very weak conditions depending on the function $c$ considered.
The problem $m(c)$ has an analogous dual representation; we refer the interested reader to Section 2.1 in~\citet{lR13} for a comprehensive summary of known results.
Since $M(c)$ and $m(c)$ can be seen as mass transportation problems, they have also been extensively treated in the more specific literature on mass transportation; see for instance~\citet{GMC96},~\citet{RR98} and the recent~\citet{bP14}.

The two formulations~\eqref{eq:optmax} and~\eqref{eq:dual} are typically used together in the so-called \emph{coupling-dual} approach, where  one has to find a joint distribution $F \in\FR$ (also called a \emph{coupling}; see~\citet{tL92}) and an admissible \emph{dual choice} $f_1,\dots,f_d$ for which
$$
\int c \, dF = \sum_{j=1}^d \int f_j \, dF_j,
$$
implying that all inequalities in~\eqref{eq:dual1} hold with $=$.
The case in which $c$ is supermodular has been extensively studied; see Sections~\ref{se:ppd} and~\ref{se:pndhd}.
This includes the case $c=\idc\{\bigtimes_{j=1}^d A_j\}$, for $A_i \subset \R, j=1,\dots,d$ which is treated in~\citet{lR81} and implies in particular the \emph{Hoeffding-Fr\'echet bounds} in~\eqref{eq:frechet}. Problems which can be linked to the maximization of a supermodular function include the minimization of a metric $d$ ($c(x_1,x_2)=-d(x_1,x_2)$; see for instance~\citet{CARTD93}) and the maximization of stop-loss functionals of the type $c(x_1,\dots,x_d)=\left(\sum_{j=1}^d x_j - k\right)_+$, for $k \in \R$; see~\citet[Chapters 3 and 4]{MS02}. For $c(x_1,\dots,x_d)=\idc\{\max\{x_i:i=1,\dots,d\} \leq s\}$, \citet{LR78} is the standard reference; the $\max$ operator is replaced by any order statistics in~\citet{tR96}. Maximization of supermodular functions is closely related to the maximization of a variety of risk measures; see~\citet{DVGKTV06}.

In general, the dual formulation in~\eqref{eq:dual} is  difficult to solve. Only partial solutions under restrictive
assumptions have been given in the above mentioned literature. Apart from the cases treated in Theorems~\ref{th:comcoh} and~\ref{th:merth}, there does not exist a general
analytical solution for $m(c)$ when $c$ is supermodular; see for instance~\citet{BJW13} and references therein.

If the dual formulation~\eqref{eq:dual} can rarely be used to obtain an analytical solution for $M(c)$ and $m(c)$,
rearrangement functions provide an easy way to reformulate the problem and compute
a numerical approximation. Using Theorem~\ref{pr:1}, problem~\eqref{eq:optmax} can be reformulated in terms of rearrangements. The following proposition is a rewriting of Lemma~1 in~\citet{lR83}.

\begin{proposition}\label{pr:rgm}
 If $U \laweq U[0,1]$, then
\begin{align}
M(c)=\sup &\left\{ \E \left[c\left(F_1^{-1} \circ f_1(U), \dots, F_d^{-1} \circ f_d(U)\right)\right]: f_j \rg \Id, ~j=1,\dots,d \right\}. \label{eq:rgmmax}
\end{align}
\end{proposition}

If the random variable $U$ is discretized and general rearrangement functions are replaced with one-to-one, piecewise continuous rearrangements as in Definition~\ref{de:som},
 the formulation in~\eqref{eq:rgmmax}  allows for a discrete representation of the corresponding problem. Denote by $U_n[0,1]$ a random variable uniformly distributed over the components of the vector $\IN:=(0,1/n,2/n,\dots,(n-1)/n)$; $\IN$ may also be chosen as $(1/(n+1),2/(n+1),\dots,n/(n+1))$ to avoid possible singularity at $0$.
A one-to-one, piecewise continuous rearrangement $f(U_n(0,1))$ implies a rearrangement
of the components of $\IN$. Therefore, a $d$-tuple of one-to-one, piecewise continuous rearrangements can be written in terms of an $(n \times d)$-matrix $\vec{X}=(x_{i,j})$ in which each column represents the implied rearrangements of the components of $\IN^t$. Any permutation of the elements within each column of $\vec{X}$ represents a different mutually complete dependence structure  among the same discrete marginals. A discretized version of the problem $M(c)$ can then be written as
\begin{equation}\label{eq:rgmprob}
M_n(c):= \frac1n \, \max \left\{\sum_{i=1}^{n} c\left(F_1^{-1}(x_{i,1}), \dots, F_d^{-1}(x_{i,d})\right) :\vec{X} \in \mathcal{P}_n\right\},
\end{equation}
where $\mathcal{P}_n$ is the set of all $(n \times d)$-matrices obtained from $(\IN^t,\dots,\IN^t)$ by rearranging the elements within a number of its columns in a different order.
Based on  approximation theorems, for example described in~\citet{DS12}, the transition from general rearrangements to the bijective, piecewise continuous ones  is justified if $n$ is large enough; formally
$$
M_n(c) \overset{n \to \infty}{\rightarrow} M(c).
$$

Though the domain $\mathcal{P}_n$ in~\eqref{eq:rgmprob} is computationally intractable, there exists an algorithm by which a very good approximation to $M_n(c)$ -- and hence to $M(c)$ -- can be computed in a relatively fast way. This is the \emph{rearrangement algorithm} first introduced in~\citet{PR12} for the computation of lower and upper bounds on distribution functions and also suitable to handle the approximation of $m(c)$ when $c$ is supermodular. The state-of-the-art of the Rearrangement Algorithm can be checked online.\footnote{At the web-page
\texttt{https://sites.google.com/site/rearrangementalgorithm/}.} 
Using $c(x_1,\dots,x_d)=(x_1+\cdots+x_d)^2$ (which leads to variance minimization problems), the algorithm is extremely effective in testing whether a Fr\'echet class admits joint mixability; see~\citet{PW14}. This application leads to a numerical answer to the more general question of whether a Fr\'echet class supports a vector $\vec{X}$ such that $X_1+\dots+X_d$ has a particular distribution; see also~\citet{WW14} on this. To fully capture the advantages of the formulations in~\eqref{eq:rgmmax} and in~\eqref{eq:rgmprob} we give an application.

\begin{figure}[h]
\begin{center}
\scalebox{.22}{\includegraphics{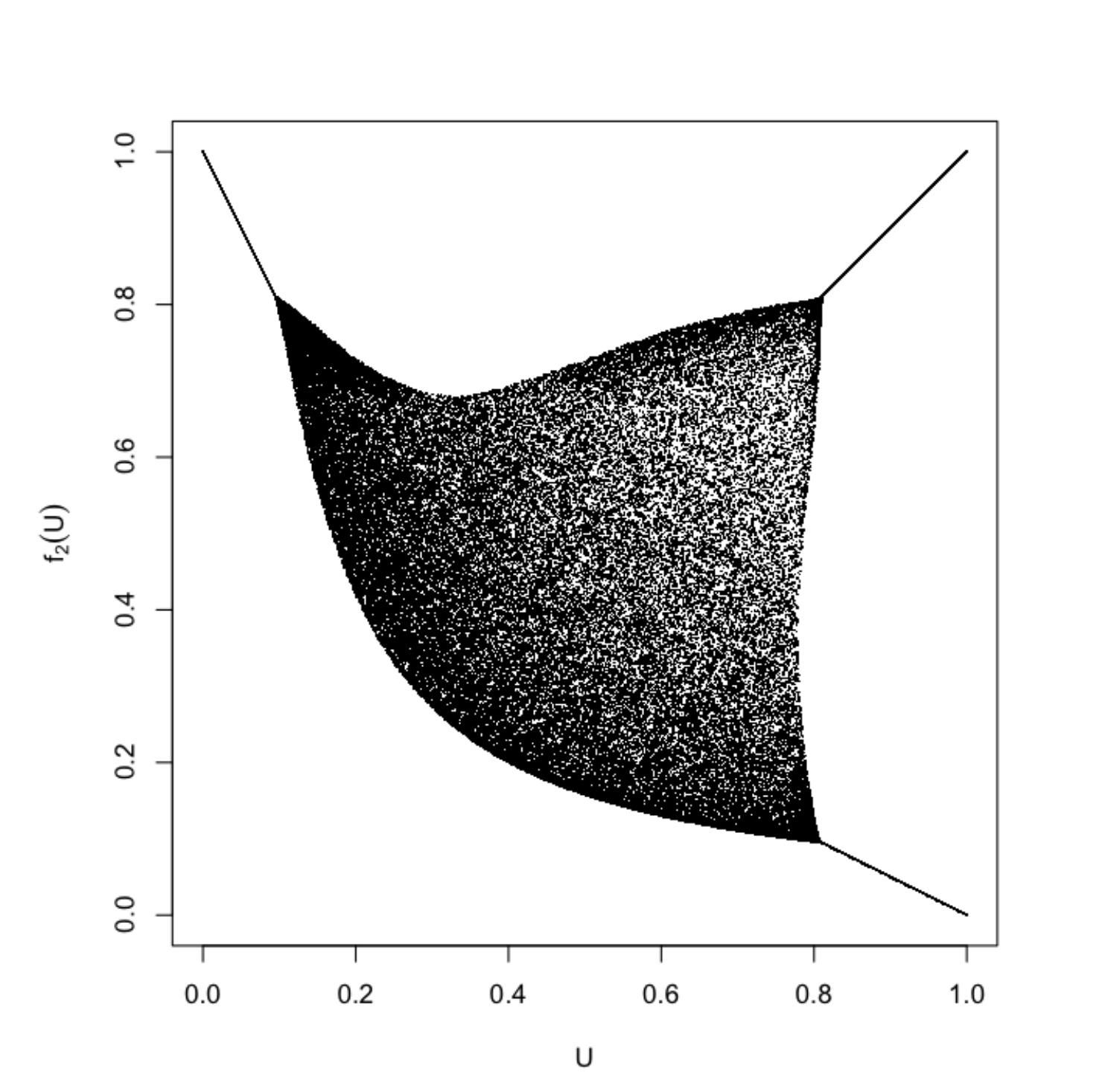}\includegraphics{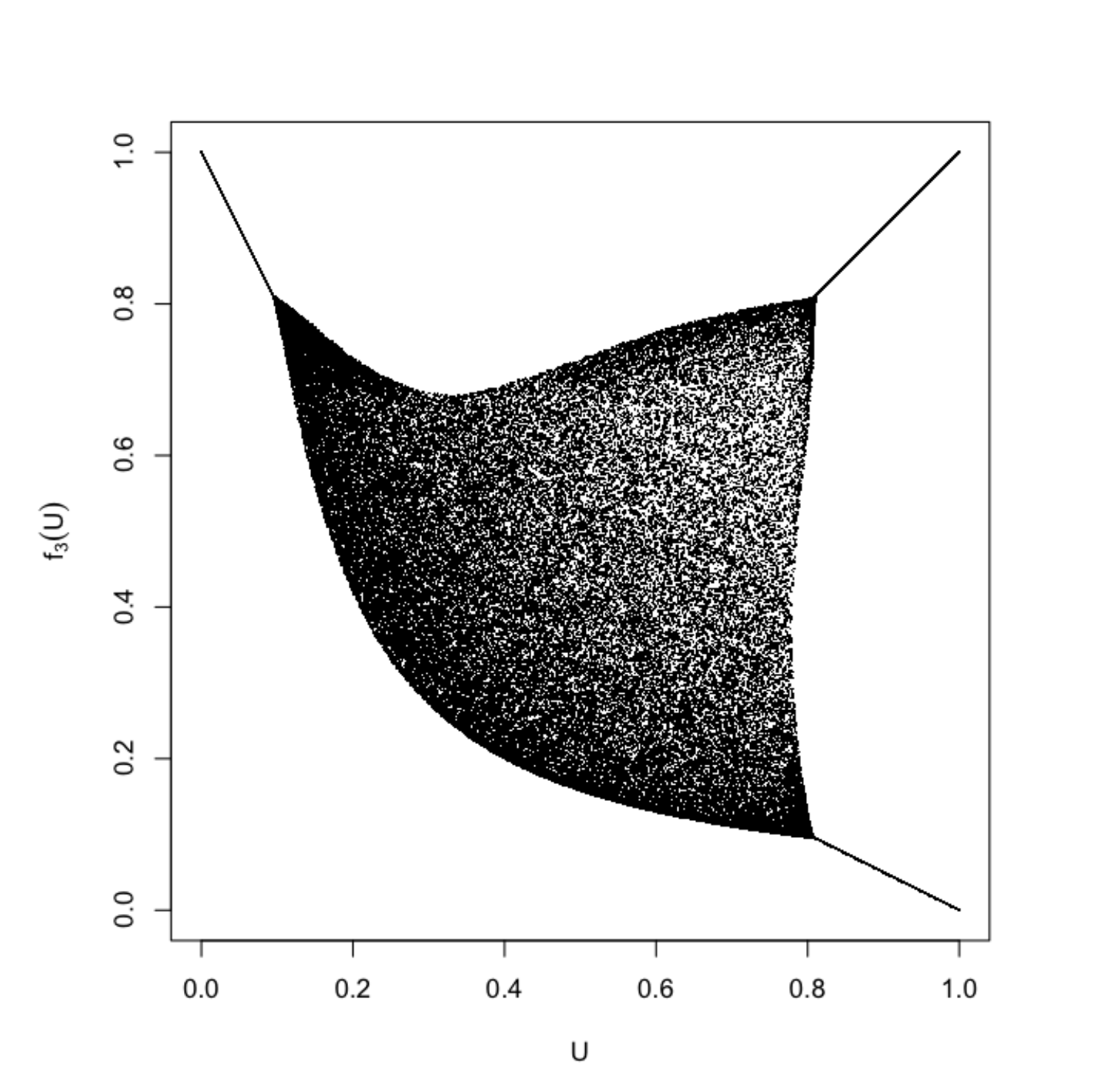}}
\caption{\small Optimal set of rearrangements $f_2,f_3$ attaining $m(\Pi)$ in~\eqref{eq:product}}
\label{fi:rgnprod1}
\end{center}
\begin{center}
\scalebox{.22}{\includegraphics{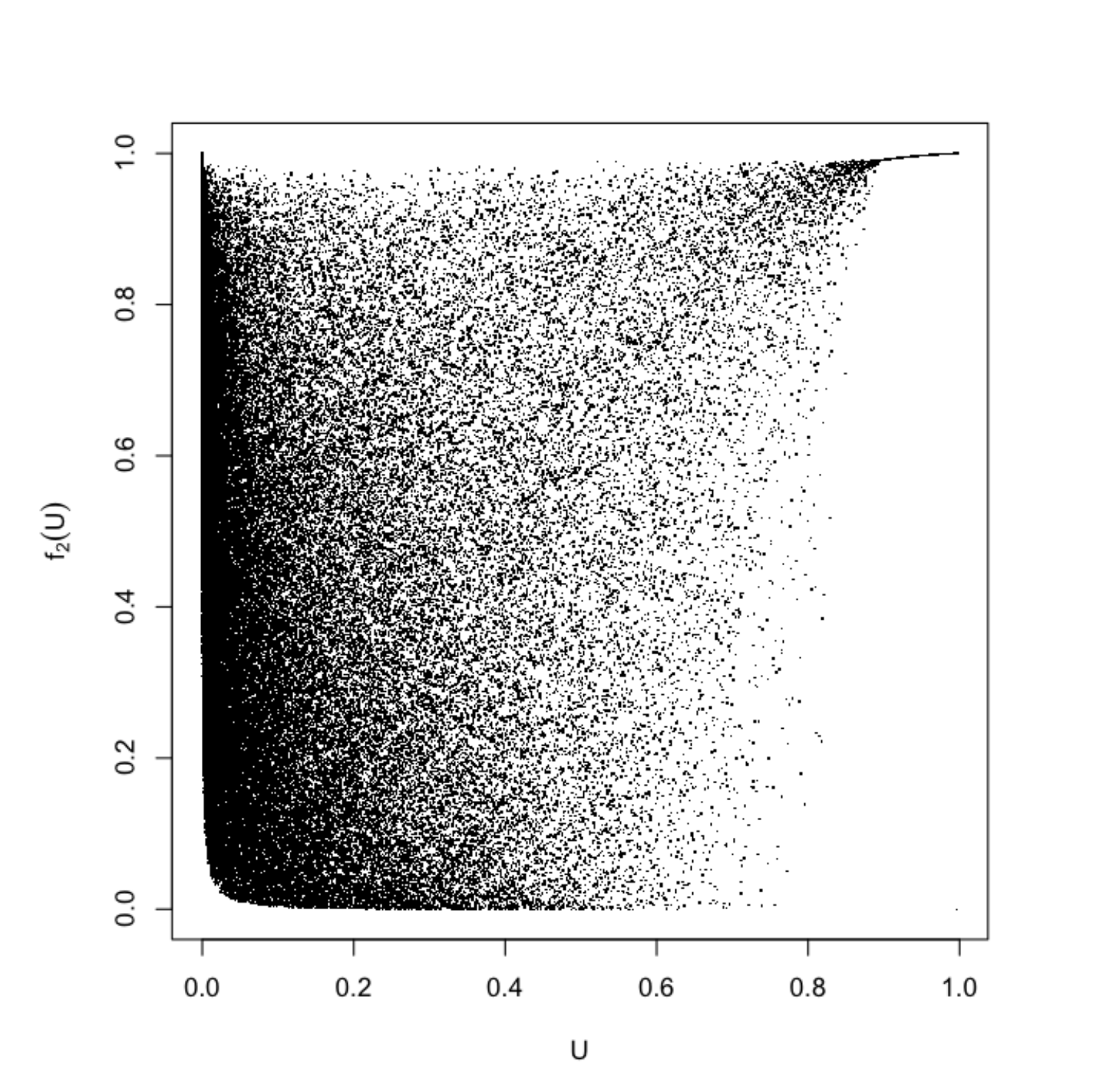}\includegraphics{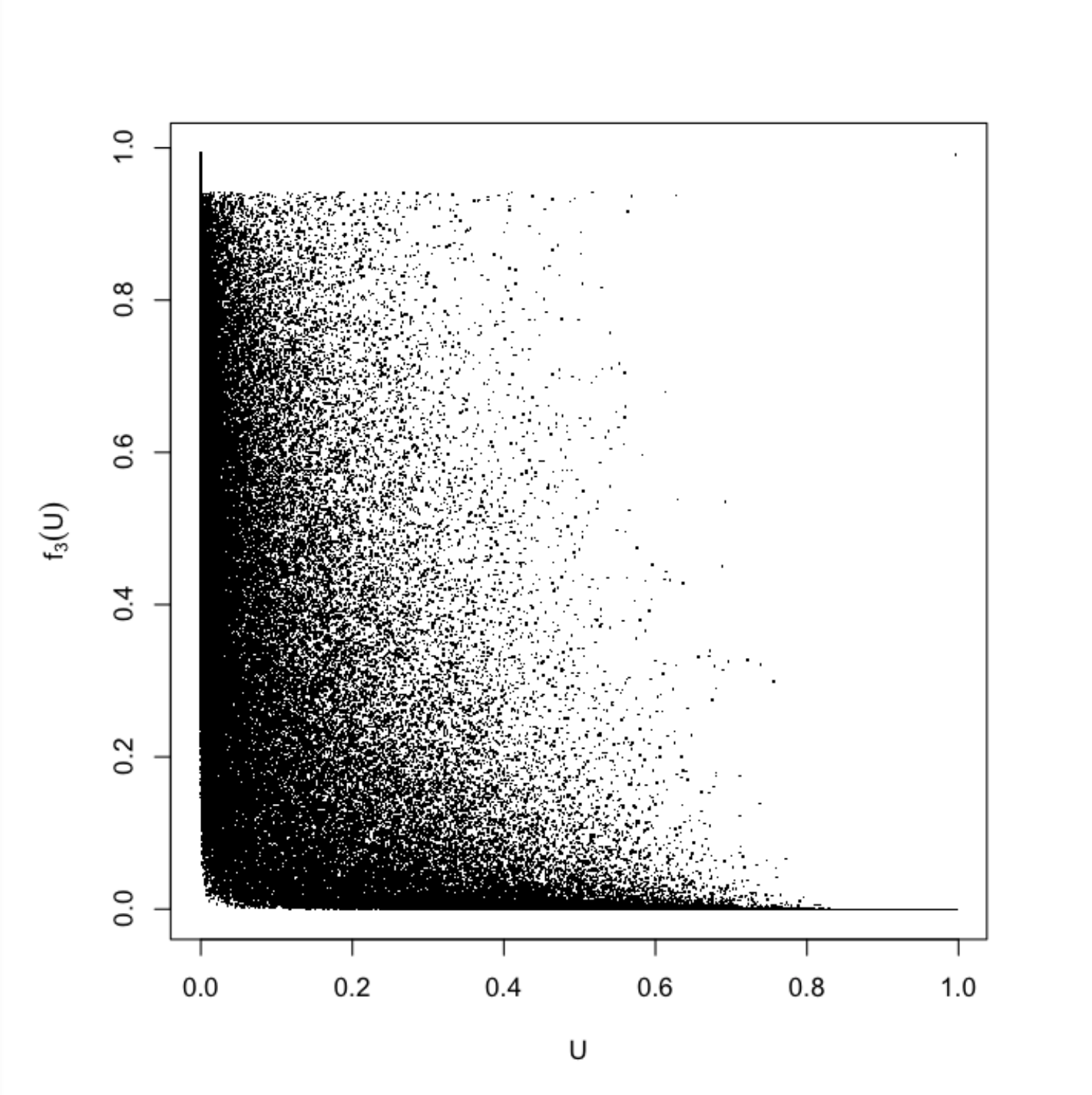}}
\caption{\small Optimal set of rearrangements $f_2,f_3$ attaining the minimal expectation of the product of three random variables with distributions $F_1=\text{Pareto(4)}$, $F_2=\text{LogN(0,1)}$, $F_3=\text{Exp}(1)$.}
\label{fi:rgnprod2}
\end{center}
\end{figure}

\begin{example}[Dependence measures]
\emph{Dependence measures} yield a scalar measurement for a pair of random variables $(X_1 X_2 )$, indicating the  strength of positive or negative dependence among its components.
Probably, the most widely known and used dependence measure is Pearson's linear correlation
\begin{equation}\label{eq:pcorr}
\rho(X,Y)=\frac{\cov(X,Y)}{\sqrt{\var(X)\var(Y)}},
\end{equation}
for $X,Y \in L^2$.
It is a measure of linear dependence that takes value in the range $[-1,1]$.
For a random vector $(X_1,X_2)$ having fixed marginal distributions, $\rho(X_1,X_2)$ is maximized by a comonotonic dependence structure and minimized by a countermonotonic one; see Theorem~4
in~\citet{EMNS02}. For a fixed pair of marginal distributions, however, the largest (smallest) value of $\rho(X_1,X_2)$ may be strictly smaller (larger) than 1 (-1); see Example~5 given in~\citet{EMNS02}.
Indeed, it is very well known that $\vert \rho(X_1,X_2) \vert=1$ if and only if $X_1$ is a.s. a \emph{linear } function of $X_2$. Pearson's linear correlation has two relevant drawbacks: it is well defined only when $X_1$ and $X_2$ have a finite variance; it does not only depend on the copula of the vector, but also depends  on the shape of the marginal distributions involved.

In order to overcome these deficiencies, \emph{copula-based} dependence measures have been developed. In contrast to ordinary correlation, these measures are functions of the copula only.
One among these copula-based dependence measures is Spearman's rank correlation coefficient,  defined as:
$$
\rho_S(X_1,X_2)= \rho(F_1(X_1),F_2(X_2)).
$$
Spearman's rank correlation takes value in the range $[-1,1]$, does not depend on the marginal distributions of a vector $(X_1,X_2)$ but only on its copula. It takes value 1 when $X_1$ and $X_2$ are comonotonic and value $-1$ when they are countermonotonic; see Theorem~3 in~\citet{EMNS02}.
A multivariate version of Spearman's rank correlation coefficient was introduced in~\citet{hJ90} (see also \cite{SS07BB} for its statistical inference). For a random vector $(X_1,\dots,X_d)$ having marginal distributions $F_1,\dots,F_d$, the multivariate Spearman's rho is defined as
$$
\rho_S{(X_1,\dots,X_d)}=\frac{d+1}{1-(d+1)2^{-d}}\E\left[\prod_{i=1}^d F_i(X_i)-2^{-d}\right].
$$
Since $(x_1,\dots,x_d)\mapsto \prod_{i=1}^d F_i(x_i)$ is a supermodular function, for a fixed set of distributions $F_1,\dots,F_d$, $\rho_S{(X_1,\dots,X_d)}$ attains its maximum value for a comonotonic random vector.
In order to find  the best-possible lower bound for $\rho_S{(X_1,\dots,X_d)}$ one has to consider the problem of minimizing the expectation of the product of $d$ uniformly distributed random variables, i.e.
\begin{equation}\label{eq:product}
m(\Pi)= \inf \left\{ \mathbb{E} \left[ \prod_{j=1}^d X_i \right]: \; X_i \sim U(0,1), 1 \leq i \leq d \right\}.
\end{equation}

An optimal coupling for $m(\Pi)$ has been found  in \citet{WW11}, where the long history of the problem is also presented. The case of the product of strictly positive uniform random variables $U(a,b)$, with $a>0$, is easier to deal with and analytical results are given in~\citet{BP15s}. For an arbitrary set of marginal distributions $F_1,\dots,F_d$, an analytical computation of the smallest expectation of the product of random variables remains unknown. However, the discretized formulation~\eqref{eq:rgmprob} used in conjunction with the rearrangement algorithm provides a numerical approximation and a discretized image of the optimal set of rearrangements for an arbitrary choice of the marginal distributions under study.

In Figure~\ref{fi:rgnprod1} we provide the set of optimal rearrangements attaining $m(\Pi)$ for the case
of $d=3$ uniform marginals (that is the original case treated in~\citet{WW11}).  The point clouds in these pictures represent a structure of   joint mix for the log-transformed variables, thus an extremal negative dependence.
In Figure~\ref{fi:rgnprod2} we show
optimal rearrangements for the analogous problem with a particular choice of non-identical marginal distributions. It is clearly visible that the optimal dependence structure heavily depends on the given marginals.
This is not true for instance in the case of the maximal expectation of a product, which is always attained by a comonotonic dependence structure (the product function is supermodular).
All the figures contained in this section represent a shuffle of min, implying that all the rearrangements illustrated are one-to-one. Furthermore, in these figures we show only the two rearrangement functions $f_2,f_3$ since we recall that for a set of bijective rearrangements the first one can always be taken as $f_1=\Id$.
In the case of identical marginal distributions it is also possible to take $f_2=f_3$ (see Figure~\ref{fi:rgnprod1}); this is a consequence of Remark~2 in~\citet{GR81}.
Summarizing, the smallest attainable value of the multivariate Spearman's rho is obtained from results based on joint mixability. Hence, it is crucial that a notion of extremal negative dependence serves as a benchmark in the modeling of dependence.
\end{example}

For some particular class of functionals, the domain of~\eqref{eq:rgmmax} can be reduced: this is especially useful when one looks for a numerical solution of $M(c)$. The following theorem is an extension, with a constructive proof, of Proposition~3(c) in~\citet{lR82}. We write $f_j^{\alpha} \rg \Id\vert [\alpha,1]$    to indicate that the function $f_j^{\alpha}:[\alpha,1] \to [\alpha,1]$  is a rearrangement of $\Id\vert [\alpha,1]$ and we denote by $U[\alpha,1]$ the law of a random variable uniformly distributed on $[\alpha,1]$. Similarly, $[\alpha,1]$ can be replaced by $[0,\alpha]$ in the above notation.

\begin{figure}
\begin{center}
\scalebox{0.42}{\includegraphics{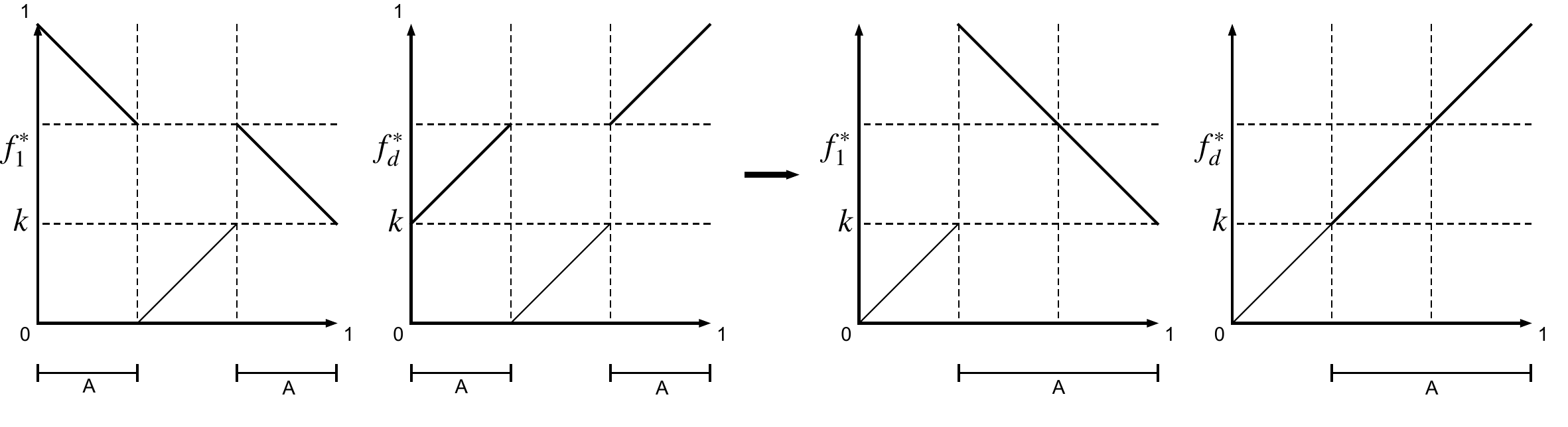}}
\caption{\small The set $A^*$ as defined in~\eqref{eq:A} (left) can always be taken as some interval $[\alpha,1]$ up to a proper rearrangement of the unit interval (right). In this figure we set $g=+$, the sum operator, and $F_j=U[0,1]$, $j=1,\dots,d$.}
\label{fi:111}
\end{center}
\begin{center}
\scalebox{0.42}{\includegraphics{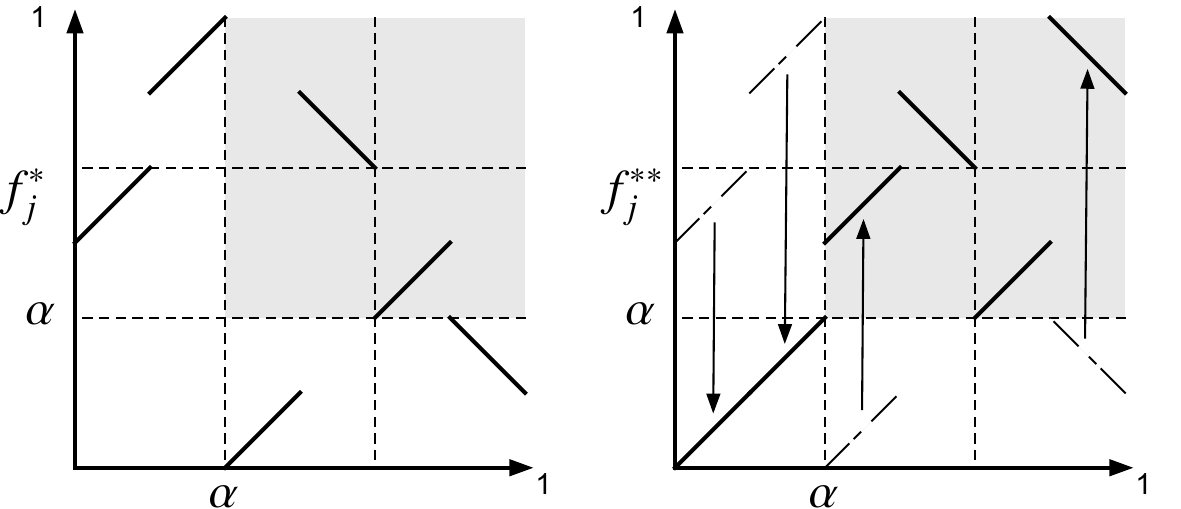}}
\caption{\small A new rearrangement $f_j^{**}:[0,1]\to[0,1]$ can always be obtained from $f_j^*$ so that  $f_j^{**}\vert [\alpha,1] \geq \alpha$ and $f_j^{**}\vert [0,\alpha) \leq \alpha$; see the proof of Theorem~\ref{th:rgma} for the notation used.}
\label{fi:2}
\end{center}
\end{figure}

\begin{theorem}\label{th:rgma}
Suppose that the function $c$ is coordinate-wise increasing and there exists a measurable, coordinate-wise increasing function $g:\R^d \to \R$ such that
\begin{align}\label{eq:first}
c(x_1,\dots,x_d)=c(x_1,\dots,x_d) \cdot \idc\{g(x_1,\dots,x_d) \geq k\},
\end{align}
for some $k \in \R$. If $M(c)$ in~\eqref{eq:rgmmax} is attained by $f_1^*,\dots,f_d^* \rg \Id$, then it holds that
\begin{align}
M(c)= (1-\alpha) \sup &\left\{ \E\left[ c \left(F_1^{-1} \circ f_1^{\alpha}(U^{\alpha}), \dots, F_d^{-1} \circ f_d^{\alpha}(U^{\alpha} )\right)\right]: f_j^{\alpha} \rg \Id\vert [\alpha,1], j=1,\dots,d \right\} \label{eq:rgmamax},
\end{align}
where $U^{\alpha} \laweq U[\alpha,1]$ and $\alpha:=1-P(g(F_1^{-1} \circ f_1^*,\dots,F_d^{-1} \circ f_d^*) \geq k)$.
\end{theorem}
\emph{Proof.}
Let $f_j^* \rg \Id$ be solutions of~\eqref{eq:rgmamax} and define the set
\begin{equation}\label{eq:A}
A^*=\left\{u \in [0,1]: g\left(F_1^{-1} \circ f_1^*(u), \dots, F_d^{-1} \circ f_d^*(u) \right) \geq k \right\}.
\end{equation}
Then, the Lebesgue measure of $A^*$ is $\lambda(A^*)=1-\alpha$ and there exists $f \rg \Id$ such that $A^*=f([\alpha,1])$. Therefore, as illustrated in Figure~\ref{fi:111}, we can assume w.l.o.g. that $A^*=[\alpha,1]$.
\\
\\
\noindent
To prove the $\geq$ inequality in~\eqref{eq:rgmamax} it is sufficient to note that any set of rearrangements $f_1^{\alpha}, \dots, f_d^{\alpha}\rg \Id\vert [\alpha,1]$ can be easily extended to a set of rearrangements $f_1,\dots,f_d \rg \Id$ for instance by
setting
$$
f_j(u):=\begin{cases}
u &\text{ if $u < \alpha$}, \\
f_j^{\alpha}(u) &\text{ if $u \geq \alpha$}. \\
\end{cases}
$$
Optimality of the $f_j^*$'s and~\eqref{eq:first} imply, for all $u \in [0,1]$, that
\begin{align*}
c \left(F_1^{-1} \circ f_1^*(u), \dots, F_d^{-1} \circ f_d^*(u)\right)
= &\,c \left(F_1^{-1} \circ f_1^*(u), \dots, F_d^{-1} \circ f_d^*(u)\right) \times \idc\{u \in [\alpha,1]\}\\
\geq &\,c \left(F_1^{-1} \circ f_1(u), \dots, F_d^{-1} \circ f_d(u)\;\right)\times \idc\{u \in [\alpha,1]\}\\
= &\,c \left(F_1^{-1} \circ f_1^{\alpha}(u), \dots, F_d^{-1} \circ f_d^{\alpha}(u)\right)\times \idc\{u \in [\alpha,1]\}.
\end{align*}

\noindent
To prove the $\leq$ inequality in~\eqref{eq:rgmamax},  for $j=1,\dots,d$, denote$$
A_j^+:=\{u \in [\alpha,1] : f_j^*(u) \geq \alpha \}, \; A_j^-:=[\alpha,1]\setminus A_j^+
$$
and
$$
B_j^+:=\{u \in [0,\alpha) : f_j^*(u) \geq \alpha \}, \; B_j^-:=[0,\alpha)\setminus B_j^+.
$$
If $A_j^-\neq \emptyset$, we can always find a new set of rearrangements $f_1^{**},\dots,f_d^{**}$ such that
\begin{equation}\label{eq:needed1}
  f_j^{**}([\alpha,1])=[\alpha,1], \quad f_j^{**}\vert[\alpha,1] \geq f^*\vert[\alpha,1] \text{ and }  f_j^{**}\vert[0,\alpha] \leq f^*\vert[0,\alpha], j=1,\dots,d.
\end{equation}
An illustration is given in~Figure~\ref{fi:2}. Formally, the functions $f_j^{**}$, $j=1,\dots,d$, are defined in $[\alpha,1]$ as
$$
f_j^{**}(u):=\begin{cases}
\phi_j(f_j^*(u)) &\text{ if $u \in A_j^-$}, \\
\xi_j(f_j^*(u)) &\text{ if $u \in B_j^+$}, \\
f_j^*(u) &\text{ if $u \in A_j^+ \cup B_j^-$}, \\
\end{cases}
$$
where $\phi_j$ denotes the unique \emph{increasing} rearrangement mapping from $C:=f_j^*(A_j^-)$ to $D:=[\alpha,1] \setminus f^*_j(A_j^+)$ and $\xi_j$ denotes the unique \emph{decreasing} rearrangement mapping from $E:=f_j^*(B_j^+)$ to $F:=[0,\alpha) \setminus f^*_j(B_j^-)$; see for instance \citet{rMC95}. This construction of $f_j^{**}$ satisfies~\eqref{eq:needed1}.
Since $f_j^*$ is measure-preserving, it is straightforward to check that $\lambda(C)=\lambda(D)$ and
$\lambda(E)=\lambda(F)$, implying that each $f_j^{**}$ is still a rearrangement of $\Id$.
Moreover, by increasingness of the function $g$ we also have that
\begin{equation*}
\left\{u \in [0,1]: g\left(F_1^{-1} \circ f_1^{**}(u), \dots, F_d^{-1} \circ f_d^{**}(u) \right) \geq k \right\}=[\alpha,1].
\end{equation*}
Finally, the assumptions on $c$ imply, for $u \in [0,1]$, that
\begin{align*}
 c \left(F_1^{-1} \circ f_1^*(u), \dots, F_d^{-1} \circ f_d^*(u)\right)
&= c \left(F_1^{-1} \circ f_1^*(u), \dots, F_d^{-1} \circ f_d^*(u)\right) \times \idc\{u \in [\alpha,1]\}\\
&\leq c \left(F_1^{-1} \circ f_1^{**}(u), \dots, F_d^{-1} \circ f_d^{**}(u)\right) \times \idc\{u \in [\alpha,1]\}\\
&=c \left(F_1^{-1} \circ f_1^{\alpha}(u), \dots, F_d^{-1} \circ f_d^{\alpha}(u)\right)\times \idc\{u \in [\alpha,1]\},
\end{align*}
where $f_j^{\alpha}:=f_j^{**}\vert [\alpha,1]$ is rearrangement of $[\alpha,1]$. \qed

An entirely analogous proof yields the corresponding theorem for $m(c)$.
\begin{theorem}\label{th:rgmamin}
Suppose that the function $c$ is coordinate-wise increasing and there exists a measurable, coordinate-wise increasing function $g:\R^d \to \R$ such that
\begin{align*}
c(x_1,\dots,x_d)=c(x_1,\dots,x_d) \cdot \idc\{g(x_1,\dots,x_d) \leq k\},
\end{align*}
for some $k \in \R$. If $m(c)$ in~\eqref{eq:rgmmax} is attained by $f_1^*,\dots,f_d^* \rg \Id$, then
it holds that
\begin{align}
m(c)=\alpha \inf &\left\{  \E\left[ c \left(F_1^{-1} \circ f_1^{\alpha}(U_{\alpha}), \dots, F_d^{-1} \circ f_d^{\alpha}(U_{\alpha} )\right)\right]: f_j^{\alpha} \rg \Id\vert [0,\alpha], j=1,\dots,d \right\} \label{eq:rgmamin},
\end{align}
where $U_{\alpha} \laweq U[0,\alpha]$ and $\alpha:=P(g(F_1^{-1} \circ f_1^*,\dots,F_d^{-1} \circ f_d^*) \leq k)$.
\end{theorem}

Equations~\eqref{eq:rgmamax} and~\eqref{eq:rgmamin}
are intuitively obvious: to maximize an increasing function which depends only on the right tail of a certain distribution
one should use in each component only the largest part of each marginal distribution. Analogously, if the increasing function to be minimized depends only on the left tail of some distribution one should use in each component only the smallest part of each marginal distribution.

\begin{figure}[h]
\begin{center}
\scalebox{.22}{\includegraphics{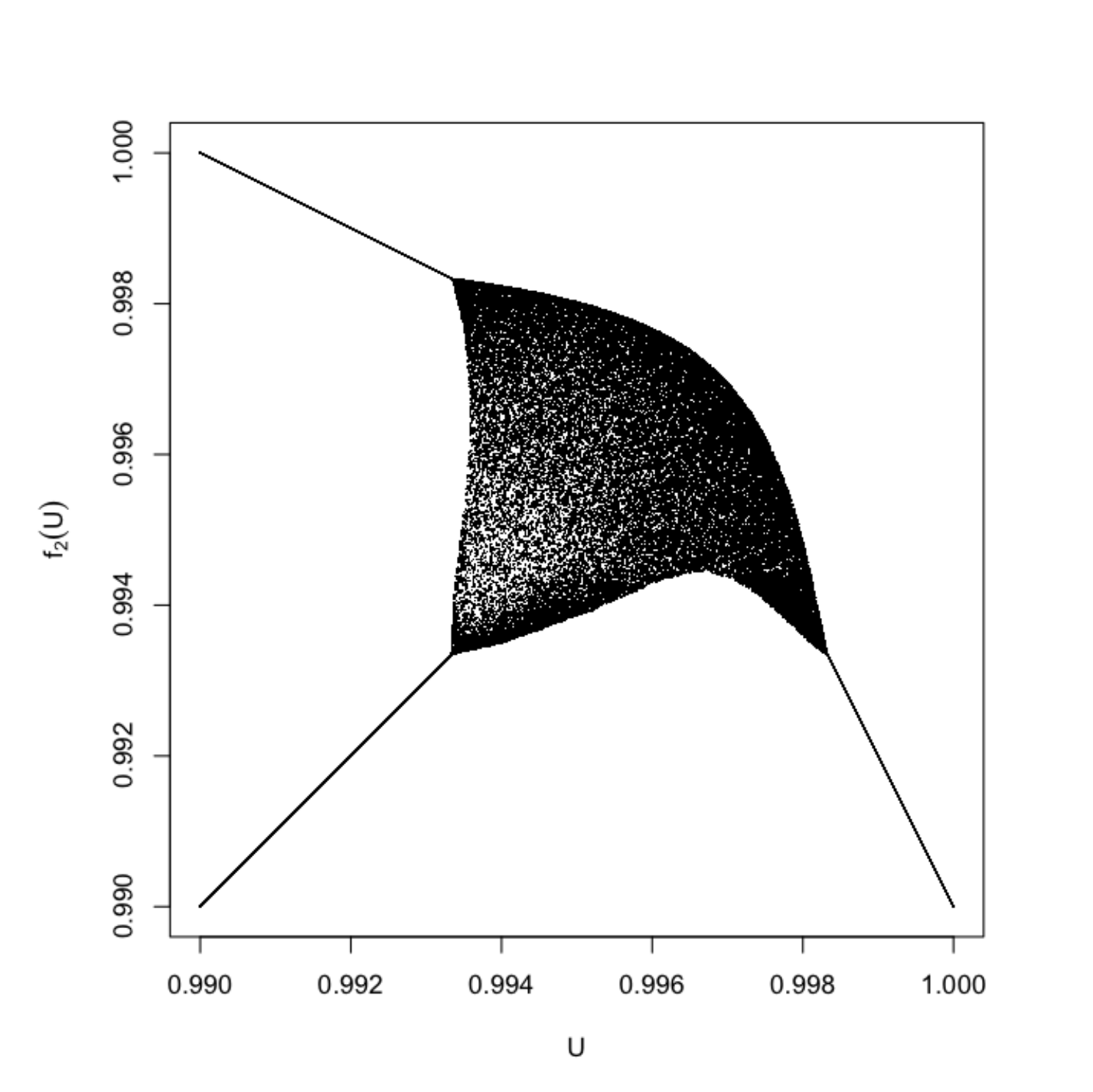}\includegraphics{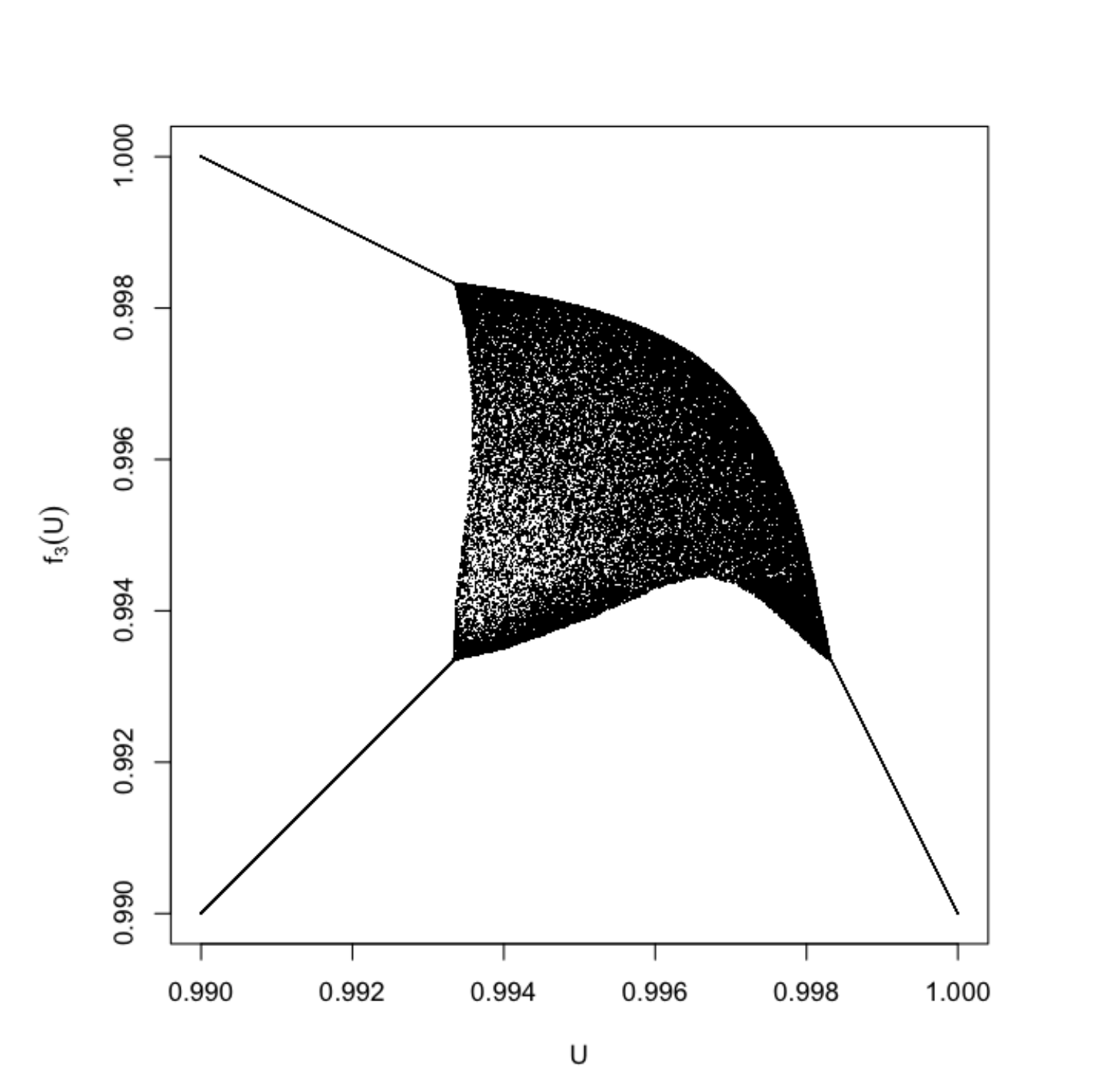}}
\caption{\small Optimal reduced rearrangements $f_2,f_3$ attaining $M(\idc\{\sum_{j=1}^{d} x_j \geq k\})$ with $k\simeq45.99$ (which corresponds to $\alpha=0.99$) $F_j=\text{Pareto(2)},1 \leq j \leq 3$; see the discussion after Theorem~\ref{th:rgmamin}.}
\label{fi:rgnprod3}
\scalebox{.22}{\includegraphics{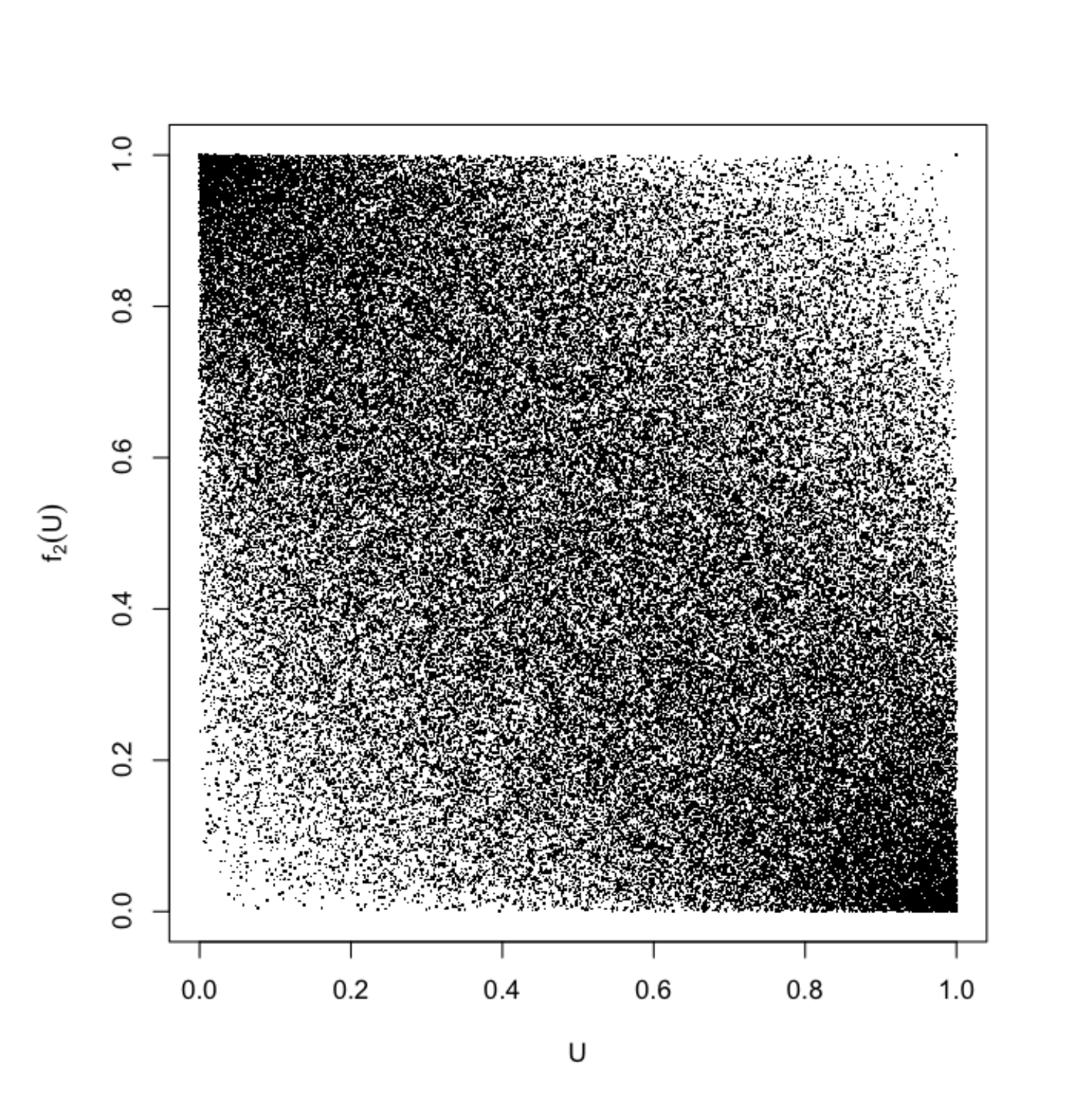}\includegraphics{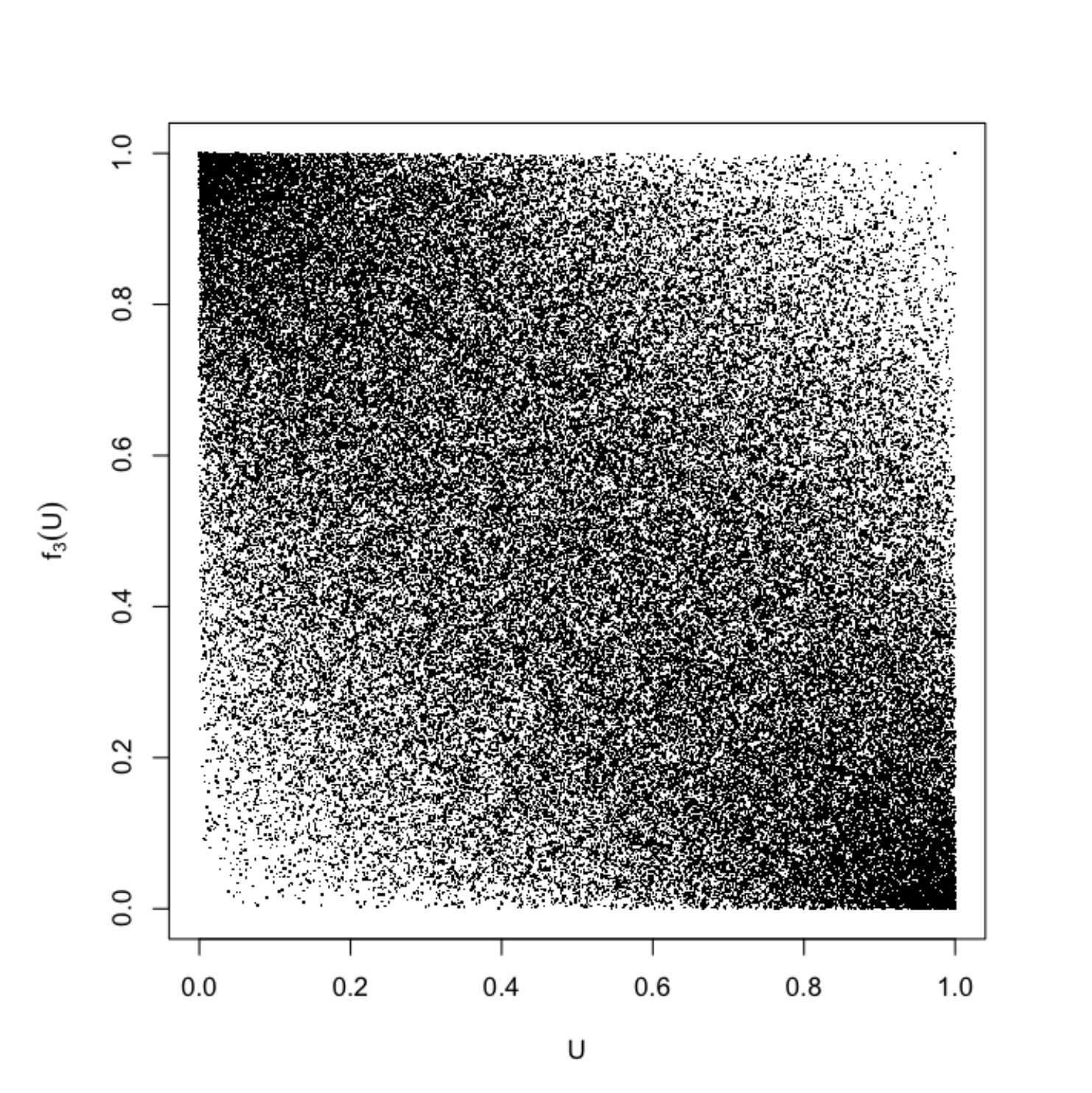}}
\caption{\small A set of rearrangements $f_2,f_3$ under which the product of three Lognormal(0,1) distributions is jointly mixable; see the open question (vi) as discussed in Section~\ref{se:conclusions}.}
\label{fi:rgnprod4}
\end{center}
\end{figure}

\begin{example}[Maximizing the distribution of a sum]
The reduced versions~\eqref{eq:rgmamax} and~\eqref{eq:rgmamin} are relevant for instance when $c(x_1,\dots,x_d):=\idc\{ \sum_{j=1}^{d} x_j \geq k\}, ~k \in \R$. This particular cost function gives lower ($m(g)$)  and upper ($M(g)$) sharp bounds on the distribution of a sum of random variables with given marginals. This problem has a long history.

During one of his walks with students, A.N.~Kolmogorov gave to G. D. Makarov  the problem of finding the lower and upper best-possible bounds on the distribution function of a sum of $d$ random variables with given marginal distributions.
 \citet{gdM81} provided the first result for $d=2$.
 Independently from Makarov's approach,~\citet{lR82} gave  an elegant proof of the same theorem using a dual result proved for a more general purpose.
 The \emph{dual} approach of R\"{u}schendorf was related to a much earlier issue, dating back to 1871: the so-called \emph{Monge mass-transportation problem};
in particular, he solved a special case of its \emph{Kantorovich version}.
A complete analysis of this kind of problem is given in~\citet{RR98}.
 Some years later~\citet{FNS87} restated Makarov's result, using a formulation of the problem based on copulas.
Introducing the use of dependence information, \citet{WD90} gave the best-possible bounds for more general aggregating operators and also in the presence of a lower bound on the copula of a two-dimensional portfolio. The extension of the above results to the case $d>2$ is non-trivial as the lower Fr\'echet bound used by Makarov in the construction of the optimal bivariate solution is not attainable by a distribution function, apart from the case in which the marginal distributions support pairwise countermonotonicity.

The above problem in arbitrary dimension was then attacked using duality theory in~\citet{EP06b}, where improved bounds were found without a sharpness condition.
Finally, the analytical computation of $M(g)$ and $m(g)$ under specific assumptions on marginal distributions has been carried out in~\citet{WPY13}. Makarov's problem has been also solved numerically in total generality using the Rearrangement Algorithm as illustrated in~\citet{EPR13}.
In Figure~\ref{fi:rgnprod3} we show a set of the three reduced rearrangements of $[\alpha,1]$ under which the distribution function of the sum of three Pareto(2) random variables attains its maximum. Again, the point clouds in these pictures represent a   structure of  joint mix, implying that even in the case of one-sided marginal distributions, a relevant part of the optimal dependence structure shows a joint  mix  (negatively dependent) behaviour.
Coherently with Theorem~\ref{th:rgma}, rearrangement functions (and thus the corresponding dependence structure) can be set arbitrarily in the remaining interval $[0,\alpha)$. The optimal dependence structure in Figure~\ref{fi:rgnprod3} has been extensively studied in~\citet[Section~3]{EPR13} and has received considerable interest in the computation of bounds on risk measures in quantitative risk management; see \cite{EPRWB14}.
 Note that Theorem~\ref{th:rgma} is particularly useful for determining bounds on any functional depending on the upper tail of the distribution of the sum; this includes a variety of \emph{risk measures} in quantitative risk management. This example clearly shows that  the maximization and minimization of a non-supermodular function call for the notion of extremal negative dependence.
\end{example}

\section{Conclusions and future research directions}\label{se:conclusions}

The dependence relationship between two or more random variables can be described using the equivalent concepts of a copula or a set of rearrangements of the identity function. Through the lens of rearrangements, this paper reviews the concept and the history of the notion of extremal positive dependence (also called comonotonicity) and surveys the various concepts available of extremal negative dependence, proposing a novel unifying notion in higher dimensions.

A  natural notion of extremal negative dependence, called countermonotonicity, is available
for bivariate random vectors. A countermonotonic vector always exists, all countermonotonic vectors have the same dependence structure and are the minimizer of the expectation of supermodular functions.

Unfortunately it is not possible to define a negative dependence concept extending all these relevant properties
to $d$-variate vectors of arbitrary dimensions. In dimensions $d>2$, different concepts of negative dependence arise:
pairwise countermonotonicity, joint mixability and $\Sigma$-countermonotonicity. These latter notions are all marginally dependent, that is, for different marginal distributions, the copula of a pairwise countermonotonic random vector/$\Sigma$-countermonotonic random vector/joint mix is not unique in general. A related challenge is that no universal solution exists for many optimization problems as for instance the minimum for a convex function of the sum of the pre-assigned marginal components of a random vector.  We believe that this is exactly the reason why more research is needed in the field of extremal negative dependence.

There are still quite a few existing open mathematical questions about extremal dependence concepts, especially concerning extremal negative dependence.   We would like to invite the interested reader to contribute to the following questions.

\begin{enumerate}[(i)]
\item \textbf{Characterization of complete/joint mixability.} Despite of some recent significant progress, a full characterization of complete/joint mixability is still open. In particular, even in the homogeneous case, necessary and sufficient conditions for complete mixability of bounded unimodal distributions is a long standing open problem. It was observed that the conditions in Proposition \ref{th:jointly mixable1} are not sufficient for such classes; see for example some numerical verifications given in \cite{PW14}. The question regarding the uniqueness of the center of a set of  $d$ jointly mixable distributions with infinite first moments is also open.
\item \textbf{Existence of the $\cx$-smallest element in a Fr\'echet class.}
A small  modification of the counterexample in Section 3 of \cite{BJW13} yields that $\FR$ may not have a  $\cx$-smallest element even  when the marginal distributions are assumed to be continuous. At the moment we do not have a clear picture of what conditions are required for the existence of a $\cx$-smallest element in a Fr\'echet class.
\item  \textbf{General solutions of $M(c)$ and $m(c)$ for non-supermodular functions $c$.}  The case where $c(x_1,\dots,x_d)=\idc\{\sum_{i=1}^d x_i\le k\}$ is only partially solved based on the idea of complete/joint mixability, as discussed, for instance, in \cite{WPY13} and \cite{PR13}. For more general $c$ the problem becomes
\begin{align*}
\hat M(\psi):= \sup\{\psi(F) :F \in \FR \},
\end{align*}
where $\psi$ is a functional which maps the set  of $d$-joint distributions to real numbers. Such $\psi$ can be interpreted as a multivariate law-determined risk measure in the context of  finance and insurance. A univariate risk measure of the sum $X_1+\dots+X_d$ is a special choice of $\psi$; see \cite{EPRWB14} for a review concerning Value-at-Risk and Expected Shortfall.

\item \textbf{Stronger extremal negative dependence concepts.} Is there a notion of extremal negative dependence which is stronger than $\Sigma$-countermonotonicity but yet is supported by all Fr\'echet classes? Till now we were not able to find stronger, reasonable concepts.

\item \textbf{Random sequences and asymptotic analysis.} The discussions on extremal dependence concepts can be naturally generalized from random vectors to random sequences. One attempt to deal with this type of question is given in~\cite{WW13} where a notion of extremal negative dependence for sequences was proposed.  Other alternative formulations of extremally negatively dependent sequences are possible, and much research is still needed, especially in the case when the marginal distributions are not identical.

\item \textbf{Different aggregating functionals.} The definition of joint mixability and $\Sigma$-countermonotonicity rely on the sum operator chosen as the aggregating functional. The extension of the concept of joint mixability and $\Sigma$-countermonotonicity to different aggregating functionals as a research problem  needs further investigation. As an illustration, in  Figure~\ref{fi:rgnprod4} we show a set of three rearrangements under which the product of three Lognormal distributions is jointly mixable.
A first step in this direction can be found in~\citet{BP15s}.

\item \textbf{Optimization problems and constrained optimization problems.} The optimization problems mentioned in Section \ref{sec:4} have important applications in operations research; see for instance \cite{uH15}. There are many theoretical as well as numerical challenges left with those optimization problems.  In particular, the problems in Section \ref{sec:4} are \emph{unconstrained} in the sense that all elements in $\FR$ are counted. However one may have more constraints than just in the margins. For the case of having an extra  variance constraint in the financial risk management context, see \cite{BRV13}.

\end{enumerate}

\setlength{\bibsep}{-1pt}
\bibliographystyle{chicago}
\footnotesize
\bibliography{mybib}
\end{document}